\documentclass[final,1p,times]{elsarticle}

\usepackage{graphicx}
\usepackage{amssymb}
\usepackage{amsmath}
\usepackage{amsthm}

\usepackage[english]{babel}
\usepackage{bm}
\usepackage[intoc]{nomencl}
\let\abk\nomenclature
\makenomenclature

\usepackage{color}

\usepackage[colorlinks=true,urlcolor=darkblue,citecolor=blue,hyperfootnotes=false]{hyperref}

\def\subsubsubsection#1{\paragraph{#1}}

\newcommand{\rhototp}{\rho_\text{tot}^+}
\newcommand{\rhototm}{\rho_\text{tot}^-}
\newcommand{\rhotot}{\rho_\text{tot}}
\newcommand{\kd}{k_\text{d}}
\newcommand{\kp}{k_\text{p}}

\newcommand{\omd}{\omega_\text{d}}
\newcommand{\oma}{\omega_\text{a}}
\newcommand{\Omd}{\Omega_\text{d}}
\newcommand{\Oma}{\Omega_\text{a}}

\newcommand{\Ntot}{N_\text{tot}}
\newcommand{\Nun}{N_\text{u}}
\newcommand{\Nbo}{N_\text{b}}

\newcommand{\rhobp}{\rho_\text{b}^+}

\newcommand{\rhobm}{\rho_\text{b}^-}
\newcommand{\rhoupm}{\rho_\text{u}^\pm}
\newcommand{\rhobpm}{\rho_\text{b}^\pm}

\newcommand{\Jb}{J_\text{b}}
\newcommand{\qc}{q_\text{crit}}
\newcommand{\alphaeff}{\alpha_\text{eff}}
\newcommand{\LCl}{L_\text{Cl}}

\newcommand{\D}{d}

\newcommand{\wdu}[0]{\ensuremath{\omega^{\uparrow}}}
\newcommand{\wud}[0]{\ensuremath{\omega^{\downarrow}}}

\newcommand{\rhou}[0]{\ensuremath{\rho^{\uparrow}}}
\newcommand{\rhod}[0]{\ensuremath{\rho^{\downarrow}}}
\newcommand{\jdu}[0]{\ensuremath{h^{\uparrow}}}
\newcommand{\jud}[0]{\ensuremath{h^{\downarrow}}}
\newcommand{\jnet}[0]{\ensuremath{h^{net}}}
\newcommand{\ju}[0]{\ensuremath{j^{\uparrow}}}
\newcommand{\jd}[0]{\ensuremath{j^{\downarrow}}}
\newcommand{\pu}[0]{\ensuremath{p^{\uparrow}}}
\newcommand{\pd}[0]{\ensuremath{p^{\downarrow}}}

\newcommand{\nug}{\nu_\text{g}}
\newcommand{\nus}{\nu_\text{s}}
\newcommand{\nur}{\nu_\text{r}}
\newcommand{\nurr}{\tilde\nu_\text{r}}
\newcommand{\nuc}{\nu_\text{c}}
\newcommand{\nud}{\nu_\text{d}}

\journal{Physics Reports}

\begin{document}

\begin{frontmatter}



\title{Intracellular transport driven by cytoskeletal
motors~:\\
General mechanisms and defects}

\author[label1,label2]{C. Appert-Rolland}
\ead{Cecile.Appert-Rolland@th.u-psud.fr}
\author[label1,label2,label3]{M. Ebbinghaus}
\ead{ebbinghaus@lusi.uni-sb.de}
\author[label3]{L. Santen}
\ead{l.santen@mx.uni-saarland.de}

\address[label1]{Univ. Paris-Sud, Laboratoire de Physique Th\'eorique, B\^at. 210, F-91405 Orsay Cedex, France}
\address[label2]{CNRS, LPT, UMR 8627, B\^at 210, F-91405 Orsay Cedex, France}
\address[label3]{Fachrichtung Theoretische Physik, Universit\"at des Saarlandes, D-66123 Saarbr\"ucken, Germany}

\today

\begin{abstract}
Cells are the elementary units of living organisms,
which are able to carry out many vital functions.
These functions rely on active processes on a
microscopic scale.
Therefore, they are strongly out-of-equilibrium systems,
which are driven by continuous energy supply.
The tasks that have to be performed in order
to maintain the cell alive require 
transportation of various ingredients,
some being small, others being large.
Intracellular transport processes are able to induce
concentration gradients and to carry objects to specific
targets. These processes cannot be carried out only by
diffusion, as cells may be crowded, and quite elongated on
molecular scales. 
Therefore active transport has to be
organized.

The cytoskeleton,
which is composed of three types of filaments
(microtubules, actin and intermediate filaments),
determines the shape of the cell, and
plays a role in cell motion.
It also serves as a road network
for a special kind of vehicles, namely the
cytoskeletal motors.
These molecules can attach to a cytoskeletal
filament, perform directed motion,
possibly carrying along some cargo, and then
detach.

It is a central issue to understand how
intracellular transport driven by molecular motors is
regulated.
The interest for this type of question
was enhanced when it was discovered that
intracellular transport breakdown
is one of the signatures of 
some neuronal diseases like the Alzheimer.

We give a survey of the
current knowledge on microtubule
based intracellular transport.
Our review includes on the one hand
an overview of biological
facts, obtained from experiments, and
on the other hand a presentation of some modeling
attempts based on cellular automata.
We present some background knowledge on the
original and variants of the TASEP (Totally
Asymmetric Simple Exclusion Process),
before turning to more application oriented
models.
After addressing
microtubule based transport in general,
with a focus on \emph{in vitro} experiments,
and on cooperative effects in the transportation
of large cargos by multiple motors,
we concentrate on axonal transport, because of its
relevance for neuronal diseases.
Some important characteristics of axonal transport
is that it takes place in a confined environment;
Besides several types of motors are involved,
that move in opposite directions. It is a challenge
to understand how this bidirectional transport is
organized.
We review several features that could contribute to
the efficiency of bidirectional transport in the axon,
including in particular the role of
motor-motor interactions and of the dynamics of the
underlying microtubule network.
Finally, we also discuss some open questions that may be
relevant for future research in this field.
\end{abstract}

\begin{keyword}
Stochastic transport, intracellular transport, exclusion processes, molecular motors, cytoskeleton, dynamic networks
\end{keyword}

\end{frontmatter}
\newpage

\tableofcontents

\newpage

\begin{center}

{\LARGE Acronyms}

\end{center}

\begin{tabular}{ll}
ADP & Adenosine diphosphate\\
ASEP & Asymmetric simple exclusion process\\
ATP & Adenosine triphosphate\\
CLIP & Cytoplasmic linker protein \\
DNA & Deoxyribonucleic acid \\
DW & Domain wall \\
EB1 & End-binding protein 1\\
EE & Early endosomes \\
GDP & Guanosine diphosphate\\
GFP & Green fluorescent protein\\
GTP & Guanosine triphosphate\\
$\gamma$TuRC & $\gamma$-tubulin-containing ring complex \\
HD & High Density phase \\
LD & Low Density phase \\
MAP & Microtubule-associated protein\\
MC & Maximal Current\\
mRNA & Messenger ribonucleic acid \\
MT & Microtubule\\
MTOC & Microtubule-organizing center\\
SEP & Symmetric simple exclusion process\\
TASEP & Totally asymmetric simple exclusion process\\
+TIP & Microtubule plus-end tracking protein \\
\end{tabular}

\newpage

\section{Introduction}
\label{sec:intro}

The cell is a perfect
example of a naturally arisen complex system. In the
course of evolution, a large number of biochemical
processes assembled forming the smallest of any living
organisms' subunits. As evidenced by the immense
diversity of life forms on earth, there exist more than
one possible combination of these processes fulfilling
the basic criteria for life such as reproduction and the
existence of a metabolism~\cite{mckay2004}. The metabolism
is the sum of processes which maintain the cell out of
equilibrium or -- to put it in the words of Erwin
Schr\"odinger~\cite{schroedinger1992} -- ``It feeds on
negative entropy.'' In fact, reproduction involves the
transmission of the genetic information stored in the
cell's DNA and as is well known from thermodynamics,
information transmission is impossible in thermal
equilibrium implying that free energy must constantly be
imported into the cell. Using the energy directly or
indirectly derived from solar radiation to keep the cell
in a non-equilibrium state therefore prevents an organism
from dying.

The sizes of cells are typically in the micrometer range
with eukaryotic cells being larger than prokaryotic ones.
The larger size of eukaryotes comes along with a
specialization of the various regions of the cell.
The spatially
distant organelles have to be connected in order to
create the complete metabolic chain~\cite{alberts2008}.
This explains the presence of a cytoskeleton in almost
all eukaryotes. The cytoplasm being rather crowded,
diffusion turns out to be a highly inefficient medium for
transportation. Instead, active intracellular transport
is necessary. Molecular motors walk
along filaments of the cytoskeleton while carrying
intracellular cargo of various types
from the site of supply to the site
of demand~\cite{schliwa2003}.
This type of intracellular transport
is crucial for the cell's survival and
makes the spatial organization possible in the first
place.
In this review, we
are particularly interested in axonal transport. 
Axons have a quasi-one-dimensional structure with a dense
microtubule network, which is a challenging environment
for bidirectional motor driven transport.

Different diseases in multicellular organisms, amongst
them human neuronal diseases, can be attributed to the
failure of intracellular transport~\cite{schliwa_w2003}.
Understanding how the transport works is thus a necessary
precondition for understanding why it fails.
Statistical physics can be a relevant tool to tackle
these questions. Indeed, intracellular transport
is a strongly out-of-equilibrium system, where permanent
currents take place, and thus meets the increasing
interest for understanding how a thermodynamics of
non-equilibrium systems could be built.
Besides, intracellular transport involves a great number
of interactions, some of which
still unknown, prohibiting a detailed modeling.
Eventually fluctuations play an important role
in the dynamics of the various components of the cell.
Statistical physics provides tools that allow one to
deal with aggregated information and with influential
fluctuations, hence rendering it
particularly suited for the analysis of the
aforementioned complex systems.
It already proved to be helpful identifying the
elements most crucial for the system's behavior,
as we will illustrate in this review.

In this review, we shall first present
the actors involved in microtubule transport.
We shall start with the ``roads'', namely 
microtubules and actin filaments (section~\ref{sec:fildyn}),
and continue with the ``vehicles'' by presenting the
most common molecular motors (section~\ref{sec:motors}).
In both chapters, a survey of the biological background and
of some modelling attempts will be given.

As exclusion processes will turn out to be useful tools
to understand collective effects,
their definition and main properties will be summerized
in section~\ref{sec:ep}.
Some variants that progressively include features
relevant for intracellular transport (multi-lane traffic,
static or dynamical defects) will be introduced
in section~\ref{sec:variants}.

Langmuir kinetics, i.e. the ability of the motors
to attach and detach from the filaments, will be
considered in section~\ref{sec:unidir}.
The models considering only uni-directional transport 
are not relevant for intracellular transport
but still can describe some \emph{in vitro}
experiments called motility assays,
which usually involve only one type of motors.
Indeed, some models based on exclusion processes
were successful
in explaining some jamming phenomena
observed experimentally in such motility assays.
Another characteristics of these \emph{in vitro}
experiments is that motors that detach from
the filament can freely diffuse in the surrounding
solution.
Confinement around the microtubule
becomes an important feature when \emph{in vivo}
transport is considered.
Section \ref{sec:unidir} shows how general classes of boundary
conditions and/or geometries of the surrounding
diffusion reservoir can modify the transport properties
in exclusion processes models.

In section~\ref{sec:vivo} we shall concentrate on axonal
transport,
which is of
special interest as its failure plays a role in several
neuronal diseases~\cite{roy2005}.
We shall present some modeling attempts to take
its specificities into account.
In particular, we shall explore the role
of microtubule dynamics and motor-motor interactions
in the global organization of microtubule-based
transport.

We shall see that transport in the cell is far from being
understood, and that both experimental and modeling
efforts are needed.
In Section \ref{sec:challenges}, we shall review some other
key ingredients for the efficiency of intracellular transport
and mention some of the numerous open questions that
still require investigation.

\section{Cytoskeleton filaments}
\label{sec:fildyn}

The cytoskeleton carries out different fundamental functions in biological 
cells \cite{alberts2008}. The cytoskeleton determines the shape of the cell and its spatial organization, 
it generates forces and serves as a "road network" for molecular motors. 
Although the structure of the cytoskeleton is quite complex and highly dynamic only 
three types of cytoskeletal polymers have been identified: actin filaments, microtubules, 
and intermediate filaments. We restrict our discussion to actin filaments and MTs since they 
carry the motor-driven transport on the cytoskeleton.  

Microtubules (MTs) are the stiffest cytoskeletal filaments and have the most complex assembly and 
disassembly dynamics as we will discuss in more detail below. MTs can span distances that are of 
the same order as the diameter of a cell. This underscores its importance as a track for long distance 
intracellular transport, which is carried out by molecular motors of the kinesin and dynein family. The 
microtubule network is structured by microtubule organizing centers (MTOC) from which the MTs are 
growing outwards. Despite its structural stability the MT-network is reorganizing itself on timescales 
that are relevant for intracellular transport. Therefore, the structure of the MT-network as well as its 
dynamics has to be considered for the analysis of motor-driven intracellular transport.  

Compared to MTs, no structure equivalent to MTOCs exists for
actin filaments. Besides, actin filaments are much more
flexible. Despite this, the 
actin network plays an important role in force generation and cell locomotion. The ability to generate forces 
on the cell membrane is related to the high density of cross linkers that promote the assembly of e.g. actin filament 
bundles, and branched networks. Also the actin filaments are very dynamic objects, which for example 
advance the leading edge of migrating cells. 

From a physical perspective the dynamics of the cytoskeleton is a process that exhibits generic non-equilibrium behavior. 
This property is intimately connected with the ability of biological cells to form complex dynamic structures, as for example in 
the course of cell division. These processes have been analyzed theoretically by macroscopic models, that describe 
the cytoskeleton as an active gel (see e.g. \cite{julicher2007} for a review). 
In this section we will not discuss the mechanical properties and the spatial organization of the cytoskeleton but rather 
focus on the structure and dynamics of single MTs and actin filaments.

\subsection{Microtubules}
\label{sec:mt}

The design of microtubules must respond
to some {\em a priori} contradictory constraints~\cite{alberts2008}:
On the one hand, their structure must resist to thermal
fluctuations and be able to sustain mechanical stress.
On the other hand, they must be able to adapt rapidly to
changes (of shape for example) and to respond to the cell
needs in a flexible way. These constraints lead to a quite complex 
structure of microtubules, which we will discuss in more detail:
Microtubules are polymers whose fundamental subunit is a
$\alpha$/$\beta$-tubulin heterodimer of 8~nm
length~\cite{alberts2008}. Each heterodimer is made of
a $\alpha$-tubulin and a
$\beta$-tubulin, which are both globular proteins that
differ slightly.

\begin{figure}[htbp]
 \begin{center}
  \includegraphics[width=0.9\linewidth]{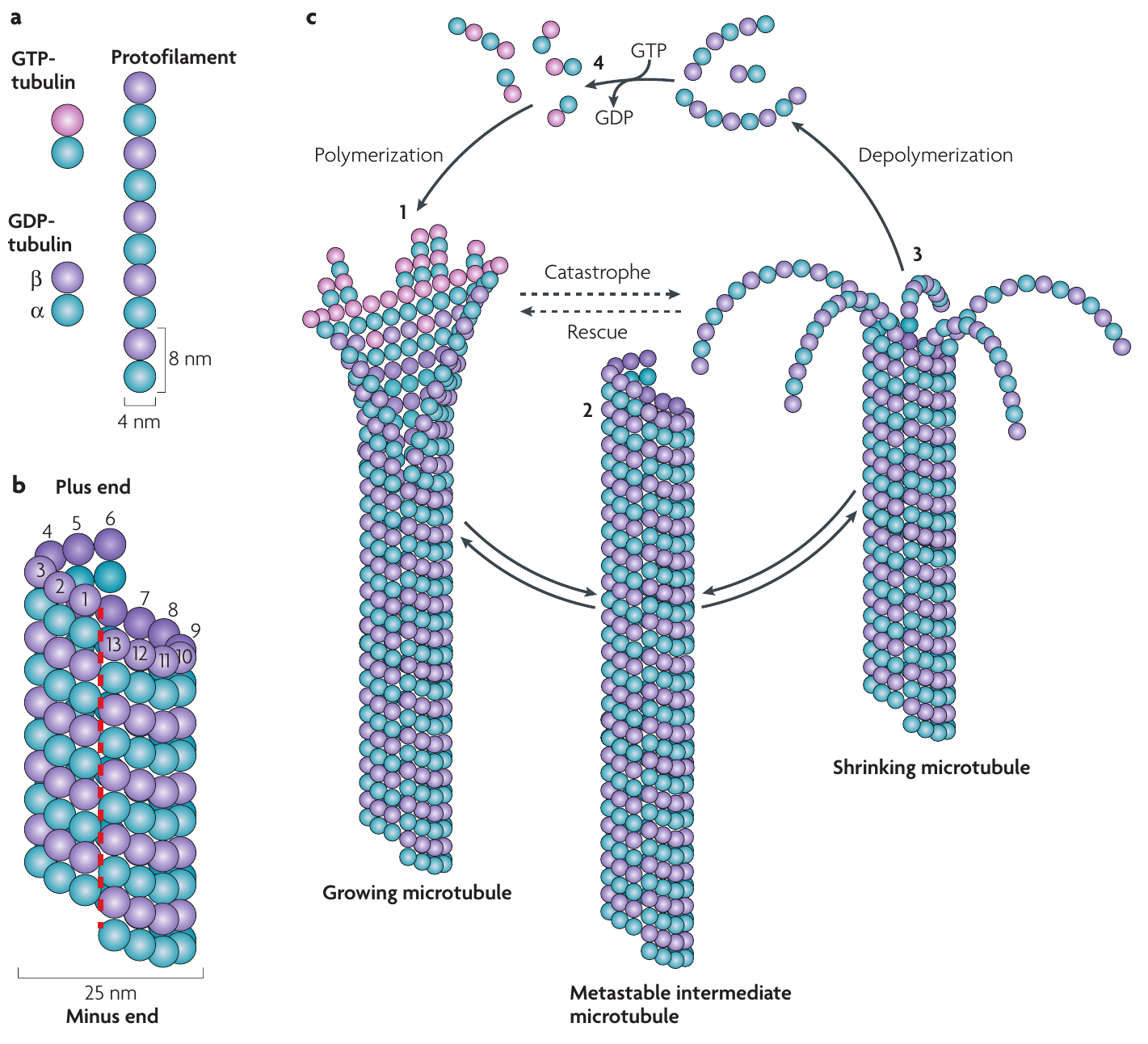}
\caption[Microtubule structure and dynamic
instability.]{Microtubule structure and dynamic instability.
{\it (A)} The $\alpha$/$\beta$-tubulin heterodimers, either
associated to GTP or GDP, assemble into protofilaments of 8~nm
periodicity. {\it (B)} About thirteen protofilament helically
wrap into a hollow cylinder. The helical pitch is 12~nm and
therefore longer than the periodicity of the MT lattice in
axial direction, resulting in a seam (red dashed line). {\it
(C)} MTs switch between phases of growth and shrinkage. A
transition from the polymerizing to the depolymerizing state
is called \emph{catastrophe} whereas the inverse transition is
called \emph{rescue}. 1. In the growing state, the plus-end is
in the form of an open sheet with GTP-bound tubulin at its
end. At further growth, the MT closes along the seam and the
GTP in the bulk of the lattice hydrolyzes to GDP. 2. The
existence of a metastable state is possible in which the MT is
completely closed and pauses. 3. The protofilaments have lost
their lateral contacts at the plus-ends and peel off. 4.
Tubulin dimers or small oligomers have their GDP replaced by
GTP in solution and are ready to be reincorporated into
another growing MT.
From~\cite{akhmanova_s2008}.
}
\protect\label{fig:MT_structure}
\end{center}
\end{figure}

The tubulin dimers are able to spontaneously assemble along
their long axis into protofilaments
(figure~\ref{fig:MT_structure}~{\it A}). Usually twelve to
fourteen of these protofilaments wrap into a helical cylinder
of approximately 25~nm diameter~\cite{akhmanova_s2008} which is
then called microtubule. The helical pitch adds up to 12~nm,
resulting in an imperfect wrapping and a distinguishable seam
in the surface lattice (figure~\ref{fig:MT_structure}~{\it
B}).
Due to the asymmetry of the dimer subunit, MTs are locally
polar but also as a whole in the sense that the two ends of
the filament are chemically different and can therefore be
distinguished. We shall see that this polarity will be
an important feature for intracellular transport, as it
determines the walking direction of molecular motors.
The common convention is to define the filament
end that exposes $\beta$-tubulin to be the plus-end of the
MT.

Microtubules exhibit a characteristic dynamic behavior, termed \emph{dynamic instability}, characterized by a stochastic
switching between phases of persistent growth (=~polymerization) or fast shrinkage
(=~depolymerization)~\cite{mitchison_k1984a}.
A transition from growth
to shrinkage is called \emph{catastrophe} and a transition
from shrinkage to growth is called \emph{rescue}.

\emph{In vitro}, early experiments indicated that both transition rates are independent 
of the time spent in the respective states~\cite{walker1988}.
Furthermore, the speed of depolymerization does not depend on
the concentration of tubulin in solution, whereas the speed of
polymerization on the other hand depends linearly on the
amount of tubulin~\cite{walker1988}. The rescue and catastrophe
frequencies \emph{in vivo} are usually rather low such that
MTs exhibit persistent growth and shortening on the typical
length scales of the cell diameter~\cite{komarova_v_b2002}. Therefore, 
the interaction between MT and the cell boundary is strongly influencing 
the MT dynamics \cite{ebbinghaus_s2011}.

Growth at the plus-end occurs in form of an open sheet
structure to which single tubulin dimers or small oligomers
are added (figure~\ref{fig:MT_structure}~{\it C}). Shrinking
MTs have single protofilaments curling outward which peel off
the filament and eventually detach from it. 
A highly idealized view of MT dynamics can be seen in the
animated cartoon \cite{viel_l_l}.

The plus- and minus-ends of microtubules obey different
kinetics~\cite{walker_i_s1989}.
The plus-end is much more dynamic that the minus-end.
Besides, on average, a minus-end, if not anchored, is depolymerizing while a
plus-end is polymerizing.
In most \emph{in vivo} systems, the minus-end of the MT is
attached to a microtubule-organizing center
(MTOC) also called centrosome, inhibiting any
dynamics at the minus-end. Then all the dynamical
activity occurs at the plus-end.

\emph{In vitro}, it is possible to observe MTs that are not anchored. As a result of the different kinetics
at the plus- and minus-ends, one may observe a very special dynamic regime of the MTs (which 
is actually much more often observed for actin filaments)
called \emph{treadmilling}~\cite{komarova_v_b2002}: A treadmilling filament polymerizes at its one end while it
depolymerizes at the other, thus moving in one direction as a whole filament while every single subunit is
immobile. Though this is not the general case, there are
also some \emph{in vivo} systems for which MTs are not anchored
and for which treadmilling can be observed. It was the case
for fish melanophores fragments~\cite{rodionov_b1997} and for
CHO and NRK cultured cell lines~\cite{komarova_v_b2002}.

In order to understand further the dynamical instability of MTs, it is necessary to mention the crucial role played in this process
by some nucleotides, the GTP (guanosine triphosphate) and GDP (guanosine diphosphate).
Free tubulin dimers are included into the MT
with GTP molecules bound to both the $\alpha$- and $\beta$-tubulin.
After polymerization, the GTP bound to the $\beta$-tubulin hydrolyzes to
GDP~\cite{david-pfeuty_e_p1977}. Therefore one observes a so-called \emph{GTP cap} 
at the growing plus-end of the microtubule, while further along the microtubule,
most $\beta$-tubulins carry a GDP molecule.
Still, `islands' of GTP can be found inside polymerized MTs
\emph{in vivo}~\cite{dimitrov2008}. The relatively long lifetimes
of these islands indicate that the
stochastic process of GTP hydrolysis occurs on different time
scales depending on whether the associated tubulin subunit is
exposed at the plus-end or whether it is buried in the MT lattice.

It turns out that GDP-tubulin subunits depolymerize
about 100 times faster than GTP subunits~\cite{alberts2008}.
The presence of a GTP cap stabilizes the plus-end of the microtubule
and stops rapidly most starting depolymerization
events~\cite{mitchison_k1984a}.
However, if by chance this protecting GTP cap disappears, the
remaining part of the microtubule can depolymerize very rapidly:
it is the aforementioned catastrophe.
Depolymerization may end when an island of GTP (a GTP remnant
inside the MT) is reached~\cite{dimitrov2008},
inducing rescue, and a new episode of growth.

From a general point of view, it is clear that the unstable nature of MT dynamics
makes it very sensitive to external factors and allows the cell to fine tune it as needed.
Apart from the role of the GTP/GDP nucleotides, MT dynamics is influenced by a wide range of microtubule-associated proteins .
\footnote{Some post-translational modifications occuring on tubulin
also play a role in specializing the microtubules into a given function.
We refer the reader interested in this last point
to a review on the subject~\cite{janke_b2011}.}

So far we focussed mainly on the (de-)polymerization of existing microtubules. 
However, as microtubules may completely depolymerize there is a need
to also nucleate new microtubules. Microtubules are able to assemble spontaneously \emph{in vitro}
in the presence of soluble tubulin and GTP~\cite{kirschner1978}. However, nucleation occurs on
time scales larger than the subsequent polymerization. It is thus the limiting step in the
formation of new microtubules~\cite{alberts2008}.

Controlling nucleation allows the cell to decide where and when MTs will grow.
\emph{In vivo}, some proteins may favor nucleation at specific
locations: In many eucaryotic cells, the minus-ends of MTs are
therefore located at a MTOC, called 
centrosome~\cite{brinkley1985,desai1997}.
A third tubulin subspecies, $\gamma$-tubulin, was shown to form a
$\gamma$-tubulin-containing ring complex ($\gamma$TuRC)
which is highly implicated in MT
nucleation~\cite{oakley1989,oakley1994,zheng1995}. This complex is formed by
approximately 13 $\gamma$-tubulin subunits which form a ring on the centrosome
that has almost the same diameter as a MT. The ring is not completely closed
but exhibits a helical pitch, which is also recovered in the polymerized MT
structure. Hundreds of $\gamma$TuRC are
found on centrosomes \emph{in vivo} where they cap the minus-end of single MTs
which nucleated from these ring complexes. They thus prevent the addition or
loss of tubulin subunits at the minus-end of nucleated
MTs~\cite{zheng1995,moritz1995}.

\subsection{Regulation of the microtubule dynamics}
\label{sec:maps}

A large number of proteins interact with the MT and are consequently classed
in the broad family of microtubule-associated proteins (MAPs). These MAPs
are very diverse in structure and function and interact very differently
with MTs. While some modify the dynamics of MTs~\cite{schlager_h2009},
others do not exert an
influence on the MTs but rather need the MT lattice to carry out their
function. An exhaustive review of MAPs is far beyond the scope of this review.
Here, we shall only mention a few of them in order to illustrate typical functions of 
MAPs which influence the microtubule dynamics.  

Certain MAPs, commonly referred to as +TIPs, are able to track the (growing) plus-ends of MTs. 
A non-negligible part of the research interest in +TIPs arises due to the fact that these proteins are 
highly conserved over large evolutionary distances and because of the question how the members 
of this protein class are able to find the end of the MT~\cite{schuyler_p2001,galjart_p2003,carvalho_t_p2003,morrison2007,akhmanova_h2005,akhmanova_s2008}. 
Several mechanisms for plus-end tracking have been attributed to different +TIPs as for example~\cite{akhmanova_s2008}:

\begin{enumerate}
\item A +TIP might be able to bind to a specific structure at the plus-end of MTs. Since a growing MT has the shape of an open sheet at its end (see figure~\ref{fig:MT_structure}), it is conceivable that interactions with MAPs are different in this part compared to the bulk of the MT lattice where the sheet has already closed.
\item In a very similar way to the previous mechanism, a +TIP might have a high affinity for a feature at the plus-end but be nevertheless able to bind to the bulk of the MT lattice on which the +TIP performs two-dimensional diffusion until it arrives at the plus-end.
\item As we shall see in section~\ref{sec:kinesin}, some family
(kinesin) of molecular motors move along the MT
in the direction of the plus-end of the MT.
These motors can thus carry +TIPs to the plus-end.
\item +TIPs might be added to the MT through a co-polymerization mechanism, i.e., tubulin dimers or oligomers associate in solution with the +TIP before these tubulin subunits are added to the filament.
\item Some +TIPs might themselves not be able to find the plus-end but bind to another +TIP which is able to do so, a scenario which is termed \emph{hitchhiking}.
\end{enumerate}

Specific +TIPs have been found for each of these mechanisms and consequently, no single mechanism for plus-end tracking can be identified. The recognition of a specific structure at the plus-end and the co-polymerization mechanism (scenarios 1 and 3) have in common that they lead to treadmilling of +TIPs: The +TIPs themselves do not move but are added at the tip and then released deeper down in the MT lattice without performing lateral motion.

One prominent example 
is the cytoplasmic linker protein-170 (CLIP-170) which has been shown to increase the rescue frequency of MTs \emph{in vivo} and therefore positively influence the average lifetime of MTs~\cite{komarova2002}. The binding process of CLIP-170 is rather complex and details can 
be found in~\cite{komarova2005,bieling2007,bieling2008a,dragestein2008,dixit2009}. Here, we just notice that CLIP-170 temporarily binds to the 
plus-end of MTs and is released with a given rate. 

Therefore both, the GTP-cap as well as the +TIPs have a strong influence on the microtubule dynamics and act in a qualitatively similar way. The key 
mechanism is that GTP as well as +TIPs are introduced at the plus-end of growing microtubles and reduce significantly the rates of catastrophes. 
The stabilizing molecules are only temporarily bound to microtubles. Either they unbind from the filament with a given rate, as for example the +TIPs
or they lose their stabilizing function, as in case of the GTP-hydrolysis. This means that fast growing microtubules are rather stable, since they are 
stabilized by a cap, which increases the rescue rates or decreases the catastrophe frequencies. By contrast, microtubules that are slowly growing 
or pausing at the cell membrane become gradually more unstable. 

Several \emph{in vitro} and  \emph{in vivo} experiments have
been carried out, that analyze the consequences of the
aforementioned stabilizing mechanisms. In
\cite{drechsel1992,walker1988} it was shown \emph{in vitro}
that the catastrophe frequency decreases with increasing
polymerization speed, which was controlled by the
GTP-concentration. The stabilization also leads to length and
lifetime distributions which differ from those one would
observe for constant polymerization/depolymerization
rates~\cite{gardner_z_h2013}.
The impact of different stabilization mechanisms has been studied \emph{in vivo}.
Some length distributions were found
experimentally to be exponentially confined near the
cell edge~\cite{komarova_v_b2002}.
However, it was also shown that these distributions
were only obtained if some cytoplasmic linker proteins
(CLIP) were present~\cite{komarova2002}.
Indeed, in the absence of CLIP, the microtubule rescue
was found to be reduced by sevenfold, with the consequence
that MTs were often completely depolymerizing.
The resulting length distribution was then found to be
flat~\cite{komarova2002} instead of exponential.

\subsection{Modeling the dynamical instability: the two-state model}
\label{sec:dyn_inst}

In this section, we discuss one of the early models introduced by Dogterom and Leibler ~\cite{dogterom_l1993} 
describing the dynamics of MTs with a fixed minus-end. In this model, the MT can either be in the growing ("+") or 
shrinking ("-") state. The MT-dynamics are characterized by four
rates: the growth rate $\nug$, the shortening rate $\nus$, the
rescue rate $\nur$ and the catastrophe rate $\nuc$. This set of
parameters is sometimes extended by a fifth parameter $c$ which
gives the concentration of free tubulin and may be used as
argument of some of the other rates. The MT is treated as a
one-dimensional chain of $l$ subunits to which subunits can be
added and removed at the plus-end.

Assuming a constant concentration of free tubulin, the concentration $c$ can be ignored in this simple model and the master equation for the probabilities to observe a filament of length $l$ in the growing or shrinking state far from any boundary is given by~\cite{dogterom_l1993} \footnote{Actually, the equations in ~\cite{dogterom_l1993} were written in a space continuous form. }:
\begin{align}
\frac{\partial p_+(l,t)}{\partial t}=\nu_gp_+(l-1,t)-\nu_gp_+(l,t)-\nu_cp_+(l,t)+\nu_rp_-(l,t)\\
\frac{\partial p_-(l,t)}{\partial t}=\nu_sp_-(l+1,t)-\nu_sp_-(l,t)+\nu_cp_+(l,t)-\nu_rp_-(l,t).
\end{align}
In this representation, the plus sign in the index denotes the growing state and the minus sign the shrinking state. Space is discretized in a natural way as MTs possess with the tubulin dimer a well-defined structural unit. 
This model presents a phase transition between bounded and unbounded growth with the transition taking place at
\begin{align}
\nus\nuc=\nug\nur.
\end{align}
The MT length distribution in the bounded growth phase is exponentially decaying with mean
\begin{align}
\langle l\rangle \approx \frac{\nug\nus}{\nus\nuc-\nug\nur},
\end{align}
while in the unbounded growth regime, the average length grows linear over time as
\begin{align}
\langle l\rangle =\frac{\nug\nur-\nus\nuc}{\nuc+\nur}t.
\end{align}
Above, it was assumed that the free tubulin concentration
$c$ was homogeneous, i.e. that free tubulin was diffusing
very fast compared to the time scales of MT dynamics, and 
time independent. 

Including spatially and temporally varying concentration
$c$ of free tubulin dimers, which are supposed to have
a finite diffusion constant, dynamical states can be created in
which two populations of MTs exist simultaneously: While one
population exhibits unbounded growth, a second one is trapped
at finite MT lengths because of a lack of free tubulin which
has been consumed by the first
population~\cite{dogterom_l1993}. This result thus shows that
the environment influences  MT polymerization and may play a role
in the regulation of the cytoskeleton.

An alternative way 
proposed in~\cite{janulevicius_p_o2006}
to account for the finite supply
of free tubulin 
is to consider that the MT dynamics takes place
in a small volume.
The authors assume that the free tubulin concentration
is homogeneous (fast diffusion), but can be globally modified
by the consumption or release of tubulin during
polymerization and depolymerization.
The fluctuations of tubulin concentration
are larger in small volumes.
Indeed, at equal initial concentration, the amount of
tubulin available in small volumes is more limited
and will be more easily depleted by the growth of the MT.
As a consequence, the distribution of MTs lengths,
which is exponential for large volumes, becomes
a gamma-like distribution in the limit of small
volumes~\cite{janulevicius_p_o2006}.
Another consequence is that
strongly confined MTs would appear more dynamic,
which might  for example be relevant for the axonal MT network
(see section~\ref{sec:axon_mt_dyn}).

In case of filaments that
are not connected to a MTOC or to any other anchoring structure,
one has to consider the dynamics of the minus-ends in a similar
fashion, but with different rates. Free minus-ends allow for
the displacements of the filament as a whole, i.e. the
treadmilling dynamics which has been previously discussed.

The dynamics of filaments is not only restricted to the plus- and minus-ends if the action of MT-severing proteins such as katanin~\footnote{Katanin forms a hexamer around the MT and uses the energy from ATP hydrolysis to break
the filament~\cite{qiang2006}} is considered. This has been introduced into the model of Dogterom and Leibler in~\cite{tindemans_m2010} by considering randomly occurring severing events. The MT is then split into two with the newly exposed plus-end being in the shrinking state. This modification could also be shown to lead to a finite average MT length with non-exponential length distributions. Interestingly, the severing does not increase the total number of MTs although the average MT length decreases.

\subsection{Models with explicit description of the GTP cap}
\label{sec:model_gtp}

In the previous discussion of the MT dynamics we underscored the 
importance of the GTP cap for the MT-dynamics, which significantly 
changes the length distribution. 

Therefore, a different class of MT models has been suggested, which
 does not postulate switching between 
a growing and a shrinking state, but describes the process
at a more refined scale~\cite{flyvbjerg_h_l1994,margolin2006,antal_k_r2007,brun2009}.
In these models, individual tubulin subunits are
allowed at all times to be added to or removed from the plus-end.
Subunits which are added always carry a GTP and have low (or
vanishing) probability to depolymerize from the plus-end. The
GTP then hydrolyzes to GDP at a certain rate after the subunit
has polymerized. The hydrolysis event modifies the subunit's
kinetics : A subunit bound to GDP (instead of GTP)
has a higher probability of
being removed when it is exposed at the plus-end.

A consequence of these evolution rules is that the density
of GTP bound subunits is highest at fast growing plus-ends.
This class of models is thus adequately reproducing the presence
of the GTP cap mentioned in section~\ref{sec:mt}.
The GTP cap has an average length which is determined
by the rate of hydrolysis of GTP to GDP
and by the attachment rate of new tubulin subunits.
Indeed, as GTP is hydrolyzed, its
density is decreasing with time, such that the concentration of
GTP is low far from the plus-end or in the case of slowly
growing or pausing filaments.

These models are thus microscopic in the sense that there is no
state variable describing the filament as a whole as with the
growing and shortening state of the two-state model.
Nevertheless, these models exhibit a macroscopically observable
behavior which reproduces the long periods of persistent growth
and shrinkage characteristic for the dynamic
instability of
MTs~\cite{flyvbjerg_h_l1994,margolin2006,antal_k_r2007,brun2009}.
The rescue and catastrophe rates assumed in
section~\ref{sec:dyn_inst} are in these
models observables which result from the kinetics of the
tubulin subunits with GDP or GTP.
For example, in~\cite{margolin2006} the
rate of catastrophes observed in the model
scales as $n^{-2}$ if $n$ is the average length of
the GTP cap.
As for the two-state model of section~\ref{sec:dyn_inst},
considering a fixed concentration of tubulin in the absence of
geometrical constraints
can lead to bound or unbound growth of the MTs depending
on the tubulin concentration~\cite{margolin2006,antal_k_r2007}.

There are still several other features that may have
to be taken into account for a full picture of the dynamic
instability. For example a dependence of the catastrophe rate
on the age of the microtubule observed \emph{in vitro}~\cite{gardner2011}
still has to be explained. Some recent molecular dynamics simulations
of the microtubule~\cite{grishchuk2014} may help to distinguish between
the various scenarios
that have been proposed.
While only spontaneous hydrolysis of the GTP was considered
here, other scenarios including induced hydrolysis
(i.e. hydrolysis favored if neighboring GTPs are
already hydrolyzed) were explored in~\cite{flyvbjerg_h_l1994} and shown to explain some dilution experiments.

The loss of the GTP cap is not the only factor that
could induce a catastrophe. Structural changes at the tip,
interactions between protofilaments,
or MAPs, could also play a role~\cite{brun2009}.
Comparing models and experiments in various 
situations could help to distinguish
between these different scenarios, as we will illustrate
below by the example of dynamic MTs in finite 
volume. 

\subsection{Dynamic MTs in finite volume}

The interaction between a growing MT and the cell 
boundary is of great importance, since catastrophes are 
induced with a very high rate
when the MTs reach the cell boundary. 

\begin{figure}[tbp]
  \begin{center}
    \includegraphics[scale=0.25, clip]{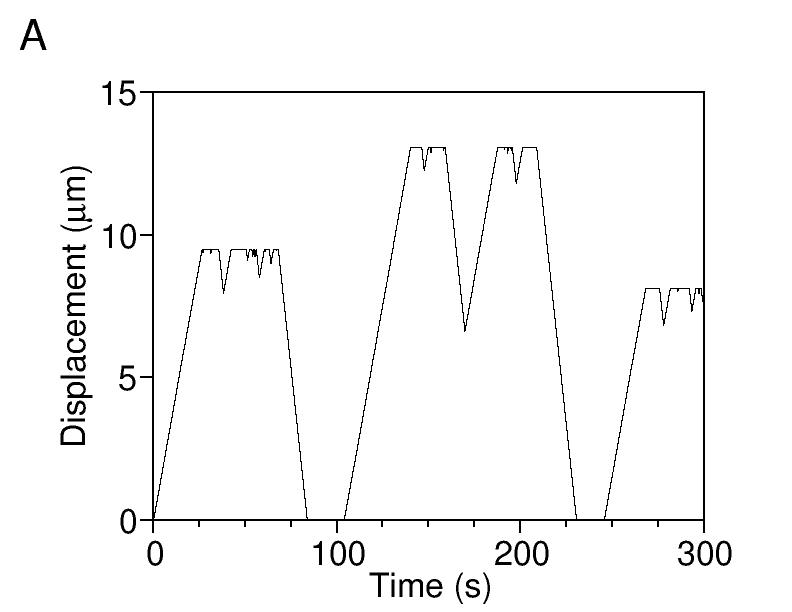}\hspace{0.5cm}
    \includegraphics[scale=0.25, clip]{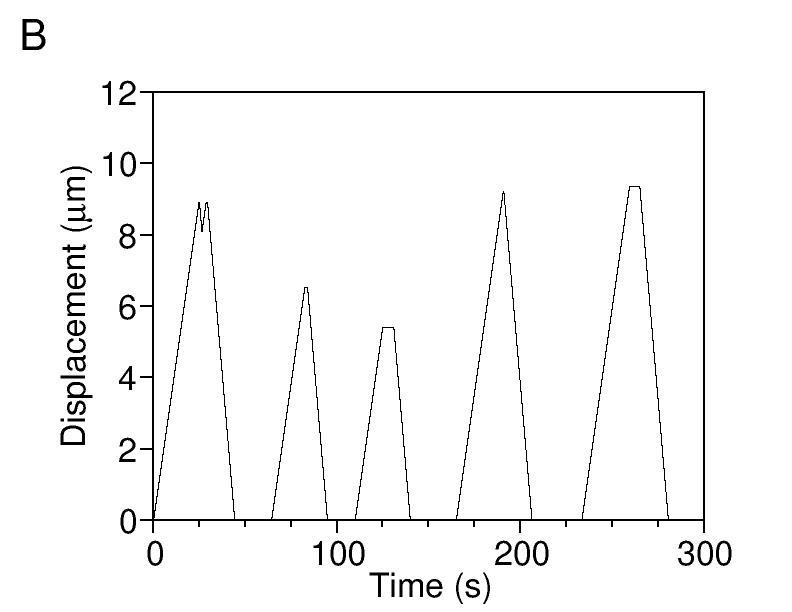}

C
    \includegraphics[width=0.4\linewidth]{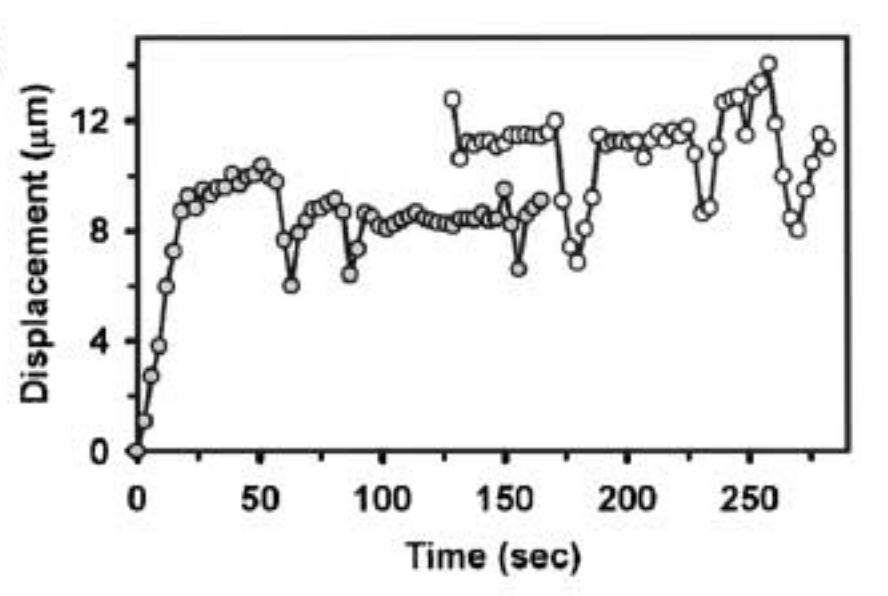}\hspace{0.5cm}
    \includegraphics[width=0.4\linewidth]{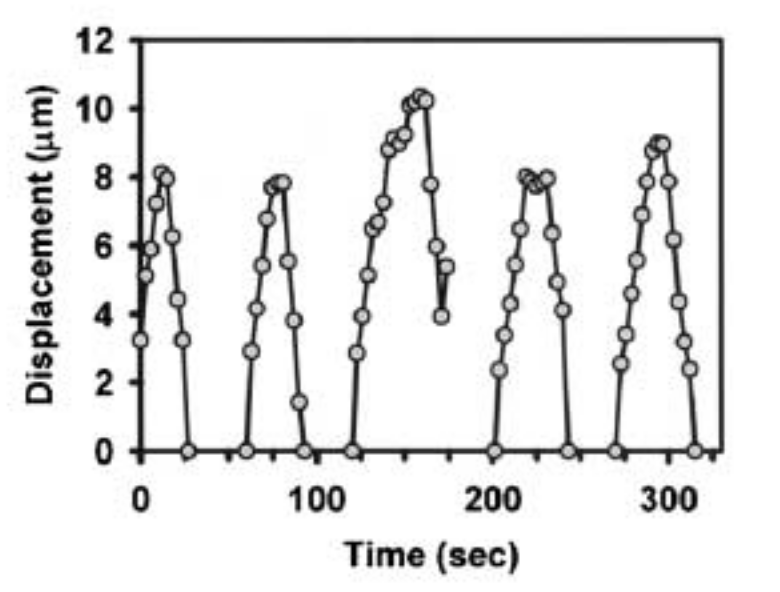}D
    \caption[Life history plots with and without rescue-enhancing +TIPs.]{Life
    history plots of {\it (A-B)} simulated or {\it (C-D)} real
    MTs {\it (A-C)} with and {\it (B-D)} without
    rescue-enhancing +TIPs. If tubulin subunits carrying a
    rescue factor are added under growth {\it (A-C)}, the MTs
    show dynamic instability at the cell boundary with
    frequent rescues. Without rescue-enhancing +TIPs {\it
    (B-D)}, the MTs depolymerize completely after a short
    contact with the cell boundary, and MT lifetimes are mostly
    determined by the time needed for polymerization up to the
    cell boundary and depolymerization back to the MTOC.
    The
    geometry for the simulations was an ellipse of half-axes
    $a=19.2~\mu \text{m}$, $b=6.6~\mu \text{m}$ with the MTOC
    being $1.2~\mu \text{m}$ off the center of the ellipse in
    $x$- and in $y$- direction which are representative
    parameters for the cell geometry discussed in
    \cite{komarova2002}.
    Figures {\it (A-B)} (simulations) are
    taken from~\cite{ebbinghaus_s2011} and figures {\it (C-D)}
    (experiments) from~\cite{komarova2002}\copyright 2002 (doi:10.1083/jcb.200208058).}
    \label{fig:mt_life_history}
  \end{center}
\end{figure}

The simplest way to explicitly consider the cell boundary is by 
imposing a maximum filament length $l^*$ such that $l\leq l^*$ as
in~\cite{govindan_s2004}, where a two-state model has been
considered. In this work it is assumed that the MT directly turns into 
the shrinking state, if the MT hits the cell boundary. For these boundary 
conditions the size of the MTs is always limited, irrespective of the model 
parameters. 
The stationary state length distribution can be calculated exactly and is given 
by $p_\pm(l)
\sim a^l$, where $a = (1+\nu_r/\nu_s)/(1+\nu_c/\nu_g)$. 
Therefore, the length distribution is exponentially  growing if $\nu_g\nu_r > \nu_s\nu_c$ 
(corresponding to unbounded growth in the open system) and  exponentially decaying
if $\nu_g\nu_r < \nu_s\nu_c$.
In the case of very small rescue
and catastrophe rates, rather flat length distributions
are obtained, corresponding to typical 
life-history plots as those 
of Fig.~\ref{fig:mt_life_history} B \& D.

Similar results were obtained with models describing
explicitely the GTP cap~\cite{gregoretti2006}.
Note that we only consider 
the case of a sufficiently high concentration
of tubulin, so that the growth of MTs is not limited
before the cell edge is reached~\cite{gregoretti2006}.

The two regimes can be obtained experimentally~:
Length distributions exponentially confined near
the cell edge were indeed found
experimentally~\cite{komarova_v_b2002}.
However, it was also shown that these distributions
were only obtained if some cytoplasmic linker proteins
(CLIP) were present~\cite{komarova2002}.
Indeed, in the absence of CLIP, the microtubule rescue
was found to be reduced by sevenfold, with the consequence
that MTs were often completely depolymerizing.
The resulting length distribution was then found to be
flat~\cite{komarova2002} instead of exponential.

\begin{figure}[tbp]
  \begin{center}
    \includegraphics[width=0.5\linewidth]{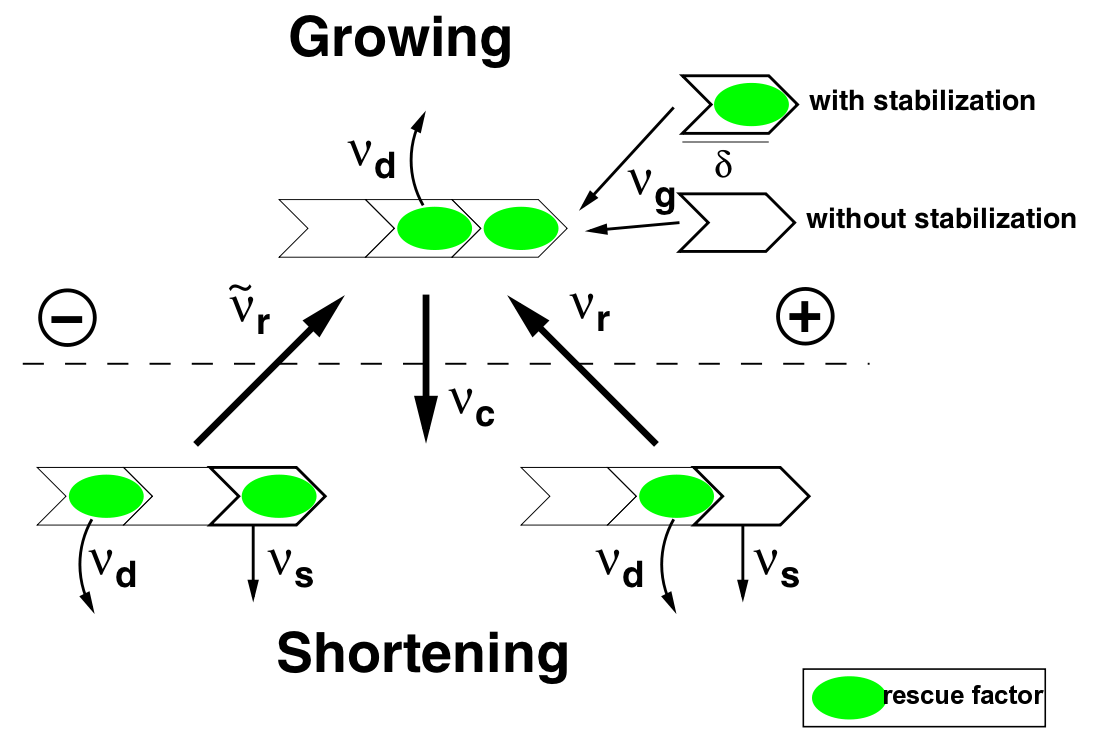}
    \caption[Schematic representation of a model for filament dynamics.]{
    Definition of a model for filament dynamics.
    The filament can be either in a growing state
    or in a shortening state.
    When it is in the growing state, a
    subunit of length $\delta$ is added at a rate $\nug$ to its
    plus-end.
    Depending on the scenario, the added subunit
    carries a rescue-enhancing +TIP or not.
    The filament will switch to the shortening state
    with rate $\nuc$. Then the filament loses
    subunits from its plus-end at rate $\nus$
    regardless of the possible presence of a rescue factor.
    However, the rate for switching back to the growing
    state (rescue) depends if a rescue factor is present
    on the last subunit at the plus tip (then the rate
    is $\nurr$) or not (then the rate is $\nur < \nurr$).
    At any time, a
    tubulin subunit can loose its rescue factor at
    rate $\nud$.  Figure modified from \cite{ebbinghaus_s2011}.
    }
    \label{fig:mt_model}
  \end{center}
\end{figure}

The effect of such a +TIP that enhances rescue - and thus
stabilizes MTs - was modeled in~\cite{ebbinghaus_s2011}.
A summary of the model dynamics is given in
Fig.~\ref{fig:mt_model}.
The key feature of this model is the following: When a tubulin dimer is added, 
it is assumed to present
a binding site to which rescue enhancing +TIPs bind at very
high frequency (taken to be infinite here).
When the last tubulin subunit at the +tip carries
a rescue factor, it modifies
accordingly the rescue rate,
which increases then from $\nur$ to $\nurr$.
A tubulin subunit included in the MT can loose its rescue factor with rate $\nud$.
This model will be compared
to a reference case without rescue factor ({\em without stabilization}
in Fig.~\ref{fig:mt_model}).

Interesting effects are
observed if the model filament is embedded into a finite
predefined geometry mimicking the shape of a cell
within which the
origin of nucleating MTs is chosen as described in the caption
of Fig.~\ref{fig:mt_life_history}. 
A MT grows in a random direction from the MTOC. The maximum
distance $l^*$ it can grow before reaching the boundary
depends on this direction.
The filament follows the bulk dynamics as described above until
it reaches the boundary.
Then no further tubulin subunit can be added.
At the boundary, switching to the shrinking state
is induced with a probability $p_\text{ind}$. 

Fig.~\ref{fig:mt_life_history} compares some life histories
obtained from the
model of Fig.~\ref{fig:mt_model}~\cite{ebbinghaus_s2011},
and from the
experiments~\cite{komarova2002}, with or without rescue factor.
The model reproduces well the observed behavior, namely
the large length fluctuations in the absence of CLIP170
(as almost no rescue will prevent the filament to
entirely depolymerize),
and the pauses and frequent switching between growth and shortening near the cell edge when CLIP170
is present.
Also typical lifetimes have the right order of magnitude.
Note that in the model, it is assumed that the +TIP is only enhancing rescue.
One could think of +TIPs that would rather prevent catastrophe
to start.
However, this would not explain the experimentally observed behavior,
 as it would not lead to the numerous rescues (see Fig. \ref{fig:mt_life_history}C).

Interestingly, the model predicts an aging
effect in the survival probability distribution of the MTs.
Indeed, the survival probability of an MT after it
hits the cell boundary for the N-th time is found to be lower
for higher values of~N~\cite{ebbinghaus_s2011}.
This aging effect, which has also been discussed later
in~\cite{gardner2011,gardner_z_h2013,akhmanova_d2011} from an experimental 
point of view, is not observed in the model if no rescue
factor is included. This aging effect is of great importance since stabilized MTs show 
a great resistance to thermal fluctuations and mechanical constraints. 
But at the same time the MTs can completely depolymerize after the stabilizing 
cap is removed from the filament. This way MTs are able to fulfill the contradictory 
constraints with respect to their rigidity and flexibility.

\subsection{Actin filaments}
\label{sec:actin}

Actin filaments are composed of only two twisted
protofilaments. 
As a result, they are much more
flexible than MTs and their radius is only 5-9nm~\cite{alberts2008}.
However, they form highly interconnected lattices
and as a whole have strong mechanical properties.
Indeed, they play a major role to give to the cell
membrane its shape and mechanical strength.

In the cell, actin filaments are usually embedded into a network,
and submitted by myosin motors to non-thermal fluctuations.
The mechanical properties of such an acto-myosin gel
can be modeled by macroscopic non-linear models
which describe the cytoskeleton as a whole~\cite{kruse2005,mackintosch_l2008}. 
A review of these approaches can
be found in \cite{julicher2007}.
Some microscopic stochastic models have also been developped
(see for example~\cite{lenz2012,lenz2014}).
Some acto-myosin active gels
have been studied in \emph{in vitro}
experiments~\cite{brangwynne2008,koenderink2009,alvarado2013,soaresesilva2014}.
We shall not address this large field of research in this review.
Instead, in this section we rather concentrate on the dynamics of
single filaments.

Like MTs, actin filaments are polar polymers, i.e.
one can distinguish a fast growing plus-end and a minus-end
where the filament disassembles rapidly if it is not
stabilized.
Actin filaments are dynamical structures,
evolving through mechanisms similar to MTs.
The GTP/GDP nucleotides embedded in MTs are replaced
by ATP/ADP that play a similar role in the case of actin.
As actin filaments are not attached to a centrosome,
they naturally undergo treadmilling
both \emph{in vitro} and \emph{in vivo}.

An important difference between MT and actin filaments is that
the dynamics of actin filaments combines both
polymerization
and depolymerization events on short time scales,
which can be averaged on mesoscopic time scales
to give effective growth and shortening rates while 
the MTs stay, as previously discussed, on much longer periods of time in one -
polymerizing or depolymerizing - state.

If we mention actin filaments in this review,
this is not because of their mechanical tasks - a subject
which is out of our scope - but because
they also serve for intracellular transport.
We shall see that a specific family of molecular
motors uses the actin filaments as tracks (see
section~\ref{sec:myosin}), and
though we shall not develop the role played by
actin lattices in intracellular transport, it should be kept
in mind that, as a complement to MT based transport,
part of the transport is performed along
the actin networks~\cite{aspengren_h_w2007} that,
for example, cover the inner
side of the cell membranes. This is in particular true
for transport along the axon (section~\ref{sec:axon}).

\subsection{Treadmilling filaments}

\begin{figure}[tbp]
  \begin{center}
 \includegraphics[scale=0.8, clip]{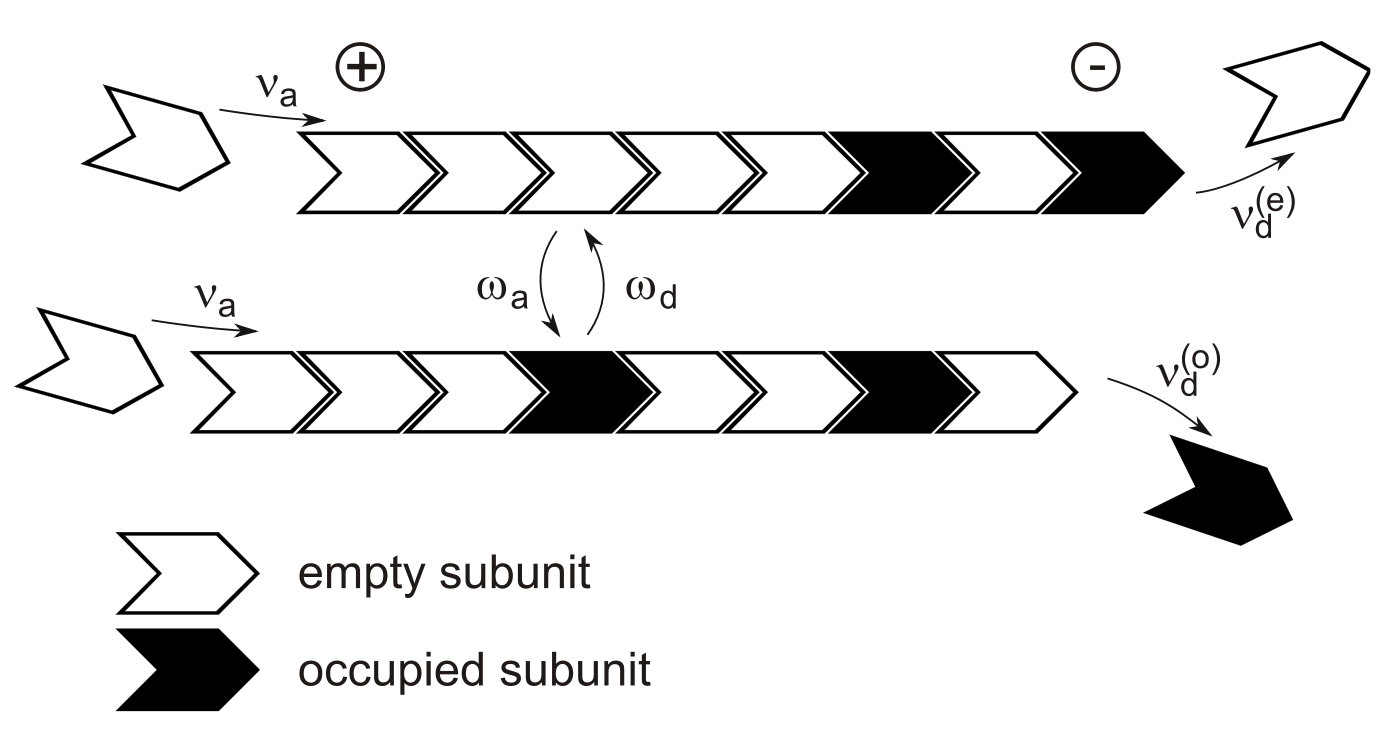}
    \caption[Treadmilling filaments]{
    Schematic representation of a treadmilling filament.
    Subunits are added at the plus-end
    with rate $\nu_a$.
    A protein can bind to an empty subunit with rate
    $\omega_a$ and detach with rate $\omega_d$.
    A subunit is removed at the minus-end with rate
    $\nu_d^{(e)}$ if it does not carry a protein
    and with rate $\nu_d^{(o)}$ if it does.
  From~\cite{erlenkamper_k2009}.
    }
    \label{fig:treadmil}
  \end{center}
\end{figure}

Actin filament dynamics can be simply described
by an average growth rate at the plus-end
and an average shrinking rate at the minus-end.
A stochastic model that realizes this dynamics
is the one of Fig.~\ref{fig:treadmil} (in the
case where only empty subunits are considered).
The filament is represented as a one-dimensional chain 
of finite length, to which subunits are added
with rate $\nu_a$ at the plus-end and removed
with rate $\nu_d$ at the minus-end.
This simple description can also apply to 
MTs that are
not anchored
(as possible for example in \emph{in vitro} experiments) when the dynamics
is considered on time scales large enough to average
over the large fluctuations due to the dynamical instability.

Unregulated polymer
assembly typically leads to exponential length distributions,
with a length scale that is given as a function of
polymerization and depolymerization rates. The length
distributions of unregulated polymers is rather broad.
This is not what is observed in certain situations,
for example for the actin filaments that are found
in stereocilia in the inner ear where length must be well
controlled~\cite{rzadzinska2004}.
Therefore,  several mechanism have been suggested that
influence the polymerization dynamics. For example,
a simple
mechanism of length regulation has been introduced in
\cite{erlenkamper_k2009}.  In this work the decoration of
tubulin subunits with a protein that can alter
the depolymerization rate at the minus-end is assumed
(see Fig. \ref{fig:treadmil}).
Depending on the
(de-)polymerization rates and the concentration of the
length-regulating protein, one observes different dynamical
regimes, i.e.  unbounded filament growth, exponential
distributions and  unimodal distributions,
which can be very sharp.  The unimodal distributions are observed in case of proteins that enhance the depolymerization rate. In this case it is possible to reduce the length fluctuations significantly, since the relative fraction of occupied subunits is increasing with the filament length, and thereby increasing the effective depolymerization velocity for relatively long filaments.

\section{Molecular motors}
\label{sec:motors}

While previous section was devoted to the ``roads''
along which intracellular transport occurs,
this section will be dedicated to the ``trucks'',
i.e., the molecular motors that
drag intracellular cargos such as organelles and vesicles
along the cytoskeletal filaments.

\subsection{Biological description of motors}
\label{sec:biol_motors}

Cytoskeletal motors -- also known as motor proteins -- are
molecules which are able to move along polar cytoskeletal
filaments such as MTs or actin filaments by using the energy
derived from hydrolysis of adenosine triphosphate (ATP):
they are active particles.
Hydrolysis of ATP induces a confirmational change of
the motor protein that results into a step along the filament.

Motor proteins can detach and attach to the filaments.
An important characteristics of motors is the number
of steps - or conformation changes - that they can
handle in a row before detaching. 
A so-called {\em processive} motor is able to perform a large
number of steps before detaching, while a non-processive
motor will usually detach after a single step.
Examples of non-processive motors are myosins II,
which are responsible for
muscle contraction and have to work as a team
to exert a continuous force.
By contrast, MT-based transport involves processive motors
which, individually, are able to interact for long
period of times with the track and to move over
large distances~\cite{schliwa_w2003}.

These motor
proteins usually consist of at least two parts: the motor
domain\footnote{The motor domain refers to the part of the
protein which is responsible for force production}
and the cargo-binding domain. The motor domain is able
to bind to the polar filament, while
the cargo domain binds to intracellular cargos such as
organelles
or other elements of the cytoskeleton~\cite{alberts2008,mallik_g2004}.

In the following, the two most important protein families for
MT-based transport are presented, kinesins and dyneins~\cite{vale2003}.
Both
families are heterogeneous in the sense that the cargo which
they are able to carry as well as the type of motion along the
filament varies within the same family despite similar
structures of the motor domains~\cite{hirokawa1998}.
In order to restrict our presentation,
we shall focus on the most important
representatives of each family for MT-based transport,
namely kinesin-1 and cytoplasmic dynein.
We shall also briefly address the case of some myosins
able to carry cargos along actin filaments.

\subsubsection{Kinesin}
\label{sec:kinesin}
The kinesin superfamily is subdivided into several tens of
kinesin families~\cite{reilein2001} which exhibit surprisingly similar motor
domains and therefore are all able to exert force on
MTs~\cite{vale_f1997}.
Due to the remaining differences in the structure of the
molecule, motion patterns vary: While most of the kinesins
move toward the plus-end of a MT (anterograde movement),
some kinesins move in the opposite or
retrograde direction. Also, many but not all kinesins perform
\emph{processive} motion.

In this review, we shall focus on
kinesin-1 (also called conventional kinesin), which
is a representative member of the kinesin
superfamily implicated in long-range transport and
which moves toward the plus-end of MTs.
Note that in the remaining of this review, we shall use
the work ``kinesin'' to refer to kinesin-I,
unless otherwise stated.

\begin{figure}[tbp]
\begin{center}
\includegraphics[width=0.9\linewidth]{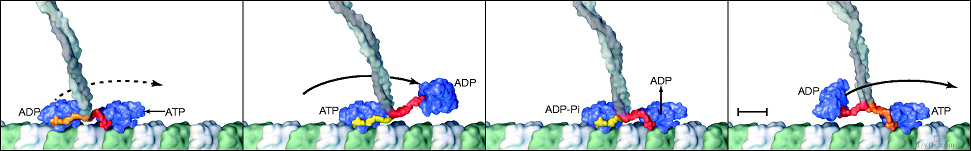}
\caption[Stepping cycle of kinesin.]{Stepping cycle of kinesin.
From left to right: An ATP-bound head induces a conformational
change which brings the ADP-bound head 
to the front. Then the ATP of the now rear head gets hydrolyzed
to ADP and inorganic phosphate.
After the release of the inorganic phosphate, the rear head
unbinds from the filament and is brought to the front as soon
as the other head exchanges its ADP by ATP.
From~\cite{vale_m2000}.
}
\protect\label{fig:stepping_kinesin}
\end{center}
\end{figure}

Kinesin-1 is composed of two heads\footnote{Note that {\em head} refers to what
would be called a foot if kinesin was considered as a walker.} that can bound
to the MT, linked by a
long neck to the cargo domain (Fig.~\ref{fig:stepping_kinesin}).
The two motor heads move in a coordinated manner,
and step in an alternating
hand-over-hand pattern on the MT~\cite{vale_m2000}.
The stepping is triggered by the hydrolysis of ATP
\cite{coy1999}
(figure~\ref{fig:stepping_kinesin}): An ATP-bound
kinesin head has a high affinity to the MT lattice
and binds strongly to it. When the ATP is hydrolyzed
into adenosine diphosphate (ADP) and inorganic
phosphate, the binding of that head gets weaker and
the motor head eventually detaches from the
filament. Meanwhile, the other motor head was able
to take up an ATP and bind to the MT, inducing a
conformational change that brings the recently detached
motor head to the front. As soon as this motor head
replaces its ADP by ATP and re-attaches to the
filament, the cycle starts over again. The step size
of kinesin, which is half of the step size of each motor
head~\cite{yildiz2004},
has been shown to be equal to 
one tubulin subunit, i.e., 8~nm~\cite{yildiz2004,fehr_a_b2008}.

Several of these stepping cycles occur
before kinesin detaches from the MT.
The length of the displacement performed along the MT
by a molecular motor between two phases of diffusion in the
cytosol is called its run length.
Typical run lengths for kinesin are of the order of one hundred
steps~\cite{seitz_s2006} which in combination with
the step size of 8~nm gives runs of about 1~$\mu$m
in length. Kinesin is thus a processive motor.

It is in general considered that conventional
kinesin has a uni-directional motion.
Indeed, without load, back-stepping is exceptional for
kinesins:
less than 1\% of the net distance covered by a single
kinesin corresponds to backward stepping~\cite{mallik2005}.
However, if kinesin is attached to a cargo, its stepping
behavior depends on the force (or load) exerted
by the cargo onto the motor.
Back-stepping becomes non negligible under high
backward load~\cite{block2003,block2007}.
Backwards steps even predominate above stall
force~\cite{carter_c2005}.
This leads obviously to a dependence of the velocity with
the force exerted on the motor.

An applied force does not only affect the stepping but
also the detachment rate of kinesin. We shall discuss in more
details the velocity-force and detachment-force relations
in section~\ref{sec:load}.

\subsubsection{Dynein}
\label{sec:dynein}

\begin{figure}[tbp]
\begin{center}
\includegraphics[width=0.35\linewidth]{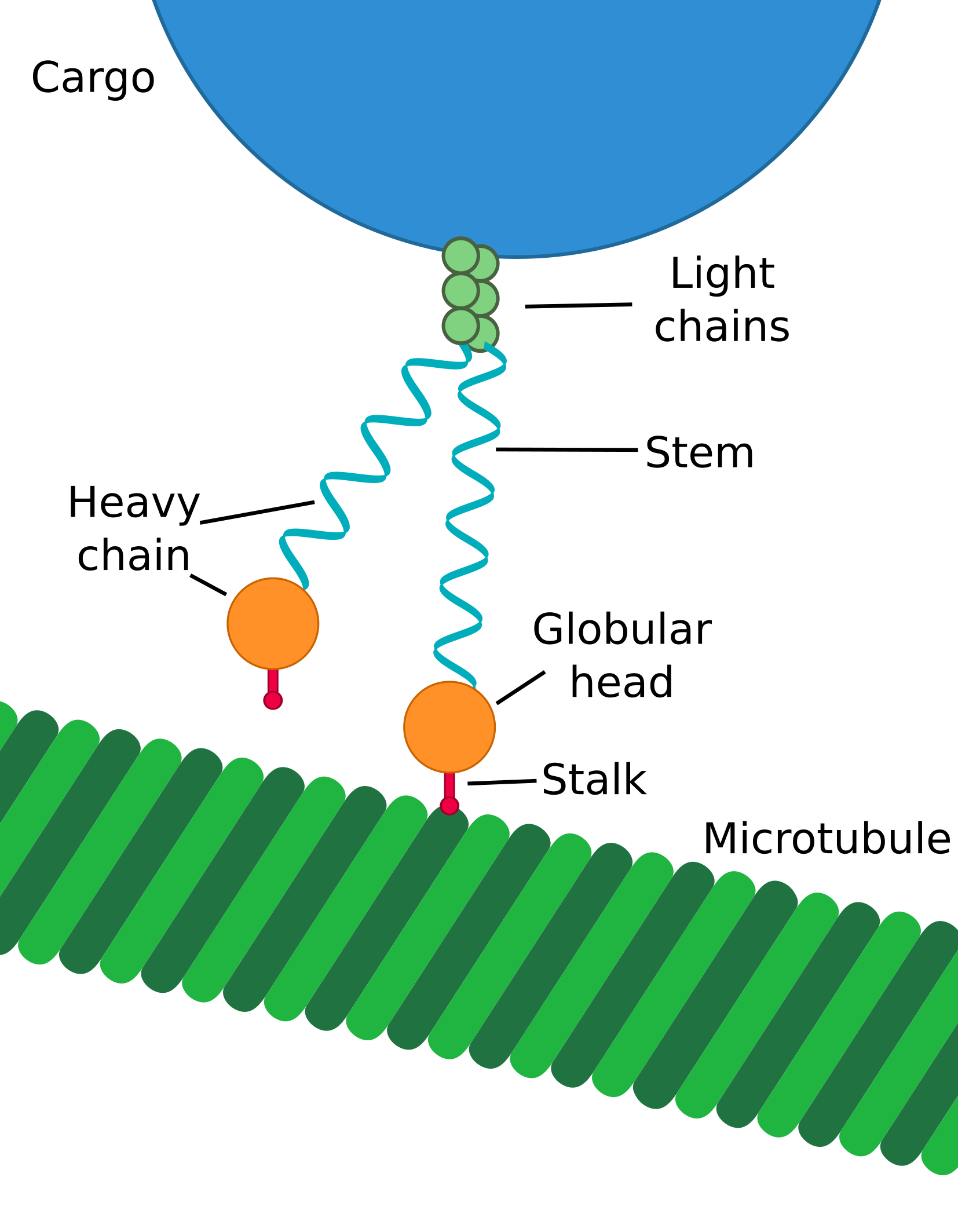}
\caption[Sketch of cytoplasmic dynein.]{
A stylized depiction of dynein carrying cargo along a microtubule.
 }
\protect\label{fig:sketch_dynein}
\end{center}
\end{figure}

The dynein family is composed of two groups: axonemal dyneins,
involved in the motion of flagella or cilia, and cytoplasmic
dyneins, which play an important role in intracellular transport.
Only the latter group will be considered here.

Cytoplasmic dynein carries out most of the minus-end directed
long-range transport along MTs. The attachment to the wide
variety of different cargos is probably assured by different
multifunctional adaptors such as dynactin which regulate dynein
function~\cite{kardon_v2009}. The structure of cytoplasmic dynein is
more complicated than those of kinesins - a fact which is also reflected in
its very high molecular weight of
$1.2~\text{MDa}$~\cite{mizuno2007}.
Indeed, among molecular motors, dyneins are the largest one~\cite{alberts2008}.
Each dynein head has the shape of a ring, with a stalk
that stretches out from the head and binds to the MT.
The cargo is attached at the end of a long tail pointing away
from the MT in the bound state.

Dynein, in the absence of nucleotide, is bound to the MT
and it is when an ATP binds that it can detach.
This is in contrast with kinesin, which associates to the MT
when an ATP is bound, and is released through the
hydrolysis of ATP~\cite{alberts2008}.
However, it was found in some experiments that
dyneins may step forward and backward under load
even in the absence of nucleotide
hydrolysis~\cite{gennerich2007}.

Dynein is able to perform processive motion if it is in dimerized
form~\cite{wang_k_s1995}.
The run lengths of the processive
motion are of the order of $1~\mu\text{m}$ and thus comparable to
those of kinesin~\cite{wang_k_s1995,reck-peterson2006}.

Contradictory results have been reported about the
step size of single dyneins.
In~\cite{mallik2004},
it was reported that under no load, dyneins were taking
predominantly 24- and 32-\text{nm} long steps, and that the
step length was decreasing under load.
Later, more direct measurements of dynein displacement
were performed through high temporal resolution fluorescence.
While some authors 
found that most steps of dyneins were 8~\text{nm} long,
whatever the load is~\cite{toba2006}, others
report a variability in the step size larger
than for kinesins~\cite{gennerich2007}.
Actually, results depend on whether it is the head or tail
of dynein that is marked~\cite{reck-peterson2006}.
While there is a peak of the step size distribution
around 16nm for the heads and 8nm for the tail,
distributions are much larger than in the case of
kinesin, and exhibit a non-negligible fraction
of backwards steps~\cite{reck-peterson2006,gennerich2007}.
Actually about 20\% of the net distance
moved by dynein is backward~\cite{mallik2005,reck-peterson2006}.

Recent measurements~\cite{qiu2012} have revealed that
the nature of dynein stepping depends on the distance
between its two heads: Stepping is stochastic and
uncoordinated for heads
close to each other, but becomes coordinated when the
distance between heads is large (indeed then
tension forces between the heads tend to bring forward
the head most behind).
It is still an open question to understand
how the uncoordinated stepping of close heads does
not alter processivity, as it should lead to higher
simultaneous detachment of both heads.
The load exerted by large cargos may contribute
to coordinate stepping, an hypothesis that has still
to be explored~\cite{qiu2012}.

By contrast with kinesin, dynein
performs easily side-steps to adjacent
proto-filaments when walking on a MT~\cite{wang_k_s1995,wang_s1999,ross2006,reck-peterson2006}.

As a general conclusion, it appears that single dyneins have a
much less regular motion than kinesins and can move faster.
The varying step size and the side steps have been hypothesized to allow the
motor to bypass obstacles~\cite{caviston_h2006,qiu2012}.
Due to its irregular stepping properties, dynein has sometimes
been compared to a drunken sailor~\cite{cameron2012}, while
kinesin would be a serious guy, walking straight and
regularly.

The binding site of dynein on the MT surface overlaps with the
binding site of kinesin~\cite{mizuno2004} (though motor heads
of dynein are much larger than those of kinesin~\cite{alberts2008}).
Both motor species thus
compete for binding sites on the filament and probably block each
other on encounters in bidirectional transport on MTs.
We shall study these collective effects later in the review.

\subsubsection{Kinesin and dynein under load}
\label{sec:load}

The processivity of a molecular motor along a MT depends on its
binding, stepping, and unbinding rates, which
determine the typical run length of the motor on the filament.
These rates depend on several external factors, and in
particular (for the stepping and detachment rate)
on the load exerted on the motor.

\begin{figure}[tbp]
\begin{center}
\includegraphics[width=0.4\linewidth]{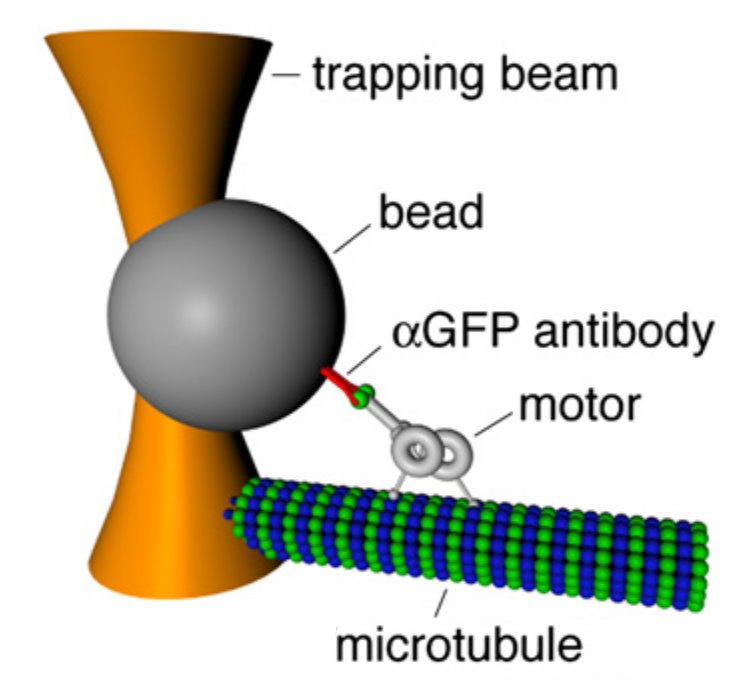}
\caption[Sketch of an optical trap.]{Sketch of an experiment to
explore the stepping of a motor under load.
 Modified from~\cite{gennerich2007}.
 }
\protect\label{fig:kin_trap}
\end{center}
\end{figure}

In this section, we shall focus on the
velocity-force and detachment rate-force relations,
which have been measured experimentally.
Typically, a coated bead is attached to the
motor~\cite{svoboda1993,jamison2010}.
The bead is placed in an optical trap, while the motor steps
on a MT (Fig.~\ref{fig:kin_trap}).
If the optical trap is immobile, the walking motor
pulls on the bead until the force exerted by the trap prevents
it to move further.
Then the motor stops, and the stall force (defined as the force that has to be applied in order to stop a motor) can
be calculated from the displacement of the bead inside the trap.
In another type of experiment, the optical trap is moved
in such a way that the applied force is constant. 
The velocity-force relation can then be directly measured.
The detachment rate can also be measured in these experiments
as a fonction of the load.

The stall force under load for kinesin was measured to be around
5 - 7pN~\cite{kojima1997,visscher_s_b1999,mallik2005}.
Contradictory results have been found for the stall
force of dynein. Most experiments indicate a smaller stall
force than for kinesins~\cite{mallik2005,kunwar2011}.
For example from {\em in vitro} data, Kunwar et al~\cite{kunwar2011} estimated $F_s = 4.7 \pm 0.04$ pN for kinesin
and $F_s = 1.36 \pm 0.02$ pN for dynein.
However in a few 
experiments stall forces similar to those
of kinesins (approximately 7pN) were
reported~\cite{gennerich2007}.
It has been sometimes mentionned that {\em in vivo}
values could differ from their {\em in vitro}
counterparts~\cite{kunwar2011}.
However, it is much more difficult to control for example
the number of motors involved in an {\em in vivo} measurement.

Though there is no general agreement on the precise form
of the dependence of velocity or detachment rate with the load,
we give here for illustration the expressions
used in particular in~\cite{kunwar2011}.

\begin{figure}[tbp]
\begin{center}
\includegraphics[width=0.4\linewidth]{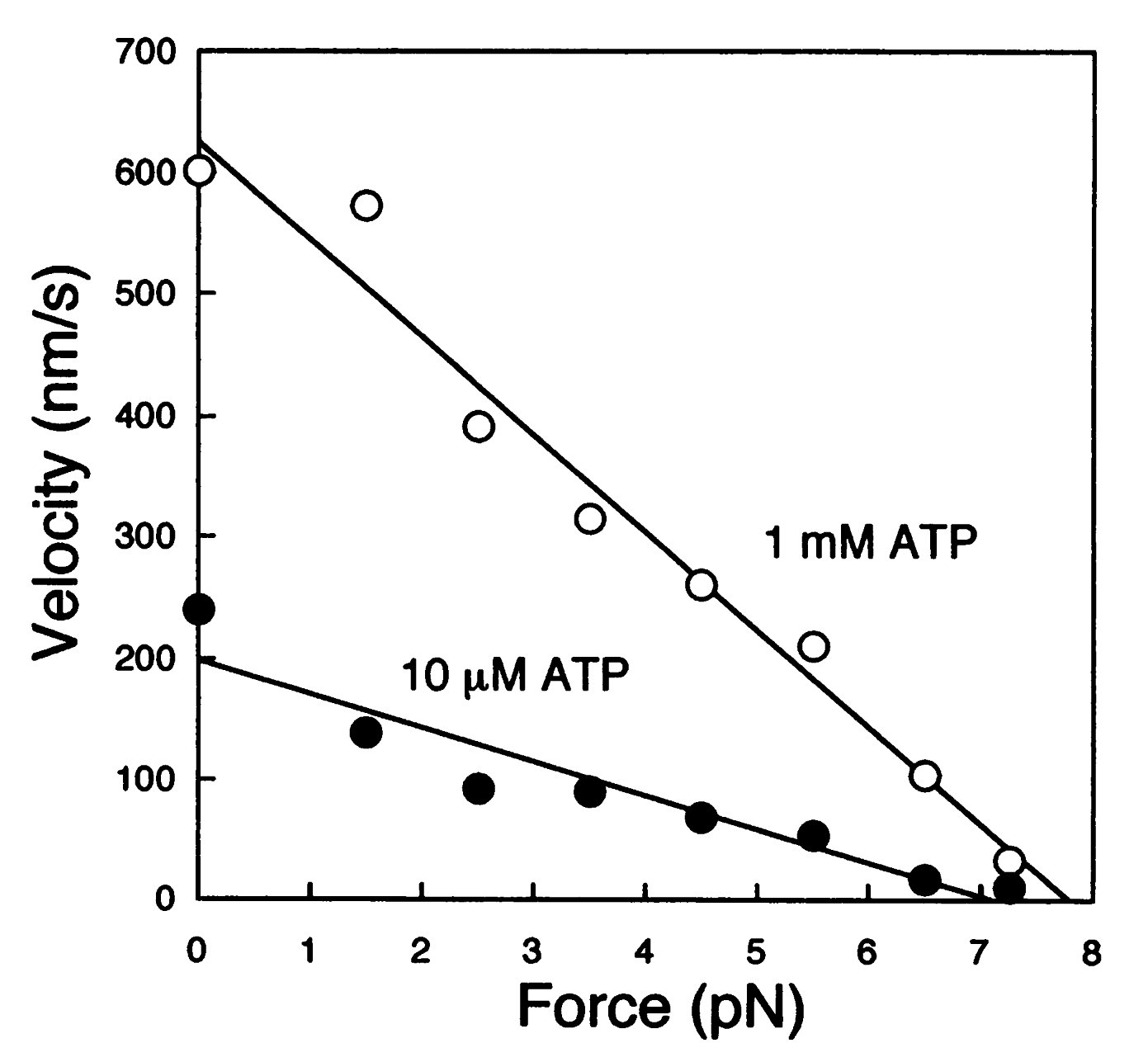}
\caption[Velocity under load]{
Velocity of kinesin as a function of the backward force,
measured {\em in vitro} using an optical trap.
 From~\cite{kojima1997}.
 }
\protect\label{fig:vload}
\end{center}
\end{figure}

In \cite{svoboda_b1994} the following velocity-force relations of single motors have been suggested:
\begin{equation}
v(F) = v_0 \left(1-\left(\frac{F}{F_s}\right)^w \right).
\end{equation}
where $F$ denotes the amplitude of the force applied to the motor and $F_s$ the so-called stall force, which is  defined as the force for which the velocity of the motor vanishes. 
The choice $w=1$ corresponding to a linear force-velocity law
is a good approximation (see the experimental results of
Fig.~\ref{fig:vload}) and has been used for example in the
model of \cite{klumpp_l2005a}. In~\cite{kunwar2011}, Kunwar et al
rather took $w=2$ for kinesin and $w=1/2$ for dynein.

\begin{figure}[tbp]
\begin{center}
\includegraphics[width=0.7\linewidth]{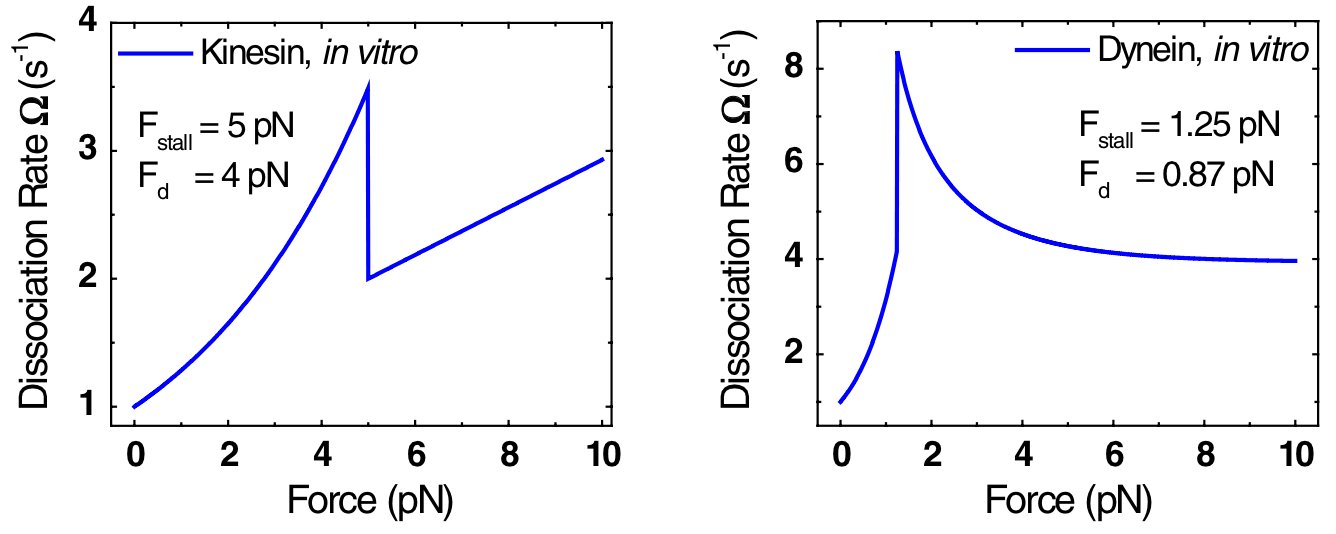}
\caption[Dissociation rate under load]{
Dissociation rate under load for kinesin (left) and dynein (right).
 From~\cite{kunwar2011}.
 }
\protect\label{fig:dissoc}
\end{center}
\end{figure}

The force-dissociation relations differ below and above stall.
Kunwar et al~\cite{kunwar2011} propose to write the detachment
rate as $\epsilon = \epsilon_0\Omega(F)$ where
$\epsilon_0$
is the detachment rate without load (typically between $0.25$ and $1$ $s^{-1}$) and $\Omega(F)$
gives the variation with the load, different for kinesins and dyneins.
Fits of {\em in vitro} data yield~\cite{kunwar2011}
\begin{equation}
\Omega(F) = \left\{ \begin{array}{ll}
\exp \left(\frac{F}{F_d}\right) & \mbox{ for } F < F_s\\
1.07 + 0.186 \frac{F}{f} & \mbox{ for } F \ge F_s
\end{array} \right. \;\;\;\; \mbox{ for kinesin}
\end{equation}
and
\begin{equation}
\Omega(F) = \left\{ \begin{array}{ll}
\exp \left(\frac{F}{F_d}\right) & \mbox{ for } F < F_s\\
\left( 0.254 \left[1-\exp \left(-\frac{F}{1.97 f}\right)\right]\right)^{-1} & \mbox{ for } F \ge F_s
\end{array} \right.
\;\;\;\; \mbox{ for dynein.}
\end{equation}
where $F_d$ is the characteristic detachment force
($4.01 \pm 0.07$ pN for kinesin, $0.87 \pm 0.04$ pN for
dynein), and $f=1pN$ determines the force scale.
These curves are represented in Fig.~\ref{fig:dissoc}.

\subsubsection{{\em In vitro} versus {\em in vivo}}

The discussion of the experimental results on the 
dynamics of molecular motors has shown that remarkable 
differences between  \emph{in vitro} and \emph{in vivo}
characteristics exist~\cite{schuster2011a}.
Some authors~\cite{kunwar2011}
report for example that stall forces
are completely different \emph{in vivo}, and in particular
that they become similar for dyneins and kinesins~\cite{leidel2012}
(which have quite different stall forces \emph{in vitro}).
The origin of these differences are so far not understood. 
It is not clear why motors should behave differently
in the cell or {\em in vitro}, and differences could
possibly by artefacts from the interpretation of measurements
which are much more difficult to do in a controlled way {\em in vivo}.
However, it is also possible that the motor activity is
strongly regulated by the cell.
The properties of 
the motors could be changed by the viscoelasticity of the cytoplasm
(in a similar way as it has been shown that mechanical
properties of MTs can be quite different in a viscoelastic
medium~\cite{brangwynne2006}).
Complementary proteins, as for example the tau protein, exist which can bind to MTs and reduce the binding affinity of molecular motors~\cite{dixit2009}. 
The attachment rate could also be influenced by regulatory proteins binding
directly to the head region of molecular motors~\cite{mallik2005}.

Another important difference between in vivo and in vitro
experiments is that the energy supply (i.e. ATP concentration)
is not controlled inside cells.
Though mitochondria provide most of the energy needed,
they are localized, while fast axonal transport requires
a constant supply of energy. Some recent
experiments~\cite{zala2013} with axons have shown that 
the transported vesicles may themselves carry
some glycolytic machinery that provides on-board
energy.

\subsubsection{Myosin}
\label{sec:myosin}

Though we shall not consider it later in the review,
we present here briefly the myosin motor family
for completeness, as some members of this family
are involved in intracellular transport.

Cytoskeletal myosin motors move along actin filaments.
The most famous one (called myosin II)
is present in muscles. The coordinated
motion of a large number of these myosins is responsible
for the contraction of muscles.
But many other myosins were discovered later, which all
move toward the plus-end of actin filaments, except
myosin VI that move towards the minus-end~\cite{wells1999}.
In this review we are interested in myosins
involved in actin-based axonal transport.
This is the case of myosin V
which transport organelles and
vesicles~\cite{tabb1998,alberts2008} and is able to take large
steps of $36$nm, which correspond to the actin helix
pitch~\cite{schliwa_w2003}. Similarly to kinesin,
it steps through a hand-over-hand mechanism~\cite{yildiz2003}.

\subsection{Motor modelling}
\label{sec:motmod}

Several modelling approaches can be pursued to model
cytoskeletal motors~\cite{bressloff_n2013}.
We shall now review a few
of them and discuss their domain of applicability.

\subsubsection{Ratchets}

A first family of models is based on ratchets.
Motor proteins live at a very small scale,
where viscous forces are important and where
thermal fluctuations play an important role.

\begin{figure}[tbp]
\begin{center}
\includegraphics[width=0.8\linewidth]{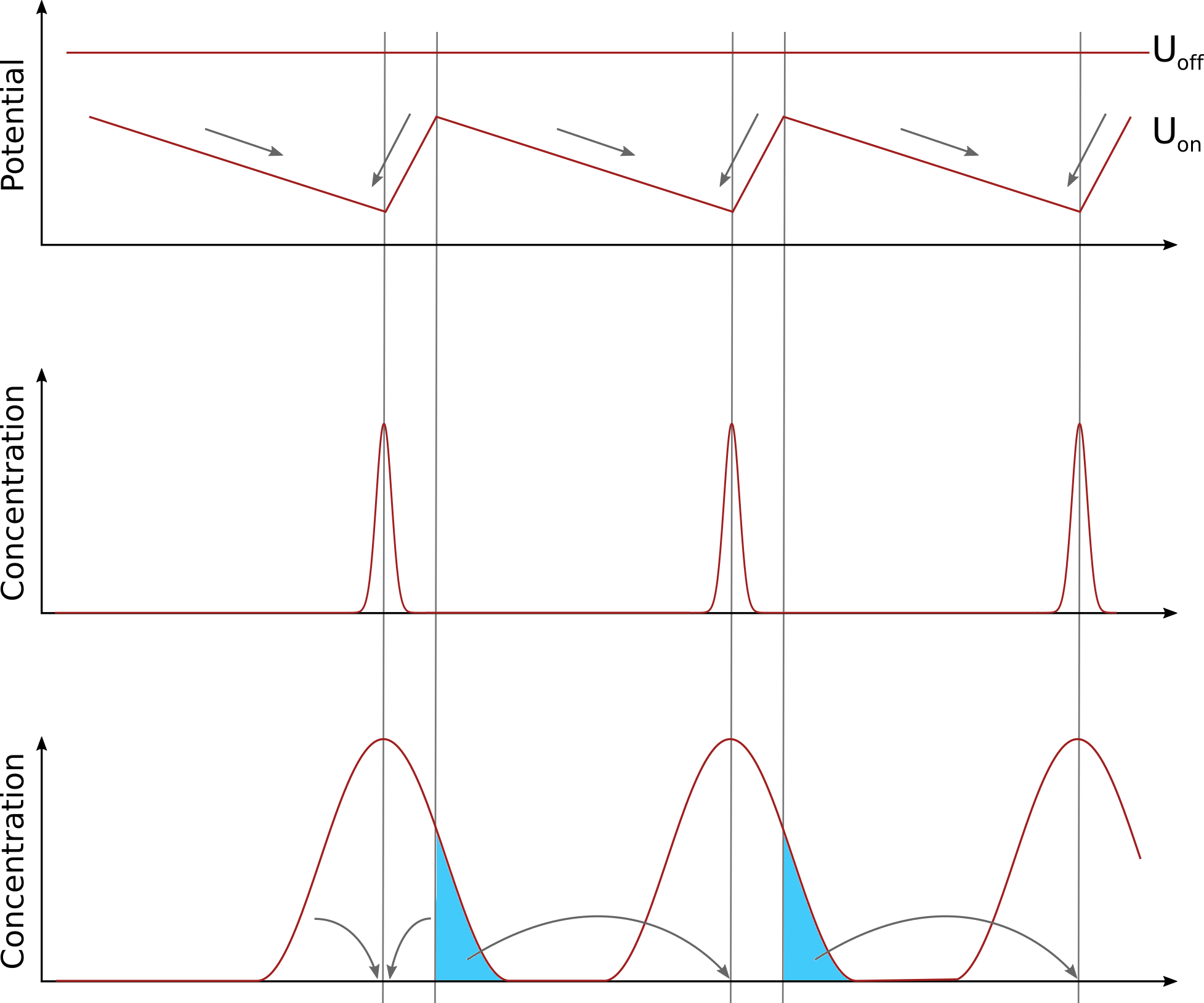}
\caption[Principle of a ratchet.]{Principle of a ratchet.
(A) on- and off-potentials. (B) Distribution of particles
after the on-period. (C) Distribution of particles
after the off-period.
}
\protect\label{fig:ratchet}
\end{center}
\end{figure}

The principle of a ratchet is to bias the
effect of the fluctuations
in order to produce some work. 
Its mechanism is illustrated in Fig.~\ref{fig:ratchet}.
Imagine a particle subject to a potential
that varies periodically with time~\cite{rousselet1994}. 
At each time period, an asymmetric potential $U_{on}$
is first applied during $\tau_{on}$. Then it is switched
off during $\tau_{off}$ (Fig.~\ref{fig:ratchet}A).
During the on-period, the probability distribution to find the
particle at a given location concentrates around the minima
of the potential $U_{on}$ (Fig.~\ref{fig:ratchet}B). During
the off-period, the distribution peaks enlarge into gaussians
due to diffusion.  When the potential $U_{on}$ is switched on
again, particles located in the hatched part of the distribution will move to the right due to the asymmetry of the potential
while the others will remain in the same minima. 
Of course the various time scales involved in this process
have to be tuned appropriately in order to obtain net motion.

Note also that this mechanism is highly out-of-equilibrium; an
external input is needed to switch between on and off states.
The transition rates between the two states are kept away
from the spontaneous values that they would have under
Boltzmann equilibrium~\cite{prost1994}.

The link with motor protein motion can be made in the
following way~\cite{julicher_a_p1997}.
Motors are considered to have two states, in which they
interact differently with the filament.
They switch from one state
to the other through the conformational change triggered
by ATP hydrolysis\cite{astumian_b1994}.
In each state, the interaction with the filament
can be modelled by a potential which is both periodic
in space and asymmetric, due to the periodic and polar
structure of the filament. If the potentials in each state
are different (different amplitude, or shifted in space),
it is possible to obtain net motion~\cite{prost1994},
even in the absence of diffusion~\cite{chauwin_a_p1994}.

Such a simple ratchet model can be used to model the
individual motion of, for example, kinesin on a microtubule.
Ratchet models are also appropriate to describe
collective motion of sliding motors, such as myosins teams
in muscle contraction~\cite{julicher_p1995,julicher_p1997,julicher_a_p1997}.
However, for collective MT-based transport, exclusion between the motors
play an important role which is not well described
by this family of models.
We shall see in the next chapter another family of models,
the exclusion processes, that naturally includes this
characteristics.

\subsubsection{Mechanochemical cycle of molecular motors}
\label{sec:cycle}

We have seen in the previous section how the switching
of motors between different internal states could result
into a net motion.
While ratchets models are useful to understand how
oriented motion is generated, we shall now present
a family of models 
that rather focus on the mechanochemical cycle itself.
These models allow in particular to explain the
variations of the motor stepping rate that are experimentally
observed when ATP concentration is varied~\cite{leibler_h1991,leibler_h1993}, or when
a force is applied to the motor.

\begin{figure}[tbp]
\begin{center}
\includegraphics[width=0.70\linewidth]{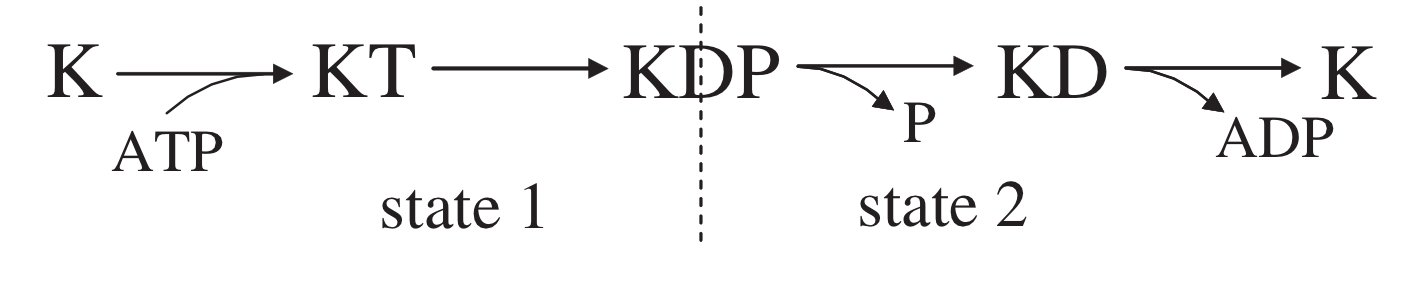}
\caption[Kinesin cycle]{Mechanochemical cycle of kinesin (K).
While the kinesin head is attached to the MT in state 1,
it diffuses freely along the MT in state 2.
From~\cite{nishinari2005a}.
}
\protect\label{fig:cycle_kin}
\end{center}
\end{figure}

One step of a motor is the result of several transitions between
different biochemical states. Depending on the degree of detail
that is needed, the number of the transitions that are
considered may vary~\cite{fisher_k1999}.
It is commonly accepted that the mechanochemical cycle of a
kinesin contains four main
transitions~\cite{fisher_k1999,block2003}, as represented in
Fig~\ref{fig:cycle_kin}:
Binding of ATP, hydrolysis of ATP (which produces
ADP + inorganic phosphate P),
release of phosphate (P), and finally release of ADP.
The model can be simplified by merging the states
corresponding to a kinesin head strongly bound to the MT
(states K and KT in Fig~\ref{fig:cycle_kin} -- state KDP has a
too short lifetime to be considered) 
and those
where the kinesin head diffuses freely along the MT (state KD).
This results into a two-state\footnote{This two-state model
for motors should not be mixed with the two-state model
for filaments of section~\ref{sec:dyn_inst}.}
model~\cite{kafri_l_n2004,nishinari2005a,lau_l_m2007} that
already captures important features of the stepping cycle.

The motor displacement is represented in a simplified way,
as successive hopping events on a discrete lattice (whose sites
would correspond to the potential minima of
figure~\ref{fig:ratchet}A).
More precisely, if we take the example of the two-state model,
and denote 
the state of a given lattice site by 0 if the site is empty,
or by 1 (resp. 2) if it is occupied by a motor in state 1
(resp. 2),
the stepping dynamics along the MT can be described by
the following transitions~\cite{nishinari2005a}:
\begin{eqnarray*}
\mbox{Hydrolysis: } & 1 \rightarrow 2  & \;\mbox{ with } \omega_h dt \\
\mbox{Ratchet: } & \left\{ \begin{array}{c} 2 \rightarrow 1 \\ 20 \rightarrow 01 \end{array} \right. & \begin{array}{l} \mbox{ with } \omega_s dt \\ \mbox{ with } \omega_f dt \end{array}  \\
\mbox{Brownian motion: } & \left\{ \begin{array}{c} 20 \rightarrow 02 \\ 02  \rightarrow 20 \end{array} \right. &\; \mbox{ with } \omega_b dt
\end{eqnarray*}
These rules implement the fact that oriented motion of kinesin
via a ratchet mechanism
can only take place when kinesin is in state 2.

The various transition rates may depend on the applied 
load and on the share of work that is carried out in each
sub-step. Some rates also depend on the ATP concentration.
Estimating these contributions allows to calculate
a number of experimentally
testable quantities, for example the stall
force and the load dependent velocity of the molecular motor,
or the dependence of velocity on the ATP
concentration~\cite{fisher_k1999,kafri_l_n2004,lau_l_m2007}.

Conversely, some informations on the mechanochemical cycle
can be extracted from experimental measurements of the
variations of stepping rates under various
conditions~\cite{kojima1997}.
As these processes are stochastic (and the stochasticity is
indeed included in the models), not only the mean value of
the motor displacement
but also the distribution of the fluctuations can bring some
information on the mechanochemical cycle,
for example to estimate the number of states in the
cycle~\cite{block2003}.

A theoretical frame for fluctuation analysis has been provided
in~\cite{lau_l_m2007}.
Due to the generic non-equilibrium features of the motor dynamics, generalized fluctuation
theorems have to be applied where the Einstein and Onsager relations are violated. It has been shown in 
this work that the degree by which the Onsager symmetry is broken is determined by the asymmetry of the 
effective potential which governs the motion of the motors. It turns out that the most efficient dynamics of 
kinesin is observed if the violation of the Onsager symmetry is maximal. 

Using the models of this section,
it is not only possible to study the
influence of an external force but also the
role of a disordered potential landscape~\cite{kafri_l_n2004},
which would model the heterogeneity of the intracellular
environment.
The resulting dynamics is reminiscent to a Sinai
walk~\cite{igloi_m2005}, where the exponent which describes
the velocity dependence on the size of the system, depends on
the slope of the potential.

\begin{figure}[tbp]
\begin{center}
\includegraphics[width=0.55\linewidth]{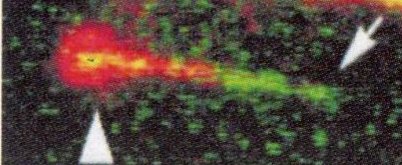}
\caption[Domain wall in kinesin motility assay]{
Domain wall in kinesin motility assay.
The triangle indicates the plus-end of the MT (green), where
kinesins (red) accumulate. The arrow points towards the
minus-end. A domain wall separates the high kinesin density
region from the low density one.
From~\cite{nishinari2005b}.
}
\protect\label{fig:nishinari2005b_fig3}
\end{center}
\end{figure}

The models of this section describe well the motion
of isolated motors, but they can also be extended
to many motor systems.
Nishinari et al~\cite{nishinari2005a} have considered
the collective motion of two-state molecular motors.
The motility of the motors depends on the
concentration of ATP, as explained above.
The interaction between motors is implement via an
exclusion rule: Only one motor can occupy a given site.
It was shown that, when attachment and detachment is included,
this two-state model 
is able to reproduce several features of \emph{in vitro}
experiments, particularly the existence 
of domain walls between high and low motor density
regions~\cite{nishinari2005a,nishinari2005b}
(see Fig.~\ref{fig:nishinari2005b_fig3}).
This model was extended to several parallel
filaments in~\cite{chowdhury_g_w2008}.

An extensive review on the stochastic mechano-chemical
kinetics of molecular motors can be found
in~\cite{chowdhury2013}.

We have seen in this section a first example
of model where the exclusion between motors is implemented.
We shall describe exclusion processes in much more details
in chapter~\ref{sec:ep}.
For the remaining of this review, we shall not consider
anymore the details of the mechanochemical cycle.
However, the dependence of the stepping rate on ATP
concentration or on load force, which can be explained with the
models of this section, will be used whenever needed.

\subsection{Motor-cargo complexes}
\label{sec:tug}

Molecular motors transport various types of cargos
\cite{welte2004,encalada2011,maeder_s_h2014} of
different sizes, such
as filament precursors, mRNA granules~\cite{vuppalanchi_w_t2009}, 
lipid droplets~\cite{welte2009}, viral capsids~\cite{lyman_e2009},
lysosomes, or even objects as large
as mitochondria~\cite{hollenbeck_s2005,ashkin1990,saxton_h2012}.

One important fact that has to be considered for a large variety of intracellular transport phenomena
is that, in most cases, cargos are not transported by single
but by multiple motors which are attached to their
surface~\cite{ashkin1990,hancock2014}.
Experimental evidence for the attachment of many motors is
based on a number of observations, in particular based on
force measurements~\cite{shubeita2008,soppina2009,kunwar2011,leidel2012}.
Another argument is that, 
for several motor-cargo systems, there is evidence for
bidirectional non diffusive motion~\cite{welte2004}. 
Note that the attached motors bind individually to the MT
and may a priori have opposite preferential directions.
It is thus not obvious to understand how they cooperate
to transport a given cargo.

In vitro, simpler systems involving only one type of motors
can be studied. For example, some experiments have been carried out where
dynein coats beads transported along immobile microtubules~\cite{mallik2005}.
In these experiments it was shown that transport is much more efficient
when more than one dynein is attached to the bead. An obvious interpretation is that, when one motor detaches,
another one can take on. In this special setting, strongly directed motion is observed, with a typical cargo displacement 
of the order of several hundreds of micrometers. 
In~\cite{mallik2005} evidence was given that the stall force
for multiple dyneins seem to be additive.

{\em In vivo}, both dynein and kinesin are present
and can attach to a given cargo.
There is a need to understand in a more refined way how
multiple motors handle cargos, for example 
in order to discriminate between coordinated and uncoordinated motion of the motors.
A first hypothesis could have been that 
only one species of motors would be attached to the cargo at a given
time: Motors would just share the load but all of them
would be pulling in the same direction (Fig.~\ref{fig:coordination_mechanism}(A)).
However, this scenario could be ruled out for several systems~\cite{ma_c2002}.
It is rather observed that motors of different types (i.e.
having opposite preferred velocities) are
attached at the same time to a given cargo~\cite{kural2005}.
There is a need to understand how the cargo displacement
emerges from these multiple interactions.

Experimentally, one has to distinguish the cargo displacements
at short time scale, along a single filament,
and displacements over large time scale, during which the cargo
may have followed different MTs, oriented
in various directions. In the latter case, the randomness
of the MT network itself may lead to superdiffusive cargo
dynamics~\cite{shaebani2014}.
In this section, we shall concentrate on cargo transport
along a single filament, while the effect of network will
be addressed in section~\ref{sec:dyn_on_networks}.

From a practical point of view, it is more easy to study
dynamics along single filaments {\em in vitro}.
However, until recently, it was not known how to handle purified
mammalian dynein 
in motility assays\footnote{Motility assays are {\em in vitro}
experiments in which usually one (but possibly more)
type of motors performs processive motion along some MTs put in
a solution or anchored on a surface.},
thus until now, most data on cargo transport
by teams of motors have been obtained {\em in vivo}.
One way to limit the aforementionned network effects
in living cells is to consider one-dimensional systems.
We shall see in chapter~\ref{sec:vivo} that in axons, the
MT network is quasi-1D.
But some other systems can also be considered.
Epithelial cells, for example, share with axons the
property of having a highly polarized MT network.
Also, in late {\em Drosophila} embryos, lipid droplets
are moved on a quasi unipolar MT network~\cite{mueller_k_l2008}.
Many studies have been devoted to the transport of
lipid droplets,
 in order to follow typical trajectories of particles
 that are driven by teams of molecular motors, and
we shall report several such experimental results in the
next section.

\subsubsection{Transport of lipid droplets}

The most extensively studied cargo transport by teams of molecular motors is the transport of lipid droplets~\cite{welte2009}, 
which can also be traced \emph{in vivo}. In order to characterize the dynamics of lipid droplets several techniques
can be used. The most widely used are based on fluorescence
microscopy~\cite{greenspan_m_f1985}.
The high contrast and resolution of this method allows to track individual lipid-droplets very accurately. Besides, it
allows to follow many droplets at a time.
Despite 
its obvious advantages there are also some limitations. First
of all the observation time (or observation frequency)
may be limited by 
photo-degradation. It has also been observed that the excitating laser is heating the cell and thereby is influencing 
the dynamics of the observed objects \cite{mueller_z2007}. 

More recently, various imaging methods have been
developped that do not require to introduce an
external label in the system.
Confocal reflection microscopy~\cite{gaspar_s2009}
measures the light reflected by the particle (e.g. lipid droplet). This technique is less invasive, and allows for higher rate
measurements. It is of course restricted to reflecting
particles.
Differential interference contrast (DIC) microscopy
is based on variations in the index of refraction.
All these techniques, if used in an optimized way,
can achieve an accuracy of the order of
5 nanometers~\cite{carter_s_g2005}.
Other techniques that do not require labels are
based on harmonic generation~\cite{watanabe2010}
and coherent Stokes Raman scattering (CARS)
microscopy~\cite{jungst_w_z11}.
Several techniques can be combined
to mark different elements of the cell.

Whatever the technique, almost all experiments report
the same type of data:
Lipid droplets motion is usually characterized
in terms of alternating runs and pauses
 (note that the precise
distinction between runs and pauses is somewhat arbitrary as it
depends on the choice of a threshold~\cite{petrov2007}).
Typical data report the length and duration of single
runs and pauses. Runs may occur in both directions (though
with possibly asymmetric distributions) and
can be quite large (up to thousands of
nanometers~\cite{gaspar_s2009}).
Pauses could be attributed to several features. Detachment
and diffusion periods could be longer than \emph{in vitro} as they take place in
a viscous environment~\cite{mallik2005}. However, estimations
of the number of attached motors (around five of each type in~\cite{kunwar2011},
$1$ or $2$ kinesins against $4$ to $8$ dyneins in~\cite{soppina2009}) 
make the full detachment of the cargo quite unlikely. Competition between different types
of motors, or obstacles could also explain these pauses.

It is commonly accepted that the lipid droplets are 
transported over long distances. But there is no
clear experimental evidence whether these large displacements
which are found in short-term measurements of uncorrelated
large bidirectional runs persist over longer time scales,
i.e. whether there is really a net drift of the overall motion.
Little is known about the net drift over
large time scales.
It would be an important information to discriminate
between models. Current techniques allow a tracking of
lipid droplets over long periods that should give
access to this data~\cite{gaspar_s2009}.

Bidirectional motion can be useful even if there is no net displacement (non biased random
walk)~\cite{wacker1997}.
Indeed, the random walk behavior of the cargo
may be useful just to explore the space,
and deliver for example vesicles at the proper place,
through a trial and error mechanism~\cite{wacker1997}.\footnote{Alternating 
slow and fast random motion has been proven as an efficient search strategy
("intermittent search"), which is realized in several biological systems (see 
\cite{bressloff_n2013} and references therein. }

Even with net drift motion, bidirectional motion
can be more efficient
than unidirectional transport in a crowded environment
as the axon: When the cargo is
blocked by an obstacle, stepping back allows to try
to find another path around.

\subsubsection{Tug-of-war}

\begin{figure}[tb]
\begin{center}
\includegraphics[width=0.3\linewidth]{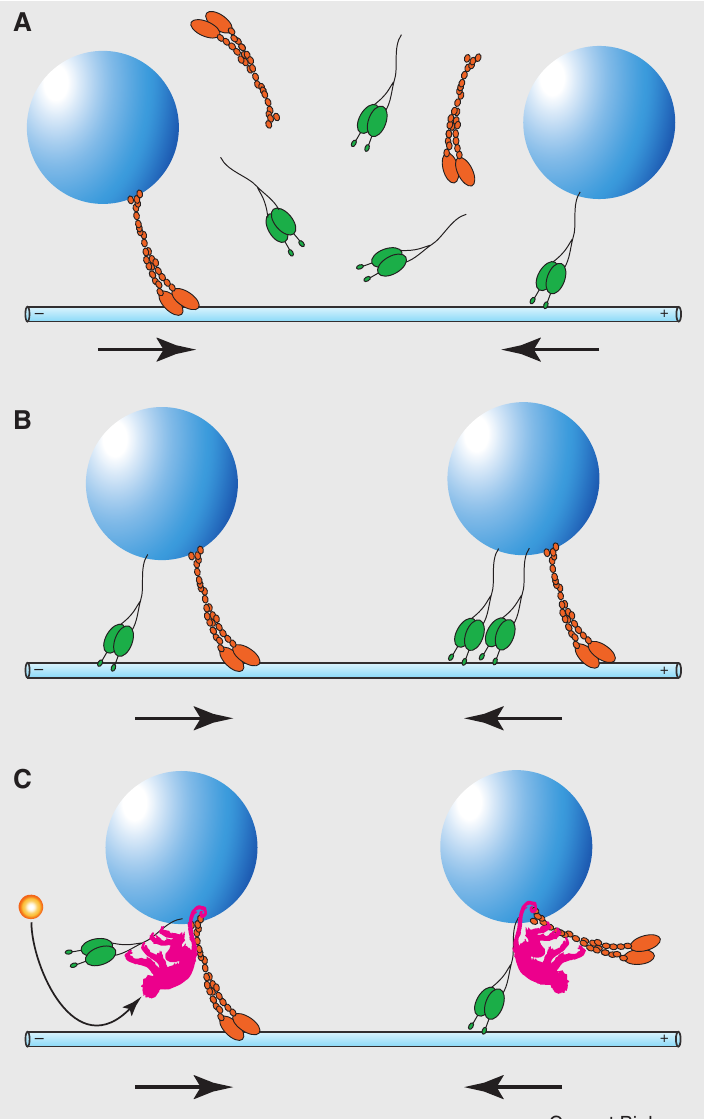}
\caption[Possible mechanisms for bidirectional transport.]{Possible mechanisms for bidirectional transport. Plus-end directed motors (orange) and minus-end directed motors (green) move a cargo (dark blue) along a MT (light blue). {\it (A)}~A single motor species can be attached.
    {\it (B)}~Tug-of-war of different motor species.
    {\it (C)}~A coordination machinery determines moving direction.
    From~\cite{welte2004}.}
\protect\label{fig:coordination_mechanism}
\end{center}
\end{figure}

It has been shown that in general, the two types of opposite motors are attached at the same time
on the cargo~\cite{welte2004,hendricks2010}.
Then two scenarios can be considered~\cite{gross2004,schlager_h2009,blehm_s2014} (Fig.~\ref{fig:coordination_mechanism}(B) and (C)).
Either there is a - yet not fully specified - mechanism that
coordinates the motion of motors, so that motors
of one type are inhibited while the others pull
the cargo and the other way round, or bidirectional
motion emerges spontaneously through stochastic fluctuations
of the number of motors attached on the cargo at
a given time. The latter mechanism is referred to as
``tug-of-war''\footnote{Tug-of-war refers to a game where two teams pull
in opposite directions on one rope, and try to force the other team into their own camp. Tug-of-war contests were organized at the Olympic Games from 1900 to 1920}.

In the tug-of-war picture, the attachment of motors of
each species results from a random procedure, and
it is only because of random fluctuations that one
or the other species will take over and carry the cargo
towards its preferred direction.
As a result, the cargo will go back and forth,
possibly performing large (compared to the typical motor
step size) excursions in each direction~\cite{hendricks2010}.
An experimental observation in favour of this scenario
was made on endosomes which were found to be elongated
when pausing~\cite{soppina2009}. This could be a consequence
of the pulling of opposite teams during tug-of-war episodes.

Some theoretical models have been proposed, in which
the number of cargo-bound motors of each type that are also attached
to the MT 
are taken as internal degrees of freedom~\cite{klumpp_l2005a,muhuri_p2008,muhuri_p2010}.
A mean-field version, in which forces are equally shared among
attached motors of the same type, showed that
drift motion could be spontaneously obtained when one type of
motors pulls the cargo strongly enough to have the motors of
the other type detach~\cite{mueller_k_l2008,zhang_f2010}.
The resulting high motility states obtained 
by forcing the minority motor species to detach
from the filament, and characterized by a symmetric bimodal
(or trimodal)
distribution of velocities,
can be seen as some kind of spontaneous
synchronization
and do not require the action of an external factor.
However in this model, though the attachment and detachment of
motors is considered, the explicit dynamics of motors along
the filament is not described, which is required 
in order to go beyond the mean-field approach and to
compute the forces individually for each motor.

Indeed, stochastic models with explicit position tracking of the motors
have been proposed, in which the individual forces felt by the
individual motors which are located at specified positions on the MT
are taken into
account~\cite{kunwar_m2010,kunwar2011,klein_a_s2014a}.
Such models were developped for cargos driven either
by one type of motors~\cite{korn2009,bouzat_f2010}
or by two types of opposite motors~\cite{kunwar2011,bouzat_f2011,bouzat_l_b2012,klein_a_s2014a} - the latter being the case
of interest here for us.
The comparison between these stochastic models and the mean-field model
revealed strong differences in the dynamics~\cite{bouzat_f2011}.
Indeed, when motor position is explicitely considered,
it turns out 
that motors which are oppositely directed to the dominant motor species are not 
automatically disrupted from the MT.
Thus the efficiency of the
tug-of-war mechanism is drastically reduced, in the sense
that the symmetric bimodal distributions
observed in the mean field model
do not exist anymore~\cite{klein_a_s2015}.
This is not necessarily a drawback, as these 
symmetric bimodal distributions
have never been clearly evidenced experimentally.
Still, the stochastic model reproduces well 
that the tug-of-war leads to anomalous diffusion
in agreement with some experimental
observations\cite{kulic2008,caspi_g_e2002,robert2010,salman2002}.
Indeed, at very short time scales and if thermal fluctuations
are taken into account, subdiffusion is observed as a result
of the trapping of the cargo by oppositely pulling
motors~\cite{klein_a_s2014c}. At larger time scales (typically
tens of ms), there is a crossover towards superdiffusion which
indicates that cargo-mediated coupling induces some cooperativity
between the motors~\cite{klein_a_s2014b}.
At even larger time scale, one expects to recover pure diffusion,
but usually network effects come into play first.

Besides, the fact that the response of motors to external factors
(load, [ATP], etc) depends on their type~\cite{mallik_g2004} provides a way for
the cell to control the bias of the cargo random walk.
Indeed, a change for example in the viscosity of the surrounding medium
can modify the bias in a non trivial way, and even reverse it, by
changing the balance between motors' teams~\cite{klein_a_s2014b}.
This mechanism could in particular
allow cargos to escape from crowded areas if the role of obstacles
could be assimilated to a large effective viscosity.

Some complementary mechanisms, for example through
hydrodynamic interactions, could also play
a role in obtaining correlated
motion~\cite{malgaretti_p_f2012,malgaretti_p_r2013}.

\subsubsection{Coordination-based scenarios}

Another hypothesis would be that, instead of resulting
from a purely random procedure as in the tug-of-war,
motion would be produced through some coordination
mechanism.
Indeed the available experimental data are consistent with a number 
of regulatory mechanisms, which we will partially review now.

The perhaps most direct option to control cargo transport 
is by the cargo itself: Through specific proteins that are
located at the cargoes' surface  it controls the docking or the 
activity of the attached motors~\cite{karcher_d_g2002}.
It has been suggested that this mechanism could orient synaptic cargos
preferentially towards MT or actin
networks~\cite{schlager_h2009}.

Another mechanism, which is also based on the selective activation 
of the different motor species, is provided  by molecules which are 
potentially able to coordinate the motor activity~\cite{encalada2011}.
Dynactin, for example,
 usually turns the  minus-directed motors off if the plus-end directed motors are active~\cite{gross2003a}. 
Thus binding of dynactin should lead to a persistent plus-end directed 
motion of the cargo~\cite{mallik_g2004}. 
A crucial regulatory factor in this process is the
hungtingtin protein, that interacts with dynein, dynactin
and kinesin. It was shown in primary cultures
of neurons that the phosphorylation of huntingtin
promotes anterograde transport, while retrograde transport
is favored when hungtingtin is not
phosphorylated~\cite{colin2008}.

Finally it is also possible to alter the properties of the MT filaments via 
decoration with so-call MT associated proteins (MAPs). MAPs as for 
example the tau protein can influence the binding affinity of molecular 
motors, which would have an influence on the run length of the cargo, 
since the average number of motors that is attached to the MT is reduced.

Whatever the mechanism is, cargos do undergo directed motion in cells.
Cooperative effects of the motors
 are observable, leading for example to increased 
processivity of a transported cargo and higher stalling forces than for individual motors 
which has been shown experimentally~\cite{mallik2005,beeg2008}.
Cooperation (spontaneous or mediated) between motors thus
enhances long-range transport of individual cargos.

\subsubsection{Transport of large cargos}
\label{sect:large_cargos}

The transport velocity of cargos may depend on their
size, the larger ones travelling at smaller velocities~\cite{smith_s2001}.

One important example of a large cargo is  mitochondria, typically
measuring $10\mu m \times 0.2\mu m$. Mitochondria are present 
in almost every eucaryotic cells and can be transported along
 microtubules. Remarkably mitochondria are also to be transported along the
axon~\cite{pilling2006,hollenbeck_s2005,otoole2008} although
their size is much larger than the normal interdistance
of MTs in the axon ($\sim 20-30nm$ \cite{chen1992}).

Several mechanisms have been proposed which enables mitochondria 
to move in this crowded environment. There may for example 
be mechanisms based on the dephosphorylation
of MAPs that would allow MTs interdistances to grow
so as to leave a free tunnel through which the vesicle
could pass~\cite{shahpasand_a_r2008}.
It is also possible that MTs locally depolymerize and 
open this way free space for giant vesicles and mitochondria.

To conclude, though much progress has been done to
understand the mechanisms underlying motor-cargo complexes,
their transport is far from being fully understood.
Since most of the experiments are carried out {\it in vivo}, the actual
configuration of the motors attached to a given cargo~\cite{blehm_s2014}
as well as the influence
of e.g. regulatory proteins is {\it a priori} not known. Therefore it would be
of great interest to carry out {\it in vitro} experiments, where it is possible
to control the configuration of molecular motors
that are attached to the cargo and the environmental parameters.
Actually, a recent discovery on how to handle mammalian
dynein in motility assays~\cite{mckenney2014c} should allow to build
in the near future {\em in vitro} experiments with two types of motors
and to study bidirectional cargo motion in much
more controled settings. Much will probably be
learned on these systems in the coming years.

\subsection{Dynamics of molecular motors on filament networks}

\label{sec:dyn_on_networks}

In this review, we have mostly considered transport
at the scale of a single filament.
However, in order to describe intracellular transport
at larger scales, 
it is necessary to take into account
the global structure and dynamics of the network.
An important set of work has been done on this subject
using non-linear continuous models. Several reviews
have been already published on this subject,
to which we refer the reader for further information~\cite{julicher2007}.

In the following, we shall rather address  this subject
using microscopic models.

In \cite{greulich2010,greulich_s2010} the dynamics of self-driven particles on regular and inhomogeneous networks embedded in a two dimensional diffusive environment has been investigated. The dynamics of the particles alternates between diffusive motion of unbound particles and driven motion if the particles are attached  
to filament like structures. The active transport of particles leads generically to the formation of particle clusters at the intersection of filaments. Interestingly the size-distribution of the particle clusters depends qualitatively on the chosen geometry of the filament network: On regular networks one typically observes exponentially decaying cluster distributions at low densities, i.e. one can attribute a typical scale to the clusters. By contrast, on inhomogeneous filament networks typically algebraic cluster-size distributions are observed in a large parameter regime. 

The origin of the scale-free cluster-size distribution can by deduced from the following consideration: Clusters nucleate if two particles encountering at an intersection of two filaments block each other such that they cannot move, while other particles running towards the intersection form queues. If the particle density exceeds a critical density $\rho^*$, queues form that cover multiple filaments and  induce branching of queues into large clusters. Now each branch contributes to the particle support of the cluster. Since the number of filaments a cluster covers is proportional to its size, the growth rate of a cluster is also proportional to its size. Dynamics of this kind can be described by a Yule process \cite{yule1924} which yields a power law distribution $P(m) \sim m^{-\gamma}$ with an exponent depending on the microscopic parameters. For disordered filaments at moderate particle densities, it turned out that 
the exponent describing the decay of the cluster-size distribution depends on the particle density  and the typical mesh-size of network, while dependence on the other model parameters is rather weak.    

We have seen that the distribution of particles on the filament
network was quite different depending whether particles were
transported through active transport on the one hand,
or random walk and aggregation processes on the other hand.
There is a need to understand these differences more
thoroughly.

In view of the application to intracellular transport it would be interesting
to study the aggregation on dynamic networks as well, since the steady
reorganization of the filament network may have a strong influence on the
formation of particle clusters  (We shall see in section \ref{sec:dyn}
such an example).
It is also possible that the motors themselves modify the dynamics
of the network. For example kinesin 8 was shown to promote
MT depolymerisation, a feature that could play a role in length
regulation of MTs.

Indeed, this would contribute to the understanding of
complex questions such as the potential toxicity of nanoparticles.
Nanoparticles are known to pass the cell membranes \cite{oberdorster_o2005}.
It is important to have some knowledge about the way
they are transported and where they are aggregated inside the cell,
to determine whether they can have an impact on cell functions.

To conclude this chapter, 
though a complete description of cargo motion requires
the full description discussed in this chapter,
at a coarse grained scale, cargo-motors complexes
could themselves be represented as single particles
(and backstepping should then be included).
Thus in the next chapters, though most of the time
single particles will be thought as single motors,
it is also possible to consider that they represent
a whole cargo-motors complex in a more coarse grained
description.

\section{Exclusion Processes}
\label{sec:ep}

When trying to model such complex issues as intracellular transport, one is usually confronted with several difficulties. First of all, there is the huge complexity of the problem. As we shall discuss in section~\ref{sec:axon} for the example of motor-driven axonal transport, transport processes 
are typically the result of the dynamics of many different proteins interacting in ways which are not fully understood. This lack of knowledge makes modeling a challenging task. But even if we had a perfect knowledge about the intracellular processes on a molecular level, it would still be hopeless to model motor-driven transport in every detail. Therefore, it is necessary to describe motor-driven transport on a coarse grained level or even by a continuous approach.

In this chapter we will discuss a stochastic approach to motor-driven intracellular transport. 
Stochastic modeling has proven itself as a powerful tool in systems with complex interactions on a microscopic scale. A convenient example is the Brownian motion of a small particle in a suspension. Although the collisions between the molecules of the liquid and the particle are not explicitly considered, the stochastic treatment of the random particle dynamics describes well the real system. By describing the dynamics of the system in terms of stochastic transitions between different states, the microscopic interactions are characterized only by giving the transition rates. This approach hence corresponds to a treatment on a coarse-grained scale and overcomes the problem of lacking knowledge or extreme complexity.

In the past, stochastic lattice gas models have been used in a wide range of
active transport phenomena, e.g., pedestrian
dynamics~\cite{schadschneider2002b,vicsek_z2012,cividini_a_h2013}, car
traffic~\cite{chowdhury_s_s2000,nagel_s1992,appert_s2001} and intracellular
processes~\cite{macdonald_g_p1968,chowdhury_s_n2005,lipowsky_k2005}. Lattice
gas models are 
 discrete in space, i.e. particles which occupy the sites of the lattice hop stochastically from one site to another. The particles move according to 
hopping rules and rates which are chosen to mimic the properties of the modeled system. In active transport models, the hopping rates may not be 
related to a potential via the detailed balance condition. Therefore  generic non-equilibrium phenomena, as for example  
macroscopic currents in the stationary state, are frequently observed in transport models. All the models which we will discuss in this review are
 non-equilibrium models.

In the case of intracellular transport, stochastic lattice gases
are an appropriate choice because the hopping of particles on the
lattice can be directly understood as a step of a molecular motor
like kinesin or dynein along the MT. The different stages of the
mechanochemical cycle during which ATP is converted to ADP
and mechanical work is produced
(see
section~\ref{sec:kinesin}) can be summarized in a single hopping rate,
hence drastically simplifying the problem
(though in a more detailed description
it is also possible to retain some features
of the mechanochemical cycle as has been discussed in
section \ref{sec:cycle}).

We shall present in this section the so-called
totally asymmetric simple exclusion process (TASEP)\footnote{The {TASEP} model was first proposed in the context of mRNA translation and synthesis of proteins~\cite{macdonald_g_p1968}.
We will not detail this application, as reviews
on the subject already exist \cite{zia_d_s2011,chou_m_z2011}.}
as a minimal model that includes 
important features of the dynamics of molecular motors~:
the stepwise, directed, and stochastic motion of single motors as well as the steric interactions between motors are described.
Nevertheless the original TASEP is an incomplete model for transport
even on the coarse grained level, since. e.g., the
attachment and detachment of
motors is not included. Therefore the TASEP should be seen in the
context of this review as a reference on which more detailed
approaches can be built.

Before presenting more sophisticated models which have been introduced in order to describe intracellular transport driven by molecular motors, we shall summarize in this section some basic properties of exclusion processes, and some variants whose properties will be useful in order to analyse intracellular transport models.
At the same time, driven
stochastic lattice gases have proven to exhibit a rich
phenomenology already in one-dimensional systems making them an
interesting object of study from the point-of-view
of statistical physics. In fact, phase transitions in
one-dimensional systems without long-range interactions can only be
observed in non-equilibrium systems.

\subsection{Totally asymmetric simple exclusion process (TASEP)}
\label{sec:tasep}

\subsubsection{Model definition}
\label{sec:tasepdef}

\begin{figure}[tbp]
  \begin{center}
  \includegraphics[scale=1.0, clip]{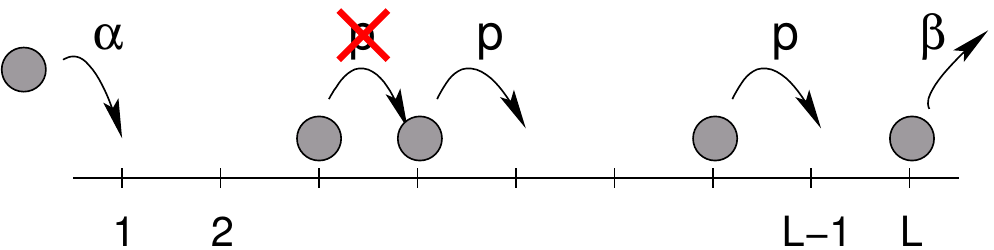}
\caption[The TASEP with open boundary conditions.]{The TASEP with open boundary conditions. A particle hops to the next site on the right if that site is empty at rate $p$. If vacant, the first site of the system is filled from the reservoir at rate $\alpha$ and a particle on the last site can leave the system at rate $\beta$.}
\protect\label{fig:TASEP}
\end{center}
\end{figure}

The TASEP is probably the most extensively studied model of
stochastic transport. It is defined on a one-dimensional lattice
with $L$ discrete sites that can either
be vacant or occupied by a particle (figure~\ref{fig:TASEP}).
The state of the system at a given time is given by 
$\{\tau_i\}_{i=1\;to\;L}$,
where $\tau_i$ gives the occupation of site~$i$ taking values $0$ or $1$ if that site is vacant or occupied, respectively.
A particle hops stochastically  to a site on its right with rate
$p$ if that site is vacant, i.e., no
site can be occupied by more than one particle at a time.
Hopping to the left is
prohibited in this model but possible in other simple exclusion
processes as the symmetric simple exclusion process (SSEP) or the
asymmetric simple exclusion process (ASEP). For the purposes of transport
by molecular motors, the totally asymmetric variant is often appropriate
because backward steps of motors as kinesin-1 are rare and can be
neglected (see section~\ref{sec:kinesin}).

Two different boundary conditions are usually considered: Under \emph{periodic boundary conditions}, the system is closed into a ring so that a particle which hops from site~$L$ arrives at site~$1$. This choice obviously exhibits conservation of particles and it can be shown that in this geometry, every allowed configuration has the same probability of occurrence~\cite{meakin1986a,derrida1998c}. Open boundary conditions are realized by coupling the chain to two boundary reservoirs. Usually one considers boundary reservoirs of infinite capacity. Particles enter the system with rate $\alpha$ at the left boundary if the first site is empty and leave the system with rate $\beta$ from the last site if it is occupied. 

\subsubsection{Update scheme}
  
In order to simulate the time evolution of the TASEP,
one has to chose an update scheme, that is one has to
define when and in which order the particles are attempting
to hop forward.

Let us first make a general remark about systems at equilibrium
(which is not the case of the TASEP).
In equilibrium statistical mechanics, lattice models are usually
defined by the different states of the lattice, the interactions between the
state variables and the geometry of the lattice.
From this definition, a partition function can be calculated,
which fully determines the equilibrium state of the system.
In order to explore numerically the
equilibrium properties of these models one can apply different Monte Carlo
schemes which sample stochastically the configuration space. It can be shown
that all ergodic Monte Carlo algorithms that fulfill the detailed balance
condition lead to the same stationary state, i.e. the equilibrium state of the
system.

For non-equilibrium systems as those considered in this review,
the model is defined by a master equation.
This implies that the update scheme is part of the definition
of the model.
In the case of the TASEP, as for many other non-equilibrium
systems, it 
can be shown that a unique stationary state exists for a given
update, but the stationary state and its properties
can differ from one update to another one~\cite{rajewsky1998,khoromskaia_h_g2014}.
We shall present
now the most widely used types of updates that have been
considered to model
stochastic transport.

\subsubsubsection{Random sequential update}

If we were considering a continuous time evolution, the distribution of time 
intervals between two events would follow an exponential law.
The most appropriate numerical method for stochastic processes continuous in 
time
would be the Gillespie
algorithm~\cite{gillespie1977,gillespie2007}, in which these
exponential time interval distributions
are realized.
However, in the special case where all rates are 
of
the same order of magnitude and where one considers
a large system, it is more convenient to use a discrete
time algorithm, namely the random sequential update.

For the TASEP, the random sequential update can be implemented as follows. At each micro-time step, a link (separating two neighboring
cells) is chosen. If there is a particle on the left and no particle on the right of this link, the particle is
moved forward with probability $p$~\footnote{Computationally it
is more efficient to to choose a particle (or a hole for
densities $\rho > 1/2$) at each micro-step, and to move it
forward (backward) with probability $p$ if the target site is
empty. For a closed system, with fixed number $N$ of particles,
the microstep is constant and lasts $1/N$. For an open system,
the microstep duration depends on the (varying) number $N$ of particles.
}. If the chosen link corresponds
to the entrance (exit), then a particle is injected on the first site (removed from the last site) with probability
$\alpha$ ($\beta$).

The duration of a microstep is 1/\#links\footnote{The notation \#links stands for the number of links.}, so that {\it on average} each link is updated once per unit time step.
The set of the \#links updates that occur during one unit time step is usually called a {\em sweep}.

Though it is not strictly identical to a continuous time dynamics, the random sequential update gives a very similar behavior for large system sizes, in particular when simple quantities such as current and densities are measured.
Still, one must be aware that discrepancies may arise
when more sensitive quantities are measured, such as
generating function or higher order cumulants (see section 6
of \cite{derrida_a1999} for an example).
An important feature of this update is that one particle may be chosen several times in a row while another one may be ignored during several time
units, which leads to non-negligible current fluctuations even at low densities. It is this update that has been considered in most 
fundamental studies, for which TASEP was taken as an archetype of non-equilibrium systems.

It should be noted that for the random sequential update, time can be rescaled. Indeed, the only effect of changing the
physical duration of the unit time step is to change the numerical values  of the rates in numerical time units - so that they remain
unchanged in physical time. Usually, one chooses the time step so that the largest
rate defined in the system is equal to one (in numerical units).
This allows to have the largest ratio of acceptance for the transitions (the time scale must
be chosen so that rates are not larger than one, as they must be interpretable as probabilities).

\subsubsubsection{Parallel update}

For some applications, a certain regularity of the particle motion
exists, which limits for example the number of steps that
can be carried out in a given time interval.
One possibility to obtain this regularity is to use a
parallel update scheme, an update which is widely used
in order to model e.g. road
traffic~\cite{nagel_s1992,eisenblatter1998}.
For these applications not only the regularisation of the speeds is of
importance but also the implicit introduction of a reaction time, which is
necessary in order to represent traffic states in a realistic way.

In the parallel update~\cite{evans1997,schadschneider_s1998,wolki_s2009}, all
sites are updated at the same time. The state of a given site at time $t+1$
depends on the state of this site and its neighbors at time $t$, through --when
applied to the TASEP-- 
the rules introduced in section~\ref{sec:tasepdef}.
In contrast to the random sequential model $p$ now is a hopping probability 
rather than a rate. This implies that for $p=1$ the bulk dynamics of the
particles is deterministic~\cite{tilstra_e1998,degier_n1999,evans_r_s1999}
in contrast to the random sequential model where the particle motion is
stochastic for any finite hopping rate.

The parallel update introduces a time step, which now cannot be arbitrarily
rescaled. The time step can be interpreted as a reaction time or as a time
which is needed in order to carry out the microscopic step. 
Another feature of this update is that not all the states of the system are
accessible through the dynamics. The non-accessible states were called "Garden
of Eden" states~\cite{schadschneider_s1998} (Note that the definition of
non-accessible states requires to include not only the positions but also the
velocities of particles in the definition of the states). 
They are related to the existence of strong short-ranged correlations. In case
of the TASEP the GoE-states introduce a particle-hole attraction, which is more
pronounced as $p$ is closer to $1$ and may lead to high current states  of
alternating empty and occupied states.

Conflicts naturally arise with parallel update if two-dimensional lattices are considered, as this is the case for pedestrian traffic: 
Two particles may attempt to hop to the same site at the same time, which is forbidden by the exclusion principle. Therefore 
 one has to introduce additional stochastic or deterministic rules, which assign the priority to one of the particles that try to occupy a given site.
One may argue that in some case, these conflicts should not be
considered as an artifact of the model but are also present in the real systems, e.g. when several pedestrians are trying to
walk through the same
door~\cite{kirchner_n_s2003,schadschneider_s2009}.
Anyhow, the presence of conflicts requires to supplement the
update rules with ad-hoc conflict-solving rules
which can be deterministic or stochastic.
These rules may not be that relevant
at intermediate densities, and their choice can be quite arbitrary.
This was enough to trigger the development of new updates
that would at the same time be sequential and have bounded
fluctuations.

\subsubsubsection{Shuffle updates}

In the {\em random shuffle update}\footnote{Note that in some communities (for example working on agent based models), the
random {\em shuffle} update is called random {\em sequential} update. This
is a source of confusion that the reader must be aware of. In this review, we
keep the definitions usually used in
physics.}~\cite{wolki_s_s2006,wolki_s_s2007,smith_w2007a,klupfel2007a}, at each
time step,
each particle is updated exactly once, but the order in which particles are updated is chosen randomly
at each time step.

If the order of update is chosen once for all at the beginning of the simulation, this update will rather be called {\em frozen
shuffle update}~\cite{appert-rolland_c_h2011a,appert-rolland_c_h2011b,appert-rolland_c_h2011c}.

Random or frozen, the shuffle update is a sequential update, and
as a consequence no conflict arise.
As in the parallel update case, the
time step cannot be trivially rescaled. 
Fluctuations of the particle
velocities are reduced compared to random sequential update
by the fact that each particle
is updated exactly once per time-step.

In the case of the frozen shuffle update, the fact
that that for a given
particle, there is exactly one time unit between two updates
reduces even further the fluctuations and
yields smooth motion in the free flow phase.

Some other updates that we will not discuss further here
have been proposed, such as sequential update ordered backward or forward in space
\cite{borodin_f_p2007,brankov_p_s2004,brankov2006,poghosyan_p2008,evans1997} or
sublattice update \cite{hinrichsen1996,fayaz2010,pigorsch_s2000,poghosyan_p_s2010}.
They can be seen as particular realisations of the frozen
shuffle update.

\subsubsubsection{Updates for intracellular traffic}

For the frozen shuffle update and parallel update, 
two hopping attempts are always separated by exactly one unit time.
This can be relevant for phenomena that proceed from an internal
clock, as stepping for pedestrians, and may have applications in biology for agents having a cyclic behavior.

In the case of molecular motors, the delay between
two steps is mainly due to the delay for the arrival
of a new ATP. Within the stepping cycle, the ATP hydrolysis is fast
enough so that it does not yield a significant incompressible
delay value. The limiting rate corresponds rather to the dissociation
of the motor head from the MT after ATP hydrolysis~\cite{gilbert1995},
but its value of $20 s^{-1}$ is usually considered to be fast
enough to be neglected.
As a consequence, the stepping behavior of molecular motors
is rather described
by a random sequential update.

Note that some papers~\cite{kunwar2011} have been using
parallel update, but with so small time steps that it
becomes equivalent to a random sequential update.
Indeed, the introduction of an intrinsic time scale will
have an impact only if it is \emph{not} negligible compared to all
the inverse rates.

In this review,
if not otherwise stated, it will always be a random
sequential update that will be considered.

\subsection{Theoretical results for the TASEP with
random sequential update}
\label{sec:tasep_theory}

As the most widely used update scheme in this review will be the
random sequential update, it is for this update that we shall
now give a non exhaustive survey of the results obtained
for the TASEP. Other complementary reviews can be found
for example in~\cite{derrida1998c,mallick2011}. Here we shall
concentrate on the properties that will be of relevance
for intracellular traffic.

\subsubsection{Time evolution of the density}
\label{sec:tasep_eq}

The evolution equation for 
the particle densities $\langle \tau_i \rangle$ 
in the bulk
are given by
\begin{equation}
\frac{\D \langle \tau_i \rangle}{\D t}= p \langle \tau_{i-1}(1 - \tau_i)
\rangle - p \langle \tau_i (1 -\tau_{i+1}) \rangle 
= J_{i-1,i} - J_{i,i+1}
\;,
\label{eq:rs_me}
\end{equation}
where $\langle \dots \rangle$  denotes the statistical average,
$\D t$ is the microscopic timestep, and $J_{i,i+1}$ is 
the average current accross bond $(i,i+1)$.
For an open system, these equations become at the boundaries:
\begin{eqnarray}
\frac{\D \langle \tau_1 \rangle}{\D t}&= 
 - p \langle \tau_1 (1 -\tau_{2}) \rangle 
+\alpha \langle 1-\tau_1 \rangle 
\;,\\
\frac{\D \langle \tau_N \rangle}{\D t}&= 
  \langle \tau_{N-1} (1 -\tau_{N}) \rangle 
- \beta \langle \tau_N \rangle
\;.
\label{eq:rs_bound}
\end{eqnarray}

\subsubsection{Current-density relation}
\label{sec:tasep_fd}
Investigating on the efficiency of motor-driven transport, the
current across a bond between two neighboring sites is a quantity
of high interest and is defined as the average number of particles
crossing that bond in one unit of time. In the stationary
state, since the TASEP obeys local conservation of particles,
the average current is also conserved throughout the system and
thus is equal for all bonds to
\begin{align}
J:=J_{i,i+1}=p\langle \tau_i(1-\tau_{i+1})\rangle,\label{eq:TASEP_current}
\end{align}
where $i$ denotes the starting site of the hopping move through
bond $(i,i+1)$.
In a product-state or mean field approximation, the averages over
products of state variables are assumed to factorize $\langle
\tau_i\tau_j\rangle\approx\langle
\tau_i\rangle\langle\tau_j\rangle$.
Let $\rho_i=\langle\tau_i\rangle$ be the average
density at site $i$.
As the periodically
closed system exhibits translational invariance, we have
$\rho_i = \rho$ $\forall i$.
Applying the mean field
approximation and translational invariance to
\eqref{eq:TASEP_current}, the unique current-density relation
\begin{align}
J(\rho)=p\rho(1-\rho)
\end{align}
is obtained which is actually exact for periodic boundary conditions in the
continuum limit~\cite{derrida1993c} and identical to the one obtained for large
systems
with open boundaries as discussed hereafter. Its maximum is at
$\rho=1/2$ where the corresponding current is $J=p/4$, a value which may serve
as benchmark to evaluate the transport capacity of other models.\\

\subsubsection{Phase diagram}
\label{sec:tasep_pd}

In the case of 
\emph{open boundary conditions}, which will be considered in the rest of this section, two infinitely large particle reservoirs are imagined at both ends of the lattice with density $\rho_0=\alpha/p$\abk{$\alpha$}{entrance rate (chapter~\ref{cha:state_of_the_art})} for the reservoir on the left and density $\rho_{L+1}=(1-\beta)/p$\abk{$\beta$}{exit rate} for the reservoir on the right. The first site of the lattice can hence be filled at rate $\alpha$ if it is empty and a particle on the last site leaves the lattice at rate $\beta$ (figure~\ref{fig:TASEP}). In the stationary state, the system with open boundary conditions has been solved by Derrida \emph{et al.}~using a matrix product ansatz~\cite{derrida1993c} and simultaneously by Sch\"utz and Domany~\cite{schutz_d1993} by solving the exact recursion relations found in~\cite{derrida_d_m1992}, confirming some
partial results obtained from a phenomenological approach by Krug~\cite{krug1991}.

An important property of the TASEP is its particle-hole-symmetry, i.e., the identical behavior of the model under certain symmetry operations:
\begin{align}
\tau_i\leftrightarrow1-\tau_i, \qquad \alpha \leftrightarrow \beta,\qquad i\leftrightarrow L+1-i.
\end{align}
These symmetry operations also imply a symmetry of the density profile:
\begin{align}
\rho_{L+1-i}(\alpha,\beta)=1-\rho_i(\beta,\alpha).
\end{align}
Thanks to this property, the whole phase diagram can be deduced from considering only half of the parameter space $(\alpha,\beta)$.

\begin{figure}[tbp]
\begin{center}
\includegraphics[height=5.5cm, clip]{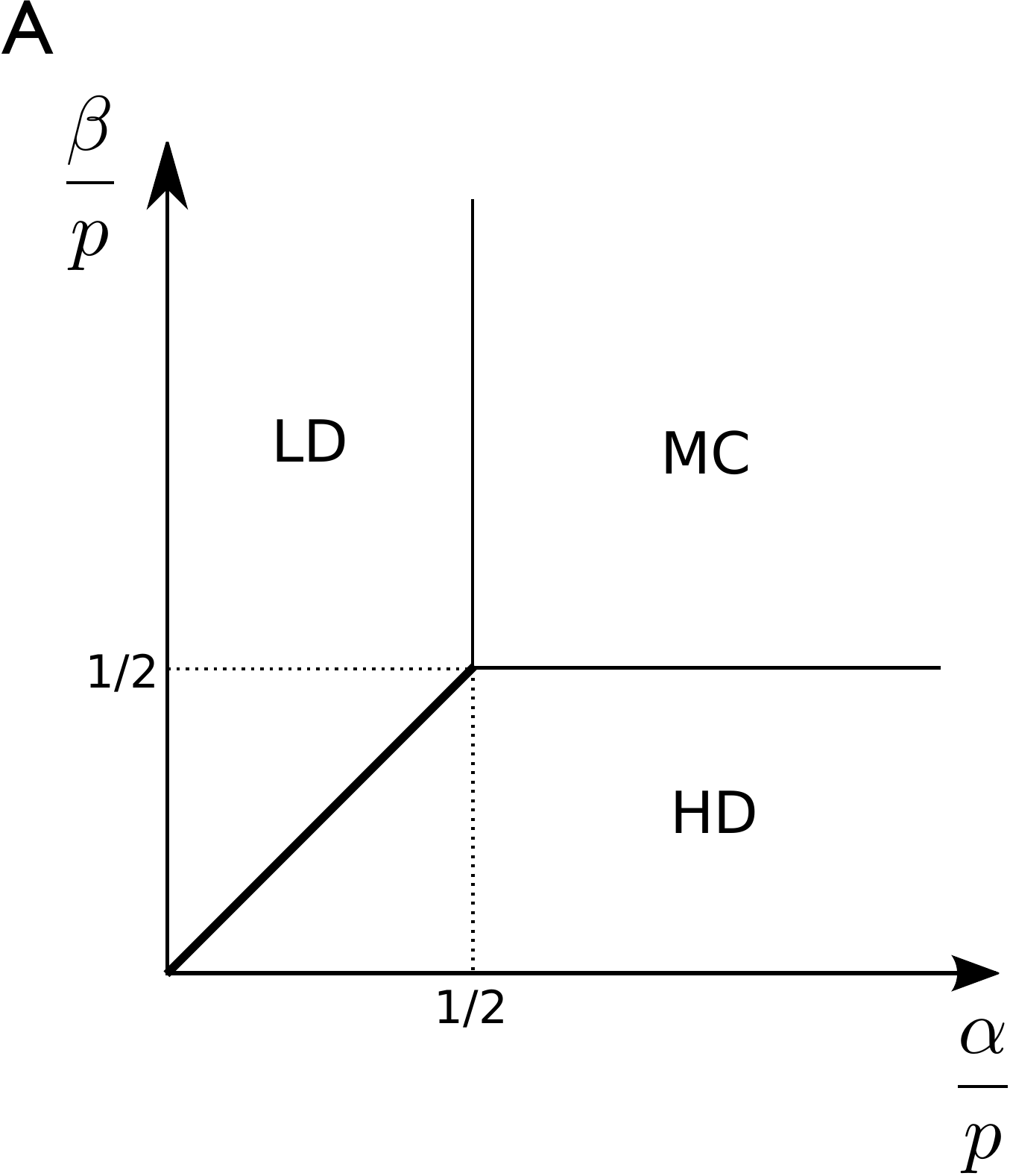}\hspace{1.0cm}
\raisebox{5mm}{\includegraphics[height=5cm, clip]{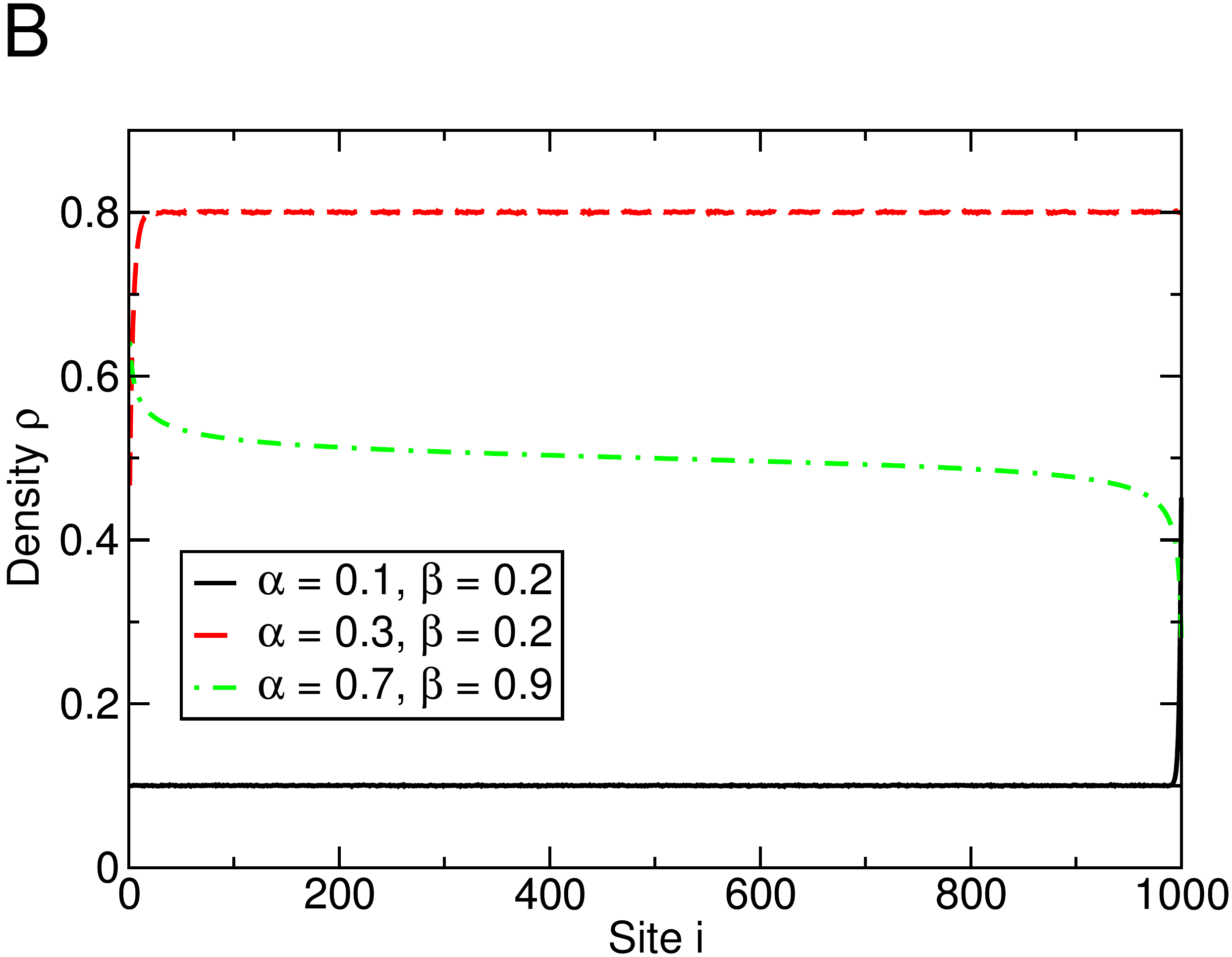}}
\caption[Phase diagram and density profiles of the TASEP with open boundary conditions.]{Phase diagram and density profiles of the TASEP with open boundary conditions. \textit{(A)} Phase diagram of the TASEP with its three phases: low-density (LD), high-density (HD) and maximal-current (MC). The thick line between LD and HD indicates a first order transition, thin lines at the HD/MC and LD/MC interfaces are second order. The thin dotted lines indicate changes of the boundary layer while the bulk behavior remains unchanged. \textit{(B)} Representative density profiles for a TASEP of system size $L=1000$ for the three phases with $p=1$. LD at $\alpha=0.1, \beta=0.2$ (solid black line), HD at $\alpha=0.3,\beta=0.2$ (red dashed line) and MC at $\alpha=0.7,\beta=0.9$ (green dash-dotted line).}
\protect\label{fig:TASEP_phase_diagram}
\end{center}
\end{figure}

In contrast to the system with periodic boundary conditions, the TASEP with open boundary conditions shows interesting non-trivial behavior. Depending on the entrance and exit rates $\alpha$ and $\beta$, three different phases can be reached~\cite{blythe_e2007} in the
stationary state:

\begin{itemize}
\item {\bf Low-density phase (LD):} For $\alpha<\beta$ and $\alpha < p/2$, the entry rate dominates the system's behavior and determines current and bulk density:
\begin{align}
J=\alpha(1-\alpha),\quad\rho=\alpha.
\label{eq:jrho_ld}
\end{align}
At the right boundary, the density profile exhibits a \emph{boundary layer} which decays exponentially to $\alpha$. If $\beta>p/2$ the exponential decay is additionally modulated by a power law.
\item {\bf High-density phase (HD):} For $\alpha>\beta$ and $\beta<1/2$, the exit rate limits the current in the system and creates a queue of particles in front of the exit which extends through the whole system except for the boundary layer on the left end of the system.
\begin{align}
J=\beta(1-\beta),\quad\rho=1-\beta,\quad\text{for }\alpha>\beta \text{ and }\beta<1/2.
\label{eq:jrho_hd}
\end{align}
Again, the left end boundary layer decays to the bulk value, with the characteristic of the decay depending on the value of $\alpha$ analogous to the low-density phase.
\item {\bf Maximal-current phase (MC):} For $\alpha>p/2$ and $\beta>p/2$, a maximal current is reached which is independent of the exact values of $\alpha$ and $\beta$.
\begin{align}
J=p/4,\quad\rho=1/2.
\label{eq:jrho_mc}
\end{align}
The boundary layers in this phase decay algebraically toward a density of {$\rho=1/2$}, leading to long-range boundary layers from both sides.
\end{itemize}
The expressions given in Eqs (\ref{eq:jrho_ld}), (\ref{eq:jrho_hd}),
and (\ref{eq:jrho_mc}), are valid in the limit of large systems.
For smaller systems, finite size corrections have to be
added~\cite{schutz_d1993}.
The phase diagram of the TASEP in the stationary state and
for large systems is shown in figure~\ref{fig:TASEP_phase_diagram}~{\it A} and typical density profiles of the three phases can be seen in figure~\ref{fig:TASEP_phase_diagram}~{\it B}.

The transitions from the low/high-density to the maximal-current phase are continuous: The bulk density approaches $1/2$ when arriving at the transition line from the low/high-density phase. In contrast, the transition line separating low-density from high-density phase is a first-order transition: The density changes abruptly from $\alpha$ to $1-\beta$ when crossing the transition line thus giving a total density change of $1-2\alpha$. The current nevertheless is continuous upon crossing of the transition line. On the transition line at $\alpha=\beta<p/2$, both reservoirs try to impose their density on the system leading to flat profiles emanating from both ends of the system which are connected by a shock front. This shock diffuses through the system, i.e., it performs a random walk along the lattice, leading to a linear averaged density profile.

In the LD and the HD phase, the shock connecting the densities of the two
reservoirs has a non-zero average velocity causing it to be driven to the right
and left boundary, respectively, where it forms the boundary layers mentioned
above.\\
We shall see in the next section how a phenomenological 
description of the motion of the shock can be developped.

In summary, due to the conservation of particles and current, the boundaries are able to determine the bulk properties in the TASEP. It is remarkable to observe this rich phase diagram despite the simple structure of the model. This is a direct consequence of the driving which keeps the system out of equilibrium.

It should be kept in mind that also partial asymmetric
dynamics can be considered (ASEP), to take into
account the fact that
 motors may have a non-vanishing probability
to back-step (see section \ref{sec:dynein}).
In most cases only minor differences are observed
compared to the TASEP.
It is only in the special case where 
the particle reservoirs drive the system against the
bias~\cite{blythe2000} that qualitatively new behavior is
observed: The net current exponentially decreases
with the system size.

In the remaining of this section, we shall consider only
the TASEP version.

\subsection{Domain wall approach}
\label{sec:dw}

We have mentionned above that in part of the phase diagram (actually
for $\alpha$ and $\beta < p/2$), each boundary reservoir
tries to impose its density to the system, resulting into
two flat domains separated by a shock.
This observation can be used to provide a coarse
grained description of the system dynamics (valid in the
aforementionned part of the phase diagram): the so-called domain wall
approach~\cite{kolomeisky1998,pigorsch_s2000}.

This phenomenological approach is complementary
to the numerous exact results that have been obtained for the TASEP:
Apart from giving a more physical intuition of the mecanisms
involved, we shall see that the domain wall description
is valid not only for stationary but also non stationary states,
and can be more easily extended to variants of the TASEP.

But first we shall describe it in the case of the TASEP.
Each boundary
condition of the system implements 
a boundary reservoir of a given density.
The actual density depends on the precise implementation of
the model.
In case of the TASEP with random
sequential update the left reservoir inserts particles with density $\alpha$
and the right reservoir holes with density $\beta$ leading to a particle
density of $1-\beta$.  If $\alpha \neq 1-\beta$ the particle density of the two
domains is different such that in the coarse grained picture the density is
discontinuous at the point where the two domains meet, i.e. the domains are
separated by a shock. 

In the domain wall description, the smooth shocks that separate
the free flow and jammed regions are considered as
sharp discontinuities. Simple considerations based
on mass conservation yield the velocity of the
shock in terms of the fluxes and densities on both
sides, as we shall explain now.
Let $\rho_i$ be the particle density at site $i$ and $j_{i+1/2}$ the current from $i$ to $i+1$. Then the conservation of mass implies
\begin{equation}
\label{eq:dw_cont}
\frac{d \rho_i(t)}{dt}
 = j_{i-1/2}(t) - j_{i+1/2}(t) \quad .
\end{equation}
\noindent The velocity of the shock far from the boundaries in the continuum limit can be obtained from the continuity equation (\ref{eq:dw_cont}) \begin{equation}
\label{eq:dw_vshock}
v_s = \frac{j_+ - j_-}{\rho_+ - \rho_-}
\end{equation}
\noindent where $\rho_\pm$ and $j_\pm$ denote the density and
current of the domain left ($-$) or  right ($+$) from the
shock. Due to the stochastic input of particles, the motion of
the shock is non-deterministic.

For the random sequential update, 
it turns out that it is
appropriate to describe the dynamics of the shock as a biased
random walk with hop rate  $D_+$ to the right and $D_-$ to the
left.
These hopping rates can be related to the drift velocity
of Eq.~(\ref{eq:dw_vshock}) by $D_+-D_-=v_s$.
This relation cannot fully determine the hopping rates.
Those can be partially inferred from limit cases
and were postulated to be given by~\cite{kolomeisky1998}
\begin{equation}
\label{eq:T1-dwp2}
D_+ = \frac{j_+}{\rho_+ - \rho_-} \quad \quad \textrm{and} \quad \quad D_- = \frac{j_-}{\rho_+ - \rho_-} \quad . 
\end{equation}
 
By mapping the dynamics of the domain wall dynamics to a
biased random walk on a finite lattice with reflecting
boundary conditions, the dynamics and the stationary
distribution of the domain wall positions can be evaluated.
The latter is given by
\begin{equation}
\label{eq:T1-dwp6}
P(i) = \mathcal{N}^{-1} \exp \left( -\frac{L-i}{\xi} \right)
\end{equation}
\noindent in the low density phase ($\alpha < \beta$)
and
\begin{equation}
\label{eq:T1-dwp7}
P(i) = \mathcal{N}^{-1} \exp \left( -\frac{i}{\xi} \right)
\end{equation}
\noindent in the high density phase ($\alpha > \beta$).
In these expressions, $\mathcal{N}$ is a normalization factor $\mathcal{N}$ 
and $\xi$ is a localization length given by 
\begin{equation}
\label{eq:T1-dwp8}
\xi^{-1} = \left| \ln \frac{D_+}{D_-} \right| = \left| \ln \frac{\beta(1-\beta)}{\alpha(1-\alpha)} \right|.
\end{equation}
\noindent 
The stationary density profile can be derived from
(\ref{eq:T1-dwp6}-\ref{eq:T1-dwp8}). It is found to be
in agreement with the exact expression valid in the
large system size limit~\cite{schutz_d1993},
with in particular the same localization length $\xi$.
This agreement can be seen as a validation {\em a posteriori}
of the choice~(\ref{eq:T1-dwp2}) for $D_+$ and $D_-$.

The first order transition between high- and low-density phase that occurs at $\alpha = \beta$ is accompanied by a divergence of the localization length.
This illustrates the non-equilibrium nature of the boundary induced phase transition.
It should be noticed that the domain wall theory is only valid if the capacity of the boundary reservoir does not exceed the capacity of the chain. For values of $\alpha/p > \frac{1}{2}$ or $\beta/p > \frac{1}{2}$ the structure of the domains is non-trivial and the simple random walk picture cannot be applied.

Despite this limitation the domain wall approach is very usefull. First  of all it gives a simple and intuitive physically based description
of the first order transition along the $\alpha = \beta$ line.
Moreover, while exact results were obtained mostly
for stationary states, the domain wall approach allowed to predict with a very good accuracy
several non-stationary features \cite{dudzinski_s2000,nagy_a_s2002,santen_a2002}.
Eventually, while the DW approach is meant to be valid only in
the large system size limit, actually it works remarkably well
for finite systems if not too close to the maximal-current
phase (see~\cite{santen_a2002} for a comparison with systems
of a few tens).

The domain wall approach can also be applied to a large set
of variants of the TASEP which have a continuous fundamental
diagram, while exact techniques usually are much more
difficult to generalize. 
We shall see later how the domain wall approach allowed
in particular to understand multilane TASEP
models~\cite{schiffmann_a_s2010} (section \ref{sec:strong_coupling})
and models without
particle conservation in the bulk (section \ref{sec:PFF}).

It should be underlined that, though mass conservation
gives a firm base to Eq.~(\ref{eq:dw_vshock}),
the expression for the hopping rates~(\ref{eq:T1-dwp2})
is a postulate.
In the case of the random sequential update,
this choice was confirmed by the agreement between the
predictions of the DW theory~\cite{kolomeisky1998} and the
exact solution~\cite{schutz_d1993} or numerical
results~\cite{nagy_a_s2002,santen_a2002}.
It was noticed in~\cite{kolomeisky1998} that the DW approach
developped for the random sequential update was also
predicting correctly the 
localization length in the case of parallel update
in the limit of small $\alpha$ and $\beta$~\cite{tilstra_e1998}.
However, in general, the DW approach as presented above
gives only approximate results
when applied to other
updates than random sequential (see~\cite{appert-rolland_c_h2011b} for an example in the case of the frozen shuffle
update~\cite{appert-rolland_c_h2011a}).
Actually the DW theory postulates that the motion
of the wall can be described by the master equation
of a biased random walk~\cite{dudzinski_s2000}
with the hopping rates~(\ref{eq:T1-dwp2}).
While this is relevant for the random sequential
update, it can be shown~\cite{cividini_h_a2014}
that in the case of a deterministic TASEP with
parallel update, a non-Markovian evolution equation
has to be used: There is a memory effect over
one time-step. Note that in this special case,
the domain wall theory is not phenomenological anymore,
but can be made exact even
at the microscopic scale.

\section{Variants of TASEP}
\label{sec:variants}

It is out of scope to review all the litterature
on variants of TASEP. Here we selected a few of them
that
contain some ingredients relevant for biological
transport.

\subsection{TASEP with on-off gates}
\label{sec:onoffgate}

\subsubsection{Single lane TASEP with dynamical gate}
\label{sec:turci}

In the previous section, the disorder was given as a characteristic
of the model, and thus independant of the evolution of the
system.
However, it may happen that the properties of the
transport network depend on the occupation state of the
network by particles.

\begin{figure}[htbp]
\begin{center}
\includegraphics[height=0.3\linewidth]{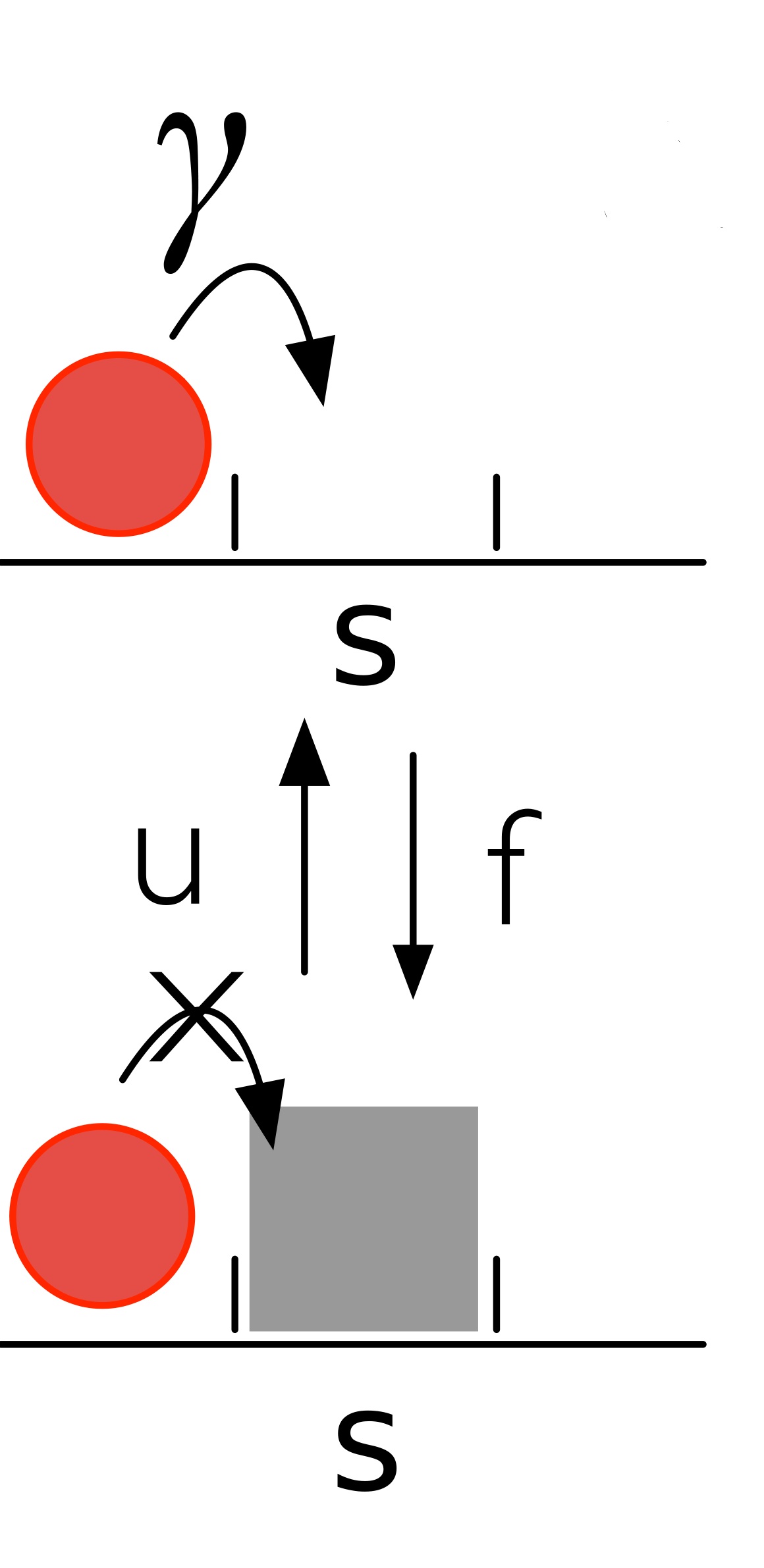}
\caption[Dynamical defect]{
Definition of the dynamics of a single site dynamical
defect located in site $s$.
Only two sites of the lattice are represented.
Particles cannot enter site $s$ when it is closed,
while the closing of site $s$ is not possible when
it is occupied by a particle.
Within this constraint, opening and closing of site $s$ occur
respectively with rates f and u (these symbols were chosen
in reference to the folded and unfolded status of the mRNA
strand represented by the one-dimensional lattice).
From~\cite{turci2013}.
}
\protect\label{fig:turci}
\end{center}
\end{figure}

One such example appears in the context of
protein synthesis control, where the lattice of the
TASEP then represents a strand of messenger
ribonucleic acid (mRNA).
Each time a ribosome
(a `particle' in TASEP) moves along the lattice, it synthesizes
one protein. One possibility to regulate the amount
of synthesized protein is based on the ability of
a given region of the mRNA
to fold (or to bind with a translation repressor protein),
and thus to prevent ribosomes to go
further along the mRNA strand.
Turci et al~\cite{turci2013}
have proposed to model this phenomenon by the following model
based on a single lane TASEP.
The rules are those of the original TASEP, except
for one site $s$ (or one region of the lattice) which can
be open or closed (see Fig. \ref{fig:turci}).
When site $s$ is closed, particles cannot hop on it,
and pile up in a waiting line until site $s$ opens.
The opening and closing of site $s$ occur with fixed rates.
However, an extra condition for closing site $s$
is that it must not be occupied.
This condition couples the dynamics of the lattice with the
dynamics of particles and leads to a rich phenomenology,
parts of which are still under study.

\subsubsection{Alternating flows in multi-lane TASEPs}
\label{sec:bottleneck}

Some other models have been proposed in which
the status of a portion of the lattice which
can be open or closed has also a dynamics coupled with the
occupation by particles.

A situation encountered in various systems
is that oppositely moving particles share a common bottleneck,
which is too narrow to allow for crossing.
The bottleneck is thus used alternatively by the
two types of particles.

Such a model was first proposed in the frame of bidirectional
pedestrian flows~\cite{jelic_a_s2012}, to model the flow oscillations that
are observed at a bottleneck (door or small corridor).
But it can also be relevant to model biological
phenomena, such as bidirectional molecular traffic
across nuclear pores~\cite{kapon08}.

\begin{figure}[htbp]
\begin{center}
\includegraphics[width=0.7\linewidth]{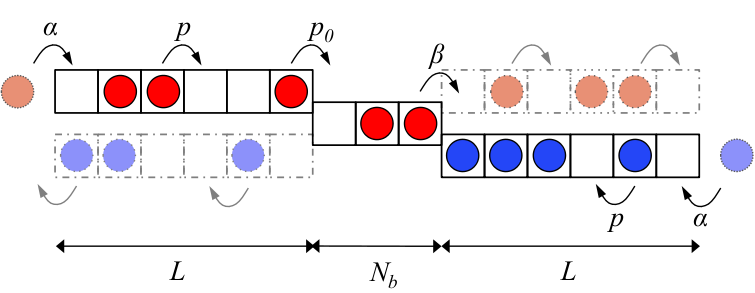}
\caption[Bottleneck model]{
Two TASEP systems interact through the sharing
of a common bottleneck.
Red particles (upper lane) and blue particles
(lower lane) move in opposite directions and
can enter the bottleneck of length $N_b$
only if there is no particle of the opposite type.
From~\cite{jelic_a_s2012}.
}
\protect\label{fig:bottleneck}
\end{center}
\end{figure}

In the bottleneck model~\cite{jelic_a_s2012}
 depicted in Fig. \ref{fig:bottleneck},
two types of particles move respectively
towards the right or towards the left.
Outside the bottleneck, particles of different types
move on different tracks and do not interact.
Inside the bottleneck, only one species can go through
at a given time, as no exchange of particles is allowed.
More precisely, a particle of a given type can enter
the bottleneck only if there is no particle of the
other type inside the bottleneck.

This constraint
is non local and assumes that the particles that
are about to enter the bottleneck have a knowledge
of the occupation status of the bottleneck.
For pedestrians, this can be obtained by visual
control. In biological systems, a conformational
or chemical change could occur somehow.

It has been shown that the dynamics of such a model is
driven by fluctuations in the bottleneck and by the transient
states that take place at each reversal of the flow in the
bottleneck~\cite{jelic_a_s2012}.
From the point of view of non-equilibrium systems,
it is interesting to have a model that gives macroscopic
signatures from local fluctuations.

Another example of on-off gate driven by particles
was given in~\cite{fouladvand_s_s2004c}, to describe
an urban trafic network with ``intelligent'' trafic lights.
There also, the coupling of the trafic lights dynamics
with the occupation of the network yields a non trivial
phenomenology.

Many features of these dynamical gates whose dynamics
depend on particle positions
are still not fully understood and raise interesting fundamental
questions, with related applications to real systems.

\subsection{Multi-lane exclusion processes}
\label{sec:multilane}

In the cell, microtubules do not offer only
one but several parallel tracks (protofilaments).
Here we explore the consequences of having
multiple tracks on the TASEP characteristics.

When multilane systems are considered, different
behaviors can be observed depending on the type of
coupling between the lanes, whether all particles
move in the same direction or not, etc.

\subsubsection{Multi-lane interactions without lane changes}

\begin{figure}[tbp]
\begin{center}
\includegraphics[width=0.7\linewidth]{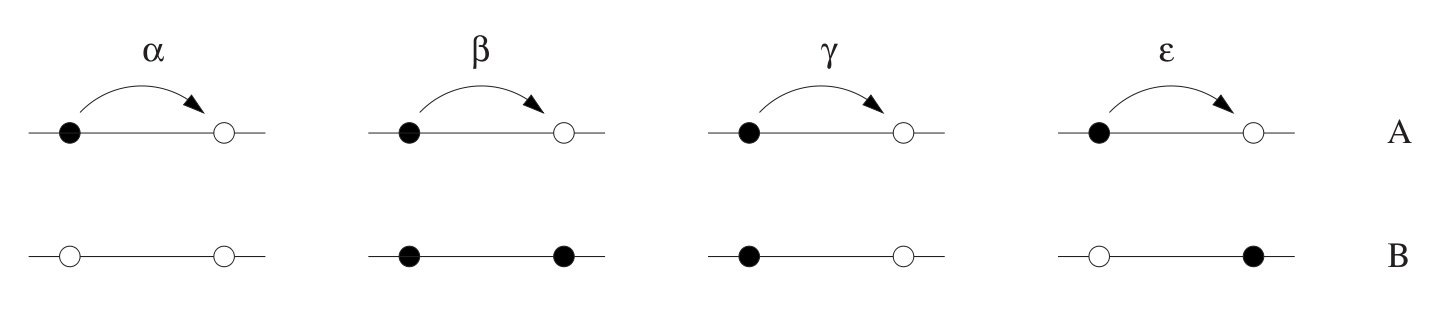}
\caption[]{Schematic representation of the
hopping rules on lane A. The hopping rates depend on the
occupation of lane B.
From~\cite{popkov2004}.
}
\protect\label{fig_popkov}
\end{center}
\end{figure}

As a simplified case of multi-lane traffic,
we consider here two parallel TASEP lanes,
labeled A and B in Fig.~\ref{fig_popkov}.
If there are no interactions between the lanes,
the system is trivial and reduces to two independent TASEPs.
In this section, we shall consider the case where
particles on different lanes can interact.
These interactions can be thought for example
as steric interactions between cargos transported by motors,
that create some friction when they pass each other.
Throughout this section, particles cannot change lane.
The case of particle exchanges between the lanes will
be considered in the next section~\ref{sec:multilane}.

In~\cite{lee_p_k1997}, particles on each lane move in
opposite directions. The interaction between the lanes is
such that the hopping rate of a given particle is reduced by
a fixed factor if there is a particle of the other species
on the site next to its target site. 
This corresponds to the case $\alpha=\gamma>\beta=\epsilon$
in Fig.~\ref{fig_popkov}.
In the special case where one lane is occupied by cars
and the other one by oppositely
going trucks (trucks' hopping rates are obtained from the cars' ones
by dividing by a fixed factor), it is shown that
the interlane interaction can result into the formation
of bound states between the trucks, i.e. a truck-truck
interaction mediated by the cars is obtained.

Another case where particles go in the same direction
on both lanes was considered in~\cite{popkov_p2001},
with the choice $\alpha=\beta=\gamma=1$ and $\epsilon<1$.
For open boundary conditions, this model exhibits
a new phase where the system oscillates between symmetry
breaking phases~\cite{popkov_p2001}, for which one lane
has a much higher density than the other one.
This behavior results
from the existence of fast and slow processes.

Eventually, for the case
$\alpha=1$ and $\gamma=\epsilon=(1+\beta)/2$,
a domain wall description of the two-lane system
shows that, while for a single lane the relaxation
towards the stationary state is obtained through
only one reflection of the shock from either of the
boundaries, multiple reflections of shocks, which now
come in pairs~\cite{popkov_s2003}, are necessary
in the two-lane system~\cite{popkov2004}.

\subsubsection{Multi-lane uni-directional transport
with bulk lane changes}

\begin{figure}[tbp]
\begin{center}
\includegraphics[width=0.7\linewidth]{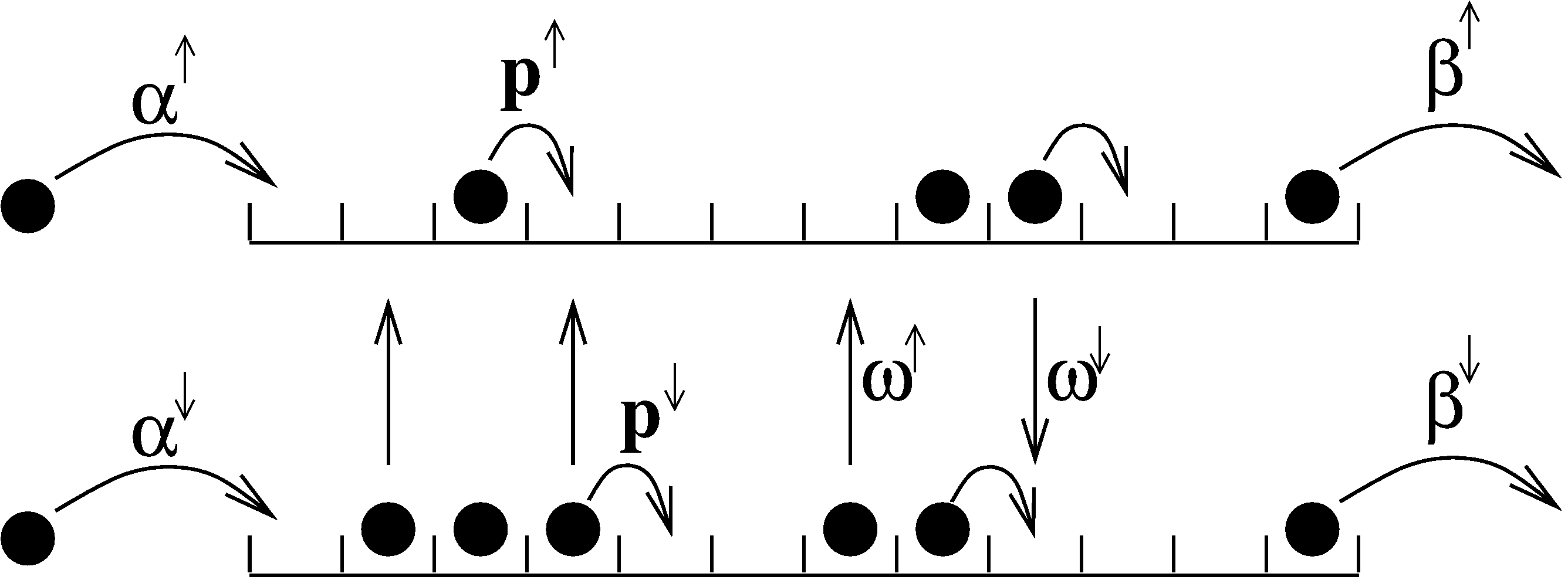}
\caption[Bulk coupling scheme]{Schematic representation of the
rules coupling two parallel TASEP lanes.
The exchange rates for the bulk coupling are called here
$\wud$ and $\wdu$. In the text, we assume that $\pu = \pd = 1$,
though a generalization to arbitrary values is straightforward.
From~\cite{schiffmann_a_s2010}.
}
\protect\label{fig_lane_exchange}
\end{center}
\end{figure}

In this section, we consider only particles going
in the same direction, on a system of parallel tracks.
When several lanes are offered for transport,
in many systems it is possible that particles
change lane. This is for example possible for 
dynein motors which are able to change the protofilament 
when walking along microtubules.
Lane changes can be simply modeled by rates $\wud$ and $\wdu$
at which a particle hops to the neighbor site
on the lower or upper lane (fig. \ref{fig_lane_exchange}). 

In order to analyze the effect of lane changes in terms of a mean-field approach, 
a set of equations has been introduced, that expresses 
mass conservation.
If we call $\ju(i+\frac{1}{2})$ (resp. $\jd(i+\frac{1}{2})$)
the horizontal flux between
site $i$ and site $i+1$ on the upper lane (resp. lower lane),
and $\jud(i)$ (resp. $\jdu(i)$) the vertical flux at position $i$ from
the upper to the lower (resp. from the lower to the upper) lane,
then mass conservation can be expressed as
\begin{eqnarray}
\rhou(i,t+dt)-\rhou(i,t) & = & \left[\ju(i-\frac{1}{2},t) - \ju(i+\frac{1}{2},t) + \jdu(i,t) - \jud(i,t)\right]dt
\label{massc1}
\\
\rhod(i,t+dt)-\rhod(i,t) & = & \left[\jd(i-\frac{1}{2},t) - \jd(i+\frac{1}{2},t) + \jud(i,t) - \jdu(i,t)\right]dt
\label{massc}
\end{eqnarray}
Replacing the fluxes by their mean-field expressions
\begin{eqnarray}
\ju(i+\frac{1}{2}) & = & \pu \rhou(i) \left[ 1-\rhou(i+1) \right] \\
\jd(i+\frac{1}{2}) & = & \pd \rhod(i) \left[ 1-\rhod(i+1) \right] \\
\jdu(i) & = & \wdu \rhod(i) \left[ 1-\rhou(i) \right] \\
\jud(i) & = & \wud \rhou(i) \left[ 1-\rhod(i) \right] \label{jMF}
\end{eqnarray}
yields, after a change of variables $x = i/L$ and
an expansion in the limit $L \gg 1$,
the following equations for the density profiles:
\begin{eqnarray}
\partial_t \rhou & = &
-  \frac{1}{L} \pu \left(1 - 2 \rhou\right) \partial_x \rhou 
+ \wdu \rhod \left(1-\rhou\right)
- \wud \rhou \left(1-\rhod\right)
\label{eqrho1}
   \\
\partial_t \rhod & = &
- \frac{1}{L} \pd \left(1 - 2 \rhod\right) \partial_x \rhod +
 \wud  \rhou \left(1-\rhod\right)
- \wdu  \rhod \left(1-\rhou\right)
\label{eqrho2}
\end{eqnarray}

Note that the two rightmost terms correspond to the
the net vertical flux $\jnet \equiv \jdu - \jud$.
In these equations, the dominating terms will be
different depending on the scaling of the coupling
rates $\wud$ and $\wdu$ (to simplify the discussion
we shall assume in the following that $\pu=\pd=1$).

\subsubsubsection{Strong Coupling}
\label{sec:strong_coupling}

If the exchange rates are of the same order as the hopping rates,
the coupling is said to be strong. This is for example the case if one
considers 
highway traffic or the dynamics of dynein in a crowded environment.
If the in- and output rates are the same on both
lanes, both lanes behave in a similar way, shocks on
both lanes are coupled~\cite{mitsudo_h2005} and
the phase diagram can be derived quite
easily~\cite{pronina_k2004}.
The dynamical characteristics of coupled lanes becomes more
complex when the lanes have different initial states,
and different in- and output rates~\cite{schiffmann_a_s2010}.

\begin{figure}[tbp]
\begin{center}
\includegraphics[width=0.7\linewidth]{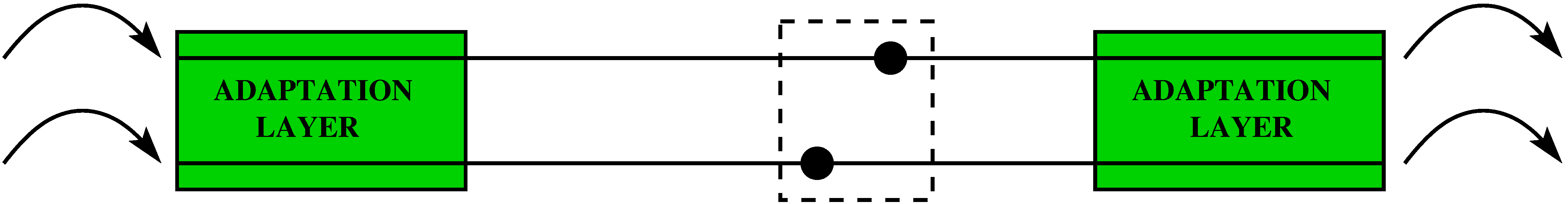}
\caption[strong coupling]{Typical structure of a two-lane system with strong coupling.
Non-adapted boundary conditions create adaptation layers
on both sides. In the bulk,
pairs of shocks are formed and
move as a whole through the system.
Then the net vertical
flux is zero everywhere except in adaptation layers
and inside the shock pairs.
From~\cite{schiffmann_a_s2010}.
}
\protect\label{fig_strong_pairs}
\end{center}
\end{figure}

We shall first discuss the transient occuring when
two initially independent TASEP lanes with different
in- and output rates are coupled
at $t>0$ according to the rules of Fig.~\ref{fig_lane_exchange}.
From the mean-field equations (\ref{eqrho1}-\ref{eqrho2}),
we see that when the coupling
rates are of order $\mathcal{O}(1)$,
strong
lane exchanges can take place over time scales of order $1$.
Indeed, this is what is observed in direct simulations: If the system is initialized
with arbitrary constant densities, very intense lane exchanges in the bulk lead
within a few time steps to a local balance between the lanes.
When the
densities on the two lanes are such that the net
flux is vanishing, we consider the lanes to be in an adapted
state.
For arbitrary in- and output rates, 
some adaptation layers localized near the boundaries
interpolate between the non-adapted densities imposed by the
reservoirs to the bulk densities which balance the vertical fluxes
(Fig. \ref{fig_strong_pairs}).
After the initial adaptation, 
the net vertical flux vanishes almost everywhere,
except in shock regions and in adaptation layers.

\begin{figure}[tbp]
\begin{center}
\includegraphics[width=0.65\linewidth]{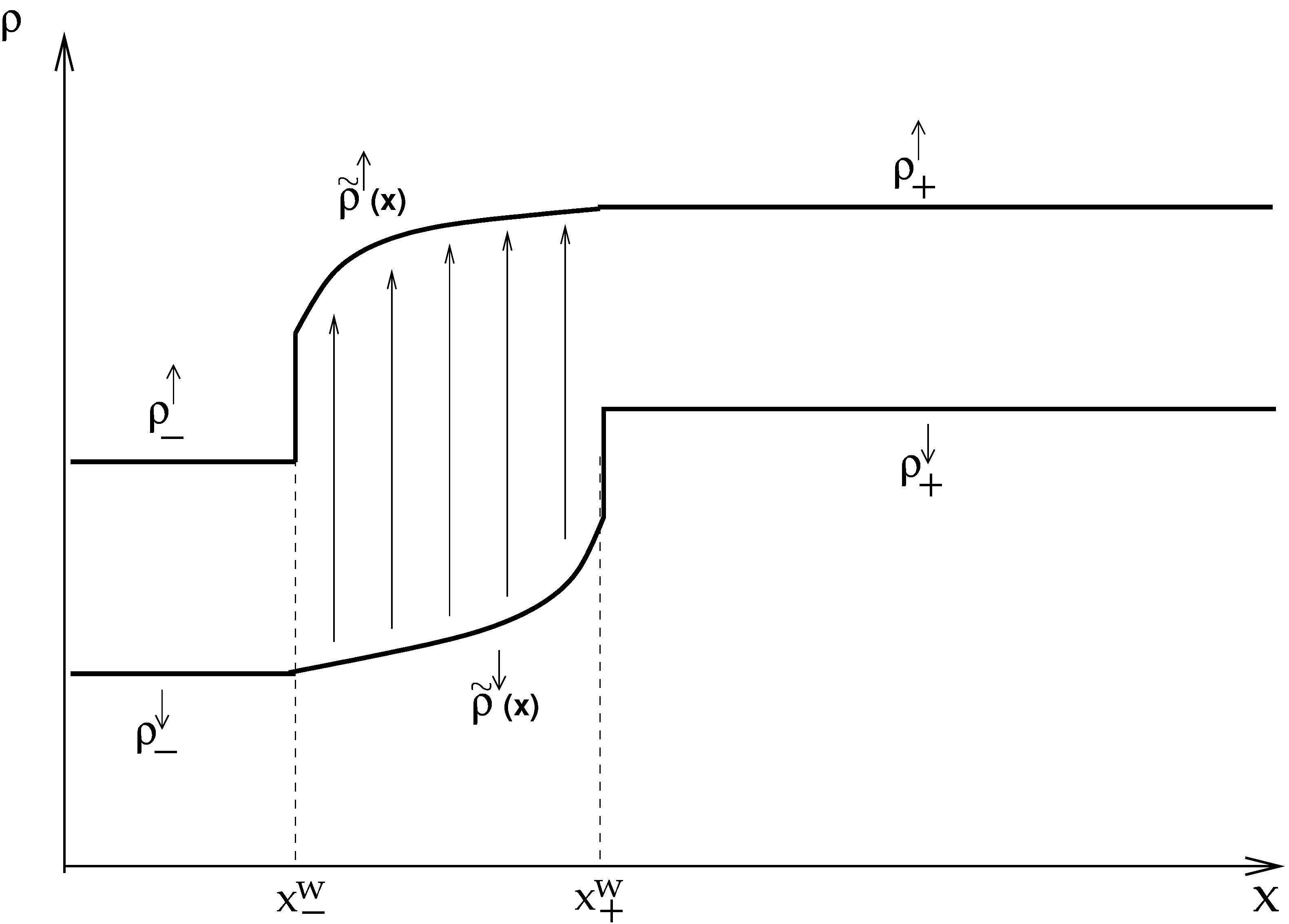}
\caption[strong coupling]{Schematic representation of the
structure of a pair of walls. The walls are represented a real discontinuities 
separated by a non-constant density profile on the two lanes.
From~\cite{schiffmann_a_s2010}.
}
\protect\label{fig_strong_shock}
\end{center}
\end{figure}

If a shock - or domain wall - exists initially on one lane in the bulk,
after the initial adaptation stage another shock will
be formed on the other lane, in the vicinity of the first one.
These two shocks
form a pair separated by a transition region
within which a persistent vertical net flux of
particles exists (see Fig.~\ref{fig_strong_shock}).
The motion of the two shocks is strongly correlated, since
their distance is controlled by the amplitude of the local
vertical flux.
Depending on the initial condition, several such shock pairs
can be formed. On much longer time scales, these pairs perform biased random walks, meet and merge,
until a unique shock-pair remains in the system.

In the long time limit, a stationary state is reached.
In most cases, the remaining shock pair localizes itself 
near a boundary, as in the one lane case.
Here however, the inner structure of the shock pair
interfere with the adaptation layer so that
the stationary state is more difficult to predict.
The characteristics of the stationary state,
and in particular the phase diagram, were studied
first in~\cite{harris_s2005} in the case of very
asymetric lane changes,
and more recently obtained for more general rules from a stability
analysis of the mean-field profiles~\cite{evans2011}.
The case of back-stepping was also
considered~\cite{evans2011,tsekouras_k2008}.

\subsubsubsection{Weak Coupling}
\label{sec:weak_coupling}

Weak coupling refers to the case where the exchange rates
(and the net current between lanes) scale as $1/L$ and was
in particular studied
in~\cite{reichenbach_f_f2006,reichenbach_f_f2007,reichenbach_f_f2008}.
It has been argued that weak coupling could be relevant for
example for kinesin based
intracellular transport: kinesins do not usually side-step,
but lane changes could be carried out
via an intermediate detachment of the molecular
motors~\cite{reichenbach_f_f2006} or under the effect
of lateral loads.

In the case of weak coupling, all terms in the mean-field
equations are of the same order. Now the net vertical flux is
not necessarily vanishing and the density profile varies
smoothly.
Contrarily to strong coupling for which shock positions on both lanes
are strongly correlated, in the case of weak coupling, shocks are only
loosely coupled.
Actually, in this scaling, it is as if the shock pairs found
for strong coupling were now separated by a distance of the 
order of the systems size. 
Thus the inner structure of a shock pair shown in
Fig.~\ref{fig_strong_shock} in the case of strong
coupling
is the analog of the
density profiles in the case of weak coupling.

If boundaries are imposing different fluxes, the shocks will
be localized at the boundaries in the stationary state.
If boundary reservoirs are imposing similar fluxes, but
different densities, then there is a need to match the solutions
coming from both ends, and shocks will form in the bulk.
While for the single lane TASEP, shocks in this case are
not localized and explore the whole system according to
a symmetric random walk, here the shocks can be localized
in the bulk if there are some adaptation layers.
Indeed, while adaptation layers were always of finite size
in the case of strong coupling (see
Fig.~\ref{fig_strong_pairs}), they may now invade the
whole system.

Though these approaches are useful to understand the
role of lane changes, 
note that in the case of molecular motors,
an analysis taking into account
the non-conservation of mass on the filament
would be more appropriate,
as developped in section~\ref{sec:PFF}.

\subsubsection{Multi-lane bi-directional transport
with bulk lane changes}

While in the previous section the lanes could represent
protofilaments on the same MT, it is also possible
to have motor exchanges between filaments (MTs or actin)
oriented in opposite directions.
We shall thus consider in this section the coupling
of two TASEP lanes,
where particles are going to the right on one lane and
to the left on the other lane. This was done
in~\cite{juhasz2007b} in the case of weak coupling.
As mentioned in~\cite{juhasz2007b}, due to the
particle/hole symmetry, one could actually replace
the holes on one lane into particles, to have again
uni-directional motion, but the interaction between
the lanes would then be that particles would be created
or annihilated in pairs.

Several types of boundary conditions can 
be considered. For example, while \cite{juhasz2007b} deals
with open boundary conditions, the case of one open and
one closed end
is studied in~\cite{ashwin_l_s2010} with a focus on the
length of the queuing line that forms near the dead end.

\subsection{Modelling inhomogeneities}
\label{sec:disorder}

Motor-driven intracellular transport is usually carried out on filament
networks which are embedded in a crowded environment. Although the bulk
structure of the filaments is surprisingly regular several mechanisms exist
which influence the local transport properties of the chain.
The environment, more or less crowded, can change locally the transport capacity
on a given MT.
Besides, several kinds of proteins can bind to the MT, a prominent
example detailed in the next subsection being tau proteins.
Tubulin itself can be locally modified: The cell can produce
different isotopes of tubulin, and various enzymes can induce
some modifications of tubulin after polymerization~\cite{janke_k2010}.
These modifications could have deep functional implications,
and even lead to ``functionally distinct microtubule types''~\cite{janke_b2011}.

The motors also can see their properties deeply modified
by external factor, e.g. dynactin on dynein.
Besides, the large variety of cargos that need to be transported
provides a very inhomogeneous crowd of moving objects.

It is therefore interesting to discuss the impact of inhomogeneities
on transport.
Actually the impact of disorder on transport has already be
largely studied in the context of the simple TASEP.
Two families of disorder have been considered:
Link or site-wise disorder, which would model the inhomogeneities
of the tracks, and particle-wise disorder, corresponding to
the inhomogeneities in cargo types.
Particle-associated disorder can be modelled by assigning different hopping rates to different particles, while for site-associated disorder, the hopping rates $p_i$ are fixed but not equal for all sites in the system.

In case of periodic boundary conditions the TASEP with particle-wise disorder
can be exactly solved \cite{evans1996} by introducing distance variables
\cite{evans1996,krug_f1996} instead of particle positions. This mapping relates
the TASEP to the zero-range process which is exactly solvable even in the
presence of disorder. The mapping is more difficult to exploit in the case of
open boundary conditions,
as the number of particles is not conserved.

When particles do not move at the same speed, they may pass each other,
taking advantage of the possibility to use parallel tracks.
K. Mallick solved a one-lane TASEP with one single slow particle that could be
passed by fast particles with some probability less than 1~\cite{mallick1996}.
As expected, jamming can occur behind the slow particle.
A two-lane TASEP model with particle-wise disorder was proposed
in~\cite{knospe1999} and showed numerically the strong tendency
to form platoons behind slow particles.
However no exact solution has been found even for periodic
boundary conditions.
Some exact results have been obtained for similar systems
in continuous space and time (kinetic models), first
for a one lane model in which passing is allowed with a certain
rate~\cite{bennaim_k1999}, or for a two lane model with
bidirectional trafic, under a mean field coupling
assumption~\cite{appert-rolland_h_s2010}.
Back in the frame of TASEP-based models, some recent
analytical results were obtained for a one-lane model
where particles have the same speed, but have different
priorities for passing each
other~\cite{ferrari_m2007,evans_f_m2009,arita2011,arita2012} - a feature
that results into different effective velocities.

By contrast with the particle-wise disorder TASEP,
the TASEP with site-disorder is not exactly
solvable even in the stationary state for periodic boundary conditions,
except in some quite special cases (for example for a single
site disorder with alternated parallel
update~\cite{schutz1993}).
However, in recent years, a number of works have approached
the site-wise disordered TASEP by numerical methods and analytical
approximations. Computer simulations have shown that even a single defect, i.e.
a site with reduced hopping rate may introduce a macroscopic high and low
density domain to the stationary state of the TASEP with periodic boundary
conditions \cite{janowsky_l1992}.

Next to a single defect site also more than one defect site in the system has
been considered. In \cite{tripathy_b1998} the site dependent hopping rates are
independently chosen from a random distribution.
As in the case of a single defect,
disorder can lead to macroscopically nonhomogeneous states, which are observed
for densities in a finite range around half filling, while at low and high
densities only local variations of the density are observed. The
self-organization into a macroscopic high and low density domain is generated
by a stretch of sites carrying small hopping rates.
Note that this stretch does not automatically include the site with the lowest hopping
rate if this site surrounded by fast links. This is in contrast to particle disorder where
the phase separated states in uni-directional systems are determined by the slowest particle.

To come closer to motor dynamics, TASEP has to be supplemented with
Langmuir kinetics. In the case of infinite diffusion in the surrounding
cytoplasm, as considered in chapter~\ref{sec:PFF},
the influence of a single stepping defect
was studied in~\cite{pierobon2006} and
for multiple stepping defects in~\cite{greulich_s2009a}. While the phases of
the defect-free system still persist, new states arise in which the system is
subdivided into parts by the defects, leading to phases which are characterized
by the coexistence of different domains.

In a generic way, stepping defects and unbinding defects
modify the properties of single motors, as the run-length along the filament or
the velocity. A run-length reduction due to defects can affect the
global motor current in different ways:
In a system with periodic boundary conditions, the current
is negatively affected by
stepping defects but may increase for unbinding defects at high enough motor
densities. This is possible due to the reduction of the density of bound
motors on the filament~\cite{chai_l_k2009}, which reduces
the tendency to form jams
on the filament.
The model of section~\ref{sec:simplified_dyn} with randomly deleted
MT sites~\cite{ebbinghaus_a_s2010a} can also be seen as an example
of such a mechanism.
The use of dynamic unbinding defects should actually
be even more efficient than static ones in order to reduce
the tendency to form jams.

Binding defects on the other hand do not affect collective motion of particles much if not present at almost all sites~\cite{grzeschik_h_s2010,chai_l_k2009}.

The consequences of site-wise defects in the models of this section have been studied for particles hopping only in one direction, i.e. representing motors. If they were representing cargos 
 carried by teams of molecular motors, bidirectional motion should
be including. In this case,
site-disorder might serve to regulate the motion of individual cargos as hypothesized in~\cite{dixit2008}.

\section{Many-particle models for uni-directional stochastic
transport: The impact of particle reservoirs}
\label{sec:unidir}

In this chapter we discuss the impact of different realizations of particle reservoirs 
on the bulk properties of the system. The importance of particle reservoirs has already 
been exemplified for the TASEP with open boundary conditions, where bulk phases 
are selected by the capacity of the boundary reservoirs. In the context of intracellular 
transport, bulk reservoirs may also be considered. Besides, spatial confinement and limited capacities of the particle reservoirs may have strong impact 
on the system's behavior.

\subsection{The TASEP with Finite Resources}
\label{sec:finiteres}

The TASEP with open boundary conditions is coupled to infinite boundary reservoirs, which impose 
constant in- and output rates of the particles. 
This situation might be realized in {\em in vitro} experiments, where the occupation of the cytoskeletal 
filaments has typically little impact on the concentration of molecular motors
in solution. In cells, however, 
the number of transported objects is limited, and this has to be considered in theoretical approaches 
as well. 

In this section, in order to single out the effect of finite resources, we discuss their impact on the standard TASEP with open boundaries,
which are then
realized by explicit particle reservoirs rather than by in- and output rates.
This has first been done in a different context\footnote{Ha
and den Nijs actually motivated their model with the \emph{garage parking
problem}.} by Ha and den Nijs~\cite{ha_n2002} who fixed the total number of particles $\Ntot$ in
the system and expressed it as
\begin{align}
\Ntot=\Nbo+\Nun,
\end{align}
where $\Nbo$ is the number  of particles on the lattice and $\Nun$
 the number  of particles outside the lattice in the
boundary reservoirs (unbound particles).
The effective entrance rate $\alphaeff$ depends on the
number of particles in the reservoir:
\begin{align}
\alphaeff(\Nun)=\alpha\, f(\Nun).
\end{align}
The function $f(\Nun)$ satisfies the conditions $f(0)=0$, $f(\infty)=1$ and
is monotonously increasing in order to respect the availability of particles in
the reservoir. In~\cite{ha_n2002}, the case $\beta=1$ was considered, whereas
Adams {\em et al.}~\cite{adams_s_z2008} had a more complete look at the modifications
brought about to the whole phase diagram by the finite resources. Finite
resources have also been considered with respect to the activity of particles,
which is limited by the supply of energy \cite{brackley_c_r2012}.

It is easy to see that for $\Ntot\to\infty$ the standard TASEP as presented in section~\ref{sec:tasep} 
is recovered with $\alphaeff=\alpha$. Interesting effects are observed if the total number of particles is chosen low 
enough in order to constrain the number of particles on the lattice and induce feedback mechanisms 
between the densities on the lattice and in the reservoir.
In particular, $\alphaeff$ can depend on time because the number
of particles attached to the lattice can vary.
Rather than discussing the particular properties 
of different finite particle reservoir realizations we will focus on their general impact on the bulk states. 

First of all it should be noted, that $f(\Nun)$ can be chosen such that the particle-hole symmetry of 
the TASEP is broken, which would lead to a modified phase diagram of the system in the stationary state.
For the open TASEP with infinite particle reservoirs a phase transition from a low density to a high density 
phase is observed at $\alpha = \beta$ ($\alpha, \beta<0.5$ ). This is also true for systems with finite resources, 
but the \emph{effective} entrance rate $\alphaeff$ has to be considered, while the exit rate $\beta$ is unchanged. 
Similarly, the transition to the MC phase is reached if  $\alphaeff\geq 1/2$ and $\beta \geq 1/2$, where the ability 
to reach a MC phase obviously depends on the total capacity of the boundary reservoirs. 

The finite reservoir may not only modify the phase boundaries but also the dynamics of the shock that appear when a jammed and free flow phase coexist. In case of 
the standard TASEP the shock velocity is independent of its position. This is not always the case for finite particle reservoirs 
since $\alphaeff$ depends on the number $\Nbo$ of bound particles which itself depends on the position of the shock~\cite{cook_z2009}. The shock position is localized via 
the following mechanism: At the mean position of the shock the condition  $\alphaeff = \beta$ holds and 
the motion of the shock is unbiased. However, if the shock is displaced towards the entry of the system the 
value of $\alphaeff$ is lowered since more particles are in the system at the expense of the reservoir. Similarly 
$\alphaeff$ is increasing for displacements of the shock towards the exit.
Therefore, this feedback 
mechanism localizes the shock at its mean position. 
This mean position can be located anywhere along the filament,
in contrast with the case of infinite reservoirs for which
shock localization can occur only at the boundaries.

Next to the localization of the shock, other non-trivial effects have been observed for the TASEP with finite boundary reservoirs 
 ~\cite{adams_s_z2008,cook_z2009}. We won't discuss this in more detail but
switch to models which consider the fact that the runs of molecular motors
along the filament have a finite length (smaller than the filament length)
due to the possibility to detach - and subsequently re-attach.

\subsection{The TASEP with Langmuir Kinetics}
\label{sec:PFF}
\begin{figure}[tbp]
\begin{center}
\includegraphics[width=0.7\linewidth]{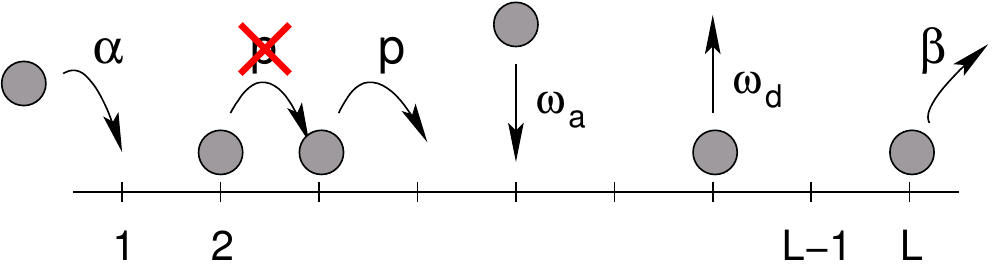}
\caption[TASEP with Langmuir kinetics.]{TASEP with Langmuir
kinetics. In addition to the usual TASEP hopping rules,
particles are also exchanged with a reservoir in the bulk of
the system: A particle detaches from the lattice at rate
$\omd$ and an empty site gets occupied by a particle at rate
$\oma$.}
\protect\label{fig:PFF_model}
\end{center}
\end{figure}

Molecular motors switch between phases of directed and
diffusive motion. During the latter, they are not attached
to filament but move randomly through the cytoplasm under
the constraints of other cytoskeletal elements.
This detachment and attachment of
motors to the MT leads to a local non-conservation of
particles and current in the bulk and has been considered in
the most simple way in a model by Parmeggiani \emph{et
al.}~\cite{parmeggiani_f_f2003} who coupled the
open-boundary TASEP of section~\ref{sec:tasep} with Langmuir
kinetics in the bulk.
More precisely, a particle detaches
from a lattice site in the bulk at rate
$\omd$\abk{$\omega_\text{d}$}{detachment rate} and
an empty site becomes occupied by a particle at rate
$\oma$\abk{$\omega_\text{a}$}{attachment rate}
(figure~\ref{fig:PFF_model}). The
model exhibits the same type of particle-hole symmetry as
the TASEP under the additional symmetry operation
$\omd\leftrightarrow\oma$.
From a physical point of view the attachment and detachment 
of particles corresponds to a coupling
of the bulk of the system to a grand-canonical reservoir
with constant density. Particles within the bulk reservoir
have an infinite diffusion rate since a detached particle
loses any memory of the site of previous attachment. 

Similarly to what was done in section~\ref{sec:multilane}
for coupled TASEP lanes,
some continuous mean-field equations
can be derived from the microscopic equations 
in the limit of large systems: 
a change of variable $x=i/L$ and an expansion using the fact
that $L\ll 1$ yields
\begin{align}
\frac{\partial\rho}{\partial t}=-(1-2\rho)\frac{1}{L}\frac{\partial\rho}{\partial x}+\omd \left[\frac{\oma}{\omd}-\left(1+\frac{\oma}{\omd}\right)\rho\right],\label{eq:PFF_mean_field1}
\end{align}
where $x=i/L$.
The dominating terms in this equation depend on the order
of the coupling rates $\omd$ and $\oma$.
In the same way as discussed for two-lane systems, we could
consider strong coupling to the bulk reservoir, for which the rates $\oma,\omd$
are of order one.
Then the second term in~\eqref{eq:PFF_mean_field1} would dominate for
large systems.
However, in the case of
intracellular transport, it is generally considered that it
is the weak coupling
in the sense of~\eqref{eq:oma_scaling}
that is relevant, since the run lengths of processive molecular
motors are of the order of the filament length, as mentionned
in section~\ref{sec:kinesin} :
\begin{align}
\oma=\frac{\Oma}{L};\qquad\omd=\frac{\Omd}{L};\qquad\text{with }\Oma,\Omd = \text{const.}\label{eq:oma_scaling}
\end{align}
The current between lanes is then of order $1/L$
and time must be rescaled as $\tau=t/L$, so that
all terms in (\ref{eq:PFF_mean_field1}) are of the same order~\cite{evans_j_s2003}~:
\begin{align}
\frac{\partial\rho}{\partial\tau}=-(1-2\rho)\frac{\partial\rho}{\partial x}+\Omd \left[\frac{\Oma}{\Omd}-\left(1+\frac{\Oma}{\Omd}\right)\rho\right],\label{eq:PFF_mean_field}
\end{align}

\subsubsubsection{Stationary states and phase diagram}

When searching the stationary state
$\partial\rho/\partial\tau=0$ of this model,
equation~\eqref{eq:PFF_mean_field} obviously leads to a
first-order differential equation. The couplings to the
boundary reservoirs on the left and right end of the lattice
impose two boundary conditions $\rho(x=0)=\alpha$ and
$\rho(x=1)=1-\beta$.
In other
words, the problem is over-determined. The overall
solution~\cite{evans_j_s2003}
consists of the two individual solutions resulting from
integration of the differential equation from the left and
from the
right side of the system. The transition point from one
solution to the other can be determined in the framework of
kinematic wave theory. If the densities of the two solutions do
not match at the point of transition, a shock (or domain wall)
occurs which depends on higher orders of $L^{-1}$.
Consequently, this shock gets sharper the longer the system is.

\begin{figure}[tbp]
\begin{center}
\includegraphics[width=0.5\linewidth]{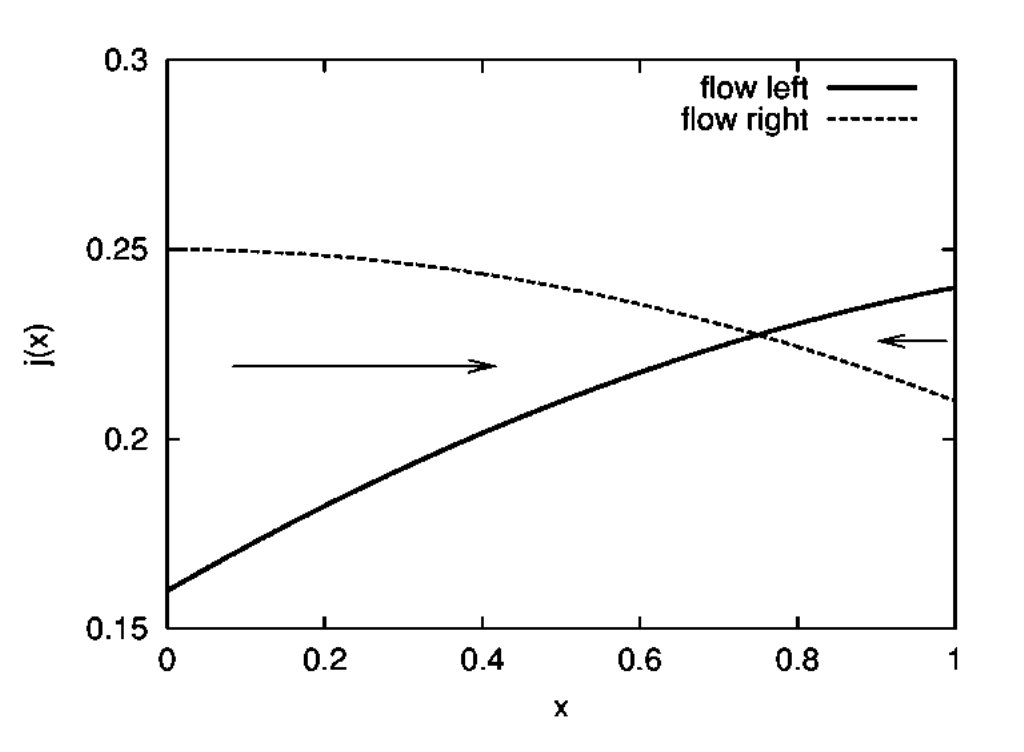}
\caption[Shock dynamics of the TASEP with Langmuir kinetics.]{Typical flow profile of the TASEP with Langmuir kinetics in the shock phase. The solid (dashed) line indicates the flow profile induced by the left (right) boundary reservoir. The arrows indicate the bias of the shock. From~\cite{evans_j_s2003}.}
\protect\label{fig:PFF_shock}
\end{center}
\end{figure}

In contrast to the TASEP, the shock front does not move within the
system, once the system has reached its stationary state. Instead,
the shock is driven to the position in the system where the mass
current through the shock front is zero (see
figure~\ref{fig:PFF_shock}), so that localization of the shock
occurs~\cite{evans_j_s2003}. This phenomenon is interesting in so
far as no defects on the lattice are implicated and the particle
dynamics is homogenous in the bulk of the system.
Surprisingly, on the line $\alpha=\beta<1/2$, the
localization effect persists 
even if the bulk
reservoir density vanishes with increasing system size, i.e.,
$\omd,\oma\propto L^{-a}$ with $1\leq a<2$~\cite{juhasz_s2004}.

In contrast to the TASEP two macroscopic boundary domains and a bulk domain can be imposed simultaneously in the stationary state. Therefore, the number of qualitatively different density profiles is much higher than for the TASEP and depends strongly on the attachment ($\oma$) and detachment  ($\omd$) rates~\cite{evans_j_s2003,parmeggiani_f_f2004} and the phase diagrams exhibit more different phases than for the TASEP.
More precisely, in addition to the three phases already known from the TASEP (see section~\ref{sec:tasep_pd}), there is also a phase corresponding to stationary states with a localized shock in the bulk of the system. Furthermore, there are several states in which two or all three TASEP phases coexist. In these phases, no shock is observed and the partial profiles are continuously connected with a cusp in the density profile. If $\Oma\neq\Omd$, the maximal-current phase is suppressed since the Langmuir kinetics tries to bring the bulk of the system to a density $\rho_\text{eq}\neq1/2$, thus driving it away from the maximal-current density.\\
Altogether, the Langmuir kinetics brings about some very interesting effects although it is an uncorrelated equilibrium process which would not lead 
to phase transitions on its own. In combination with a non-equilibrium process, however, the Langmuir kinetics modifies the phase transitions found in the TASEP and leads to qualitatively new behavior. 

Similar results are obtained if two parallel lattices are considered, which exchange particles with a bulk reservoir via Langmuir kinetics and are additionally able to exchange particles between filaments~\cite{wang2007}. The qualitative behavior also remains unaltered, if particles are considered to be dimers whose two heads individually bind to two neighboring sites and walk in the hand-over-hand fashion described for kinesin in section~\ref{sec:kinesin}~\cite{pierobon_f_f2006}.

The model also allows for comparison with \emph{in vitro} experiments, where the concentration of the molecular motors and of ATP can be controlled. The motor concentration in the reservoir is proportional to the attachment rate while the ATP-concentration has strong influence on the stepping rate of the motor proteins (see ~\cite{nishinari2005b} and our discussion in section~\ref{sec:cycle}).
. 

In order to improve the realism of the theoretical model the substructure of the 
filaments has been taken into account. In~\cite{chowdhury_g_w2008}, Chowdhury {\em et al}
extend the model to several parallel lanes, allowing
for lane-changes. These lanes could represent the
13 protofilaments that compose the MT.
Their model includes not only Langmuir kinetics
but also a minimal description of the mechanochemical
cycle responsible for motor stepping (see section
\ref{sec:cycle}).
They show that including lane changes may increase
or decrease the flux per lane, depending on the motor
concentration and the rate of hydrolysis of ATP.
They suggest that this could be checked by \emph{in vitro}
experiments.

\subsection{Explicit Realizations of Particle Reservoirs }
\label{sec:model_lipowsky}

\begin{figure}[tbp]
\begin{center}
\includegraphics[width=0.7\linewidth]{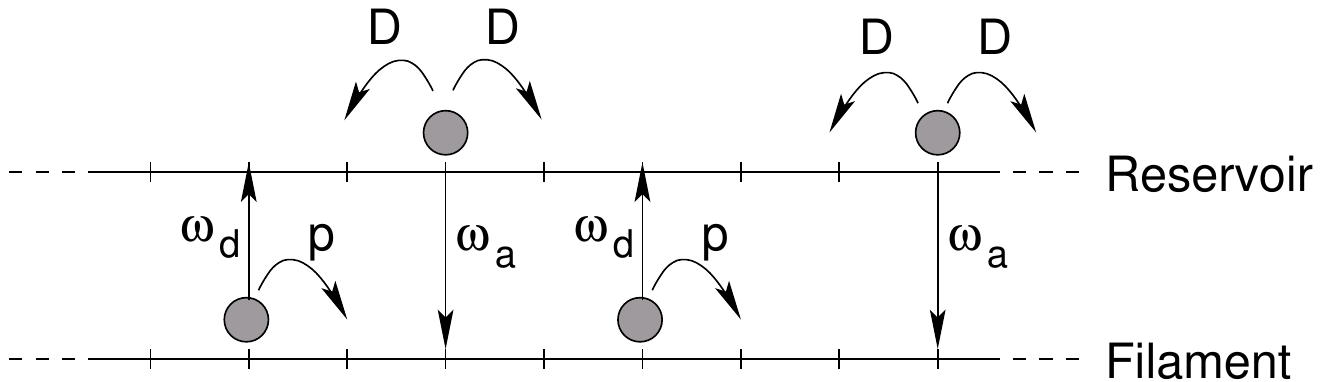}
\caption[One-species TASEP coupled to a diffusive
lane.]{One-species TASEP coupled to a diffusive lane. The model
consists of two one-dimensional lattices. The
lower lattice represents the filament and serves as track
for directed motion of the particles (represented as disks)
which interact via hard-core
exclusion on this lane.
The upper lane is a diffusive environment
with unbiased motion and without any interaction between
particles. As a result, one site of the diffusive lane
can be occupied by several particles at the same time.}
\protect\label{fig:standard_model1}
\end{center}
\end{figure}

As a first step toward a realistic description of the particle reservoir, one may
represent the particle reservoir by an additional lane, parallel
to the filament, as sketched on
figure~\ref{fig:standard_model1}. 
On the diffusive lane, there is no exclusion principle\footnote{Actually the diffusive lane represents the whole volume around the filament, and thus several particles can be present at the same time at a given position.},
and particles diffuse freely by hopping with equal
rates to their right or to their left.
On the filament lane, particles are subject
to directed motion and exclusion.
This model is relevant for filaments that are embedded in 
a small volume such that the motors do not drift too far 
away from the filament. Moreover the density of motors 
should be low since exclusion effects are not considered 
in the diffusive lane. {\em In vitro}, this confinement could be achieved
in micro-channels or nanotubes.
We shall discuss this confinement assumption for an {\em in vivo}
situation (axonal transport) in section~\ref{sec:confin}.

The properties of the stationary state for such a two-lane
model, and in particular the density profile,
have been calculated by M.R.~Evans \emph{et al} in~\cite{evans2011}.

\begin{figure}[tbp]
\begin{center}
\includegraphics[width=0.6\linewidth]{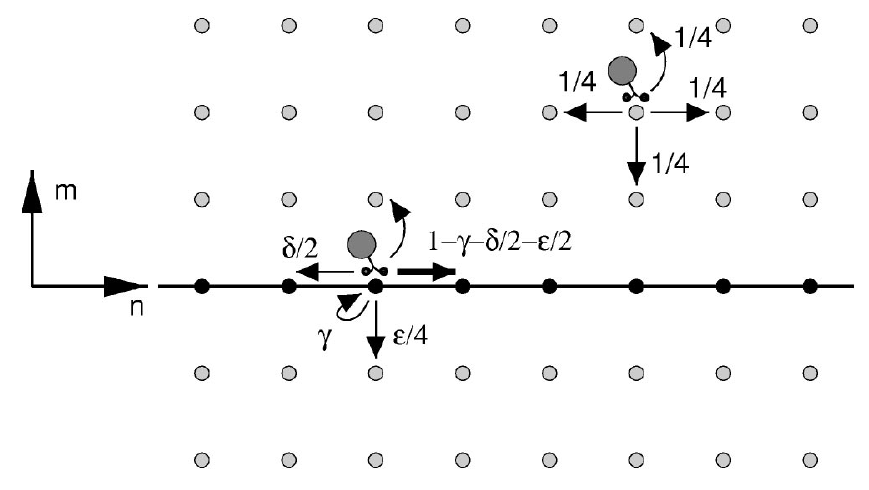}
\caption[Model with explicit consideration of the bulk
reservoir.]{Model of~\cite{nieuwenhuizen_k_l2004} with explicit consideration of the bulk
reservoir. Particles hop
between neighboring nodes on a cubic lattice, where the filament
is represented as a line of black nodes and the bulk reservoir by
grey nodes. Hopping within the bulk reservoir is unbiased, but
hopping to and away from filament sites depend on the attachment
and detachment rates. Please note that the notation of the rates
in this figure does not correspond to the notation in the remaining
of this review paper.
From~\cite{nieuwenhuizen_k_l2004}.}
\protect\label{fig:KL_model}
\end{center}
\end{figure}

Lipowsky \emph{et al.}~\cite{lipowsky_k_n2001} have introduced a many particle 
model for motor-driven transport which describes not only the motion of the molecular 
motors on the filament  but also in the reservoir.  
The model is defined in a three-dimensional geometry
with discretized space. Particles hop between nodes of a cubic
lattice with spacing $l$ which corresponds to the step size of
the molecular motor on a filament, e.g., $8~\text{nm}$ for
kinesins on MTs. As in the previous models, the filament is
represented
by a linear arrangement of lattice nodes while all other nodes
represent the surrounding volume (the cytoplasm in a cell,
the solution for an \emph{in vitro} experiment) where motors undergo
diffusion.
The filament nodes differ from the
non-filament nodes by the hopping rules of the particles.
Filament-bound particles undergo directed motion.
They hop forward with rate $p$ provided
the target site is empty and never step back.
Unbound particles undergo an unbiased random walk
on a discrete three-dimensional lattice while interacting with
each other via hard-core exclusion (figure~\ref{fig:KL_model}).
Because of the cubic lattice structure, every filament site can be
reached from four reservoir nodes. The ratio of the hopping
rates from these four sites to the filament and back, i.e., the
attachment and detachment rates, determines the affinity of a
particle to the filament. The model enforces the conservation
of the number of particles and the hopping between reservoir
nodes determines the strength of diffusion.

This model includes more features of intracellular transport than
the previously presented models by explicitly considering a
three-dimensional geometry within which a filament is placed, so
that different cellular or experimental environments can be
modeled.
Lipowsky \emph{et al.}~\cite{lipowsky_k_n2001}
have tested the outcome of different geometries while estimating
the rates from real experiments.
This should allow a more direct comparison with experiments.

The explicit consideration of the particle reservoir shows that the 
attachment and detachment rates are not constant since strong 
concentration gradients of the particles in reservoirs are observed. 
The actual distribution of the particles depends strongly on 
the chosen reservoir geometry~\cite{nieuwenhuizen_k_l2002,nieuwenhuizen_k_l2004,lipowsky_k_n2001,klumpp_n_l2005,klumpp_l2003,ciandrini2014}.

In order to illustrate the influence of the system's geometry we 
briefly discuss the case of a mixed choice of an open boundary 
at the minus-end of the
filament and a closed boundary at the plus-end~\cite{mueller_k_l2005}. 
This choice mimics for example the situation in an
axon (see section~\ref{sec:axon}) more accurately
where the closed end represents the synapse and
the open end the connection to the cell body which provides the axon
with a constant density of molecular motors.
Using this geometry,
M\"uller \emph{et al.}~showed
 that the density on the filament in the stationary
state either approaches zero or one far from the open end, i.e.,
either the long tube is empty or all transport is blocked due to
a jam extending through the whole system. The transition between
these two states obviously depends on the strength of the
diffusion which drives particles back to the open end against the
preferential motion of particles on the filament.
M\"uller \emph{et al.}~concluded in the article that additional
regulatory
mechanisms are necessary in order to maintain efficient transport
in long tube-like geometries as the axon at intermediate
densities.

\subsection{Enhanced diffusion on filament networks}

In this section we consider the case where multiple filaments
can serve as tracks for the motors.
Klumpp and Lipowsky
used the framework of the three-dimensional lattice gas to show
that certain arrangements of filaments lead to non-directed
diffusive movements which can be used for transport over large
distances and exhibit very large diffusion coefficients compared
to passive diffusion in the absence of
filaments~\cite{klumpp_l2005b}. This diffusive motion is driven
by phases of directed transport along individual MTs and is thus
dependent on the active stepping of individual motion. Klumpp and
Lipowsky therefore refer to this phenomenon by the term
\emph{active diffusion} as the overall motion is undirected due
to the uncoordinated attachment to filaments of different
orientation. They show this type of motion not to depend on the
diffusion coefficients in the cytoplasm surrounding the
filaments.

Klumpp and Lipowsky discuss the relevance of this enhanced
diffusion for several technical applications, but it may also
play a role in dendritic transport. Dendrites are very
elongated with a parallel MT network. But in contrast to axons,
dendritic MTs are not uniformly oriented so that the structure of
the MT network might indeed enhance the diffusion of
intracellular cargo. However, the same is not true for axons
that will be considered in the next section~\ref{sec:axon},
and which have an almost uniformly oriented MT network.

\subsection{Continuous models}
\label{sec:continuous}

We have seen that in the frame of a mean-field
approximation, it is usually possible to derive
some continuous equations as
(\ref{eq:PFF_mean_field})
from the microscopic dynamics.
But it is not mandatory to build continuous
equations from microscopic models.
Some authors have proposed to describe
intracellular transport directly at the level
of reaction-diffusion-transport continuous
equations.

For example, Smith and Simmons~\cite{smith_s2001}
have written a set of differential equations for
a multi-lane system.
They distinguish some lanes with oriented motion (filaments)
and a lane for simple diffusion.
Only one type of motor was considered, moving
on either on a fully
polarized network,
or on bi-polar filament networks. The latter
case would
correspond for example to myosin V on an
actin network.

We shall not describe further this class
of models,
as we are interested in stochastic models in this
review.
One advantage of stochastic transport
models is that it is often possible
to relate the model parameters directly to experimental
observation.

\subsection{Membrane tubes}
\label{sec:tubes}

In this section we address a phenomenon quite different from the intracellular transport
mechanisms described above, namely the formation of membrane tubes, i.e. the formation
of one-dimensional cylindrical objects that are pulled or pushed out of the cell (or organelle)
body. This kind of process is for example of great importance for several biological system
including the growth of axons out of the soma of nerve cells.
Another example is the tubulation of Golgi
membranes~\cite{sheetz1996}
which is known to be also driven by MT-based motors.

A cooperative effort is required from molecular
motors to pull a membrane nanotube from a
vesicle. \emph{In vitro} experiments have shown that
if a vesicle is coated with kinesin and then
brought near a MT, the kinesins were able to
pull a nanotube~\cite{roux2002}.
This phenomenon has been modeled
with a two-lane exclusion
process~\cite{campas2008,tailleur_e_k2009}:
the particles represent the kinesins, which all
have their bounding domain permanently
attached to the nanotube.
Kinesins which are also bound to the MT, step
on it with a constant rate $p$, always in the
same direction.
Kinesins may detach from the MT and then
perform a diffusive motion on the other lane,
without any exclusion rule.
They become bound again to the MT with a certain
attachment rate.
Up to this level, the model is strictly equivalent
to the model for uni-directional intracellular
transport with finite diffusion described in
Fig.~\ref{fig:standard_model1}.

What makes this model specific for pulling nanotubes
is that the nanotube as a varying length. Similarly,
at the end representing the tip of the nanotube,
the two lanes of the model may extend and retract
with rates $v^+$ and $v^-$.
Tailleur, Evans and Kafri~\cite{tailleur_e_k2009}
propose a mean-field
approach that predict reentrant phase
transitions, as a result of the competition
of the densities imposed by the two ends of
the tube and of the dynamics of the shock
that separates these densities.
They indicate a certain number of signatures
of their theory that could be experimentally
tested.

This model gives a first example of 
coupling  between the dynamics of
the underlying track and transport.
We shall see other examples in the next chapter.

\section{Bi-directional transport and spatial confinement}
\label{sec:vivo}

Throughout this section, we shall refer more precisely
to axonal transport, as it brings some specific questions
and gives a precise frame for our discussion.
However, it must be kept in mind that, though
the axon is the biological system that motivated the
present discussion, a fully realistic
description of axonal transport is still lacking,
and this section will be followed by a list of open
questions (section~\ref{sec:challenges})
that still need further investigations, both on the point
of view of experiments and models.
We shall present in this section attempts to understand
the basic phenomena that could underly axonal transport
and more generally MT-based transport.
But first we shall describe the biological system
under consideration, namely the axon.

\subsection{Biological description of the axon}
\label{sec:axon}

\begin{figure}[tb]
\begin{center}
\includegraphics[width=0.6\linewidth]{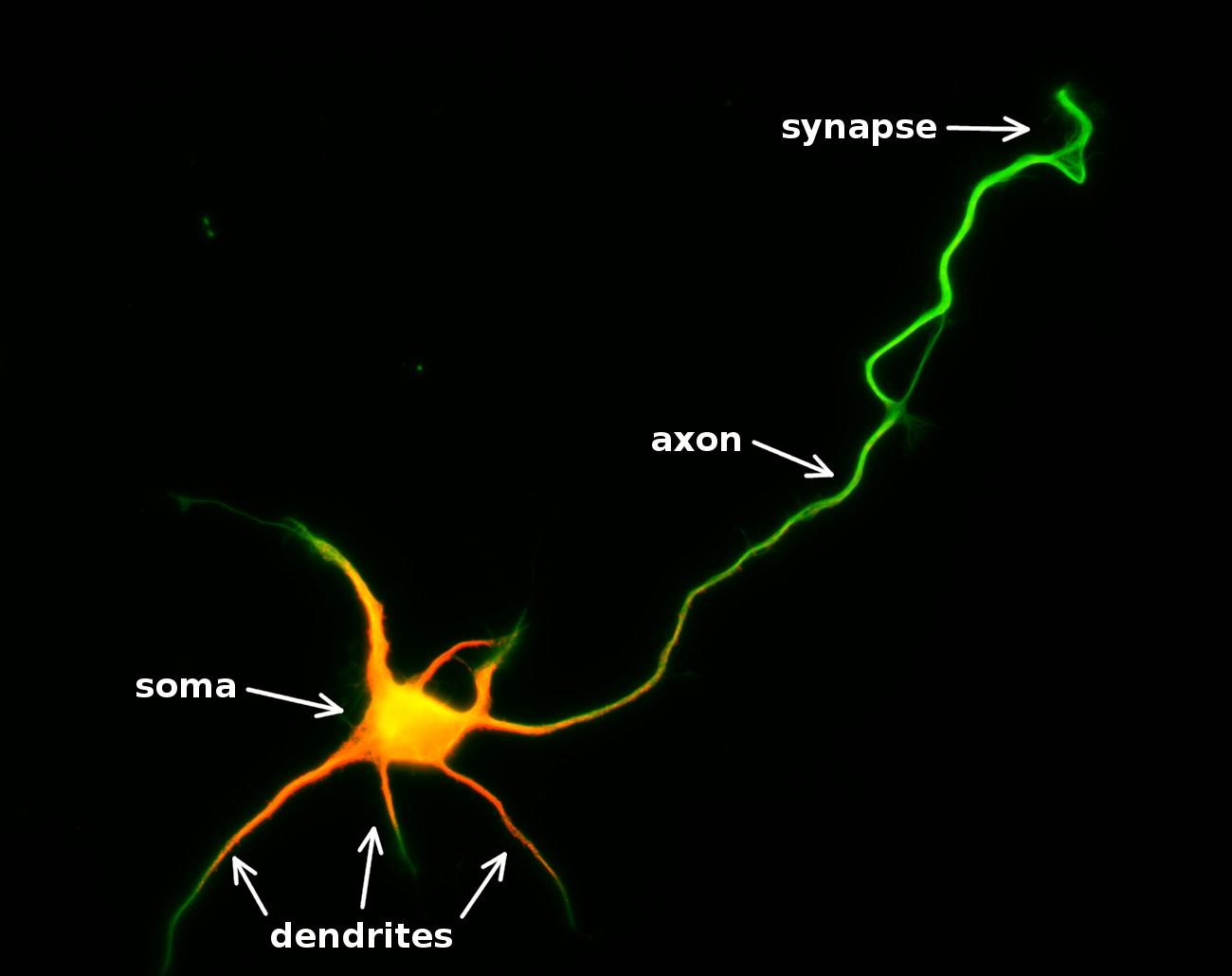}
\caption[Neuron.]{
Cultured hippocampal neuron
after 2 days in vitro.
One can distinguish the axon, the dendrites and the soma.
Modified from~\cite{brandner_w2010}.
}
\protect\label{fig:neuron}
\end{center}
\end{figure}

The axon is one of the neurites of a neuron and as such
connecting the distant synapse with the cell body or soma
(figure~\ref{fig:neuron}). In vertebrates, axons are
among the longest structures at the level of individual cells
with lengths of up to $1~\text{m}$ in humans while having
thicknesses typically not exceeding the micrometer
scale~\cite{gray2008,gov2009}. The axon therefore presents a
quasi-one-dimensional system within which intracellular cargos
are bidirectionally transported from the soma to the synapse and
vice versa~\cite{morfini2012}.

\begin{figure}[tb]
\begin{center}
\includegraphics[width=0.7\linewidth]{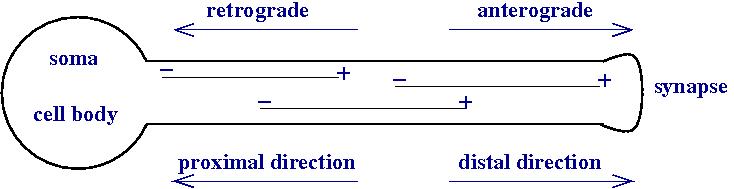}
\caption[Vocabulary axon]{Schematic representation of a
neuron and of the directions associated with various terms
used to describe axonal transport.
}
\protect\label{fig:scheme_axon}
\end{center}
\end{figure}

\begin{figure}[tb]
\begin{center}
\includegraphics[width=0.25\linewidth]{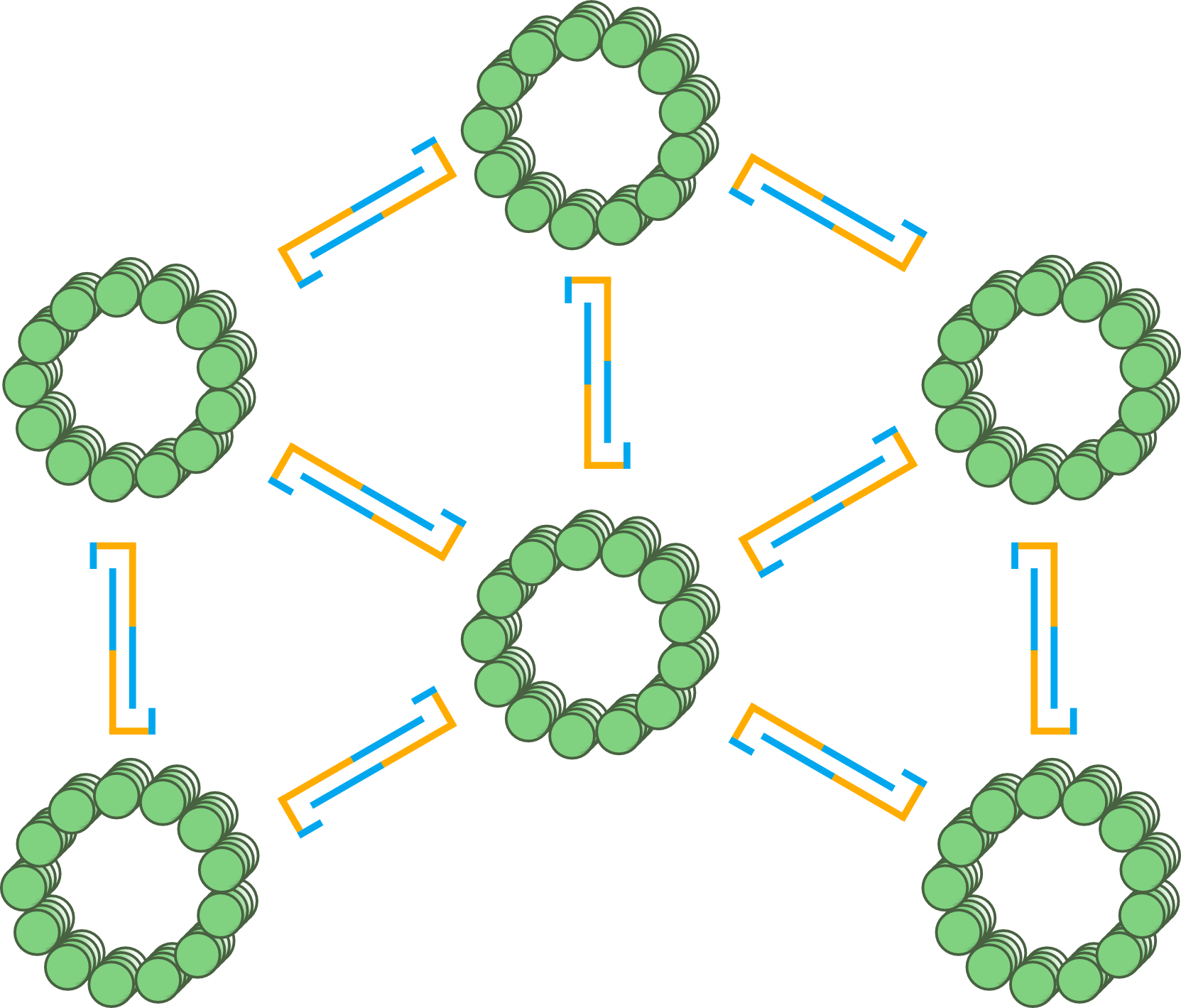}\hspace{3 mm}
\caption[MT spacing by MAPs in axons.]{
MT spacing by MAPs in axons:
Sketch of the cross-section of an axon.
Tau dimers (blue and orange)
bundle MTs (green rings) and induce a hexagonal structure of the
axonal MT network.
}
\protect\label{fig:neuron_section}
\end{center}
\end{figure}

\subsubsection{Structure of the axonal microtubule network}
\label{sec:axon_mt_struct}

The axonal MT network is highly polarized and consists of many
MTs of different lengths. The MTs are longitudinally arranged and
a large majority
point
with their plus-end in distal direction,
i.e. away from the soma~\cite{baas1988,kapitein_h2011}.
Transport towards the synapse or the soma have different
characteristics, obtained in particular through this high
polarisation of the MT network, and some vocabulary has been
developped to distinguish among these two transport directions
(see sketch in Fig.~\ref{fig:scheme_axon} for a summary of the
terminology).
It seems for example that organelles that move towards the distal
end are on average smaller but more numerous than those moving
towards the soma~\cite{morfini2012}.

 Individual MTs may be as long as
$\sim100~\mu\text{m}$~\cite{brown1993,yu_b1994}.
Some protein called tau
(see section~\ref{sec:tau})
 has a bundling function and leads to a parallel arrangement of
 the axonal MTs in an imperfect hexagonal lattice
 (see Fig.~\ref{fig:neuron_section})
 with spacing of
 $26.4\pm9.5~\text{nm}$~\cite{chen1992,rosenberg2008}.
An electron micrograph showing the arrangement of
MTs under the influence of tau 
can be found in~\cite{chen1992}.

The MT network in the axon is complemented by
actin networks, mainly located close to the cell
surface.
Actin filaments could also provide tracks for transport
in regions with few MTs, as the axon terminal~\cite{mallik_g2004}.

There are also neurofilaments present between the MTs,
especially in axons of large diameter~\cite{chen1992,brown_j2012}.

\subsubsection{Dynamics of the axonal microtubule network}
\label{sec:axon_mt_dyn}

\begin{figure}[tb]
\begin{center}
\includegraphics[width=0.7\linewidth]{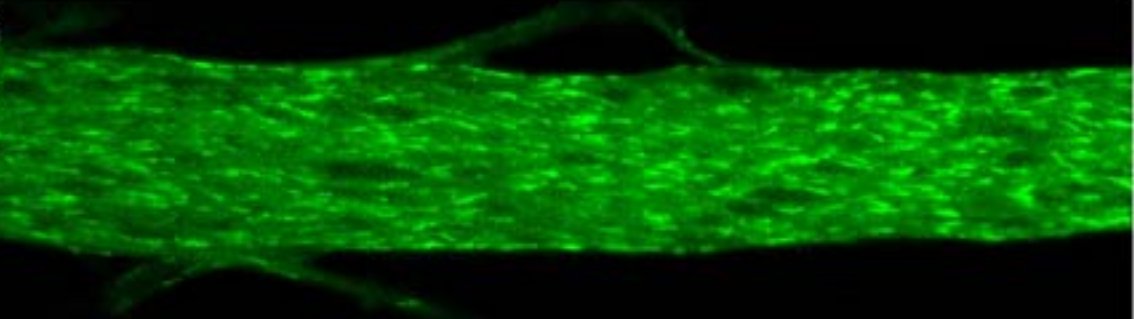}
\caption[Confocal microscopy imaging of the polymerization of MT plus-ends]{
Confocal microscopy imaging of the dynamics of MT plus-ends in the
axon of cultured Aplysia neurons.
Some fluorescent plus-end-tracking protein allows
to mark the plus-ends of MT that are {\em polymerizing}.
The resulting `comet tails' show the orientation
of the MT, with a majority of MT plus-ends pointing towards
the distal tip of the axon.
The width of the axon is approximately 10 microns.
The video provided in the supplementary materials of
\cite{shemesh_s2010a} (Movie S1A) shows the very dynamical
behavior of MT tips.
    Reprinted from~\cite{shemesh_s2010a}.
}
\protect\label{fig:dyn_MT}
\end{center}
\end{figure}

In contrast to non-neuronal cells, the minus-ends of axonal MTs
are not attached to the centrosome or one of the other common
nucleating structures. 
Still, it seems that, in general, only the plus-end of
axonal MTs is dynamic while the minus-end seems to be
stable~\cite{baas_b1990,baas_a1992,li_b1996}, probably due to some capping
protein~\cite{dent_g2003}. 
For example, $\gamma$-tubulin has been recently found to be present in the axon and
might both play a role in the nucleation of axonal MTs and serve
as cap on the minus-end~\cite{stiess2010}\footnote{It should however be
kept in mind that, though minus-ends of axonal MTs are usually
supposed to be stable, treadmilling has also been observed in
vivo~\cite{waterman-storer_s1997}, and it cannot be excluded that
it also occurs in axons.}.
Recently were discovered some protein families that can
associate with free MT minus-ends and stabilize them against total
depolymerization~\cite{jiang2014}.

Polymerization and depolymerization
of the plus-end is about a factor two slower in mature axons than
in non-neuronal cells and occurs along the main axis of the
axon~\cite{stepanova2003,shemesh_s2010a}.

It recently became possible to visualise the dynamics
of MTs directly in living neurons using fluorescence
and confocal microscopy 
techniques (see for example Movie S1 in ~\cite{shemesh2008}).
\cite{shemesh2008,shemesh_s2010a,shemesh_s2010b} report
some experiments in which
some mRNA encoding for the plus-end-tracking protein
EB3 tagged by GFP is injected in the cell body of
cultured Aplysia neurons.
EB3-GFP has the property that it binds transiently
to dynamically polymerizing MT plus-ends.
Images show comet-tail-like traces (see Fig. \ref{fig:dyn_MT}),
which can be interpreted as the growing part of fixed MTs
(though it can also be a signature for the
motion of short MTs transported along longer MTs).
The number of comet tails gives an estimate
for a lower bound of the number of
polymerizing MTs (not all polymerizing plus-ends
are visualized).

These images renew the classical view of a static
highly interconnected MT axonal network.
Instead, plus-ends of MTs are polymerizing and depolymerizing
very actively,
a feature that allows to reconcile the antinomic
requests for structure survival and adaptation capacity.
However, while MT dynamics is quite well known
in \emph{in vitro} conditions, we are still lacking a full picture of
MT dynamics in \emph{in vivo} environments.

\subsubsection{Confinement}
\label{sec:confin}

\begin{figure}[tbp]
\begin{center}
\includegraphics[width=0.6\linewidth]{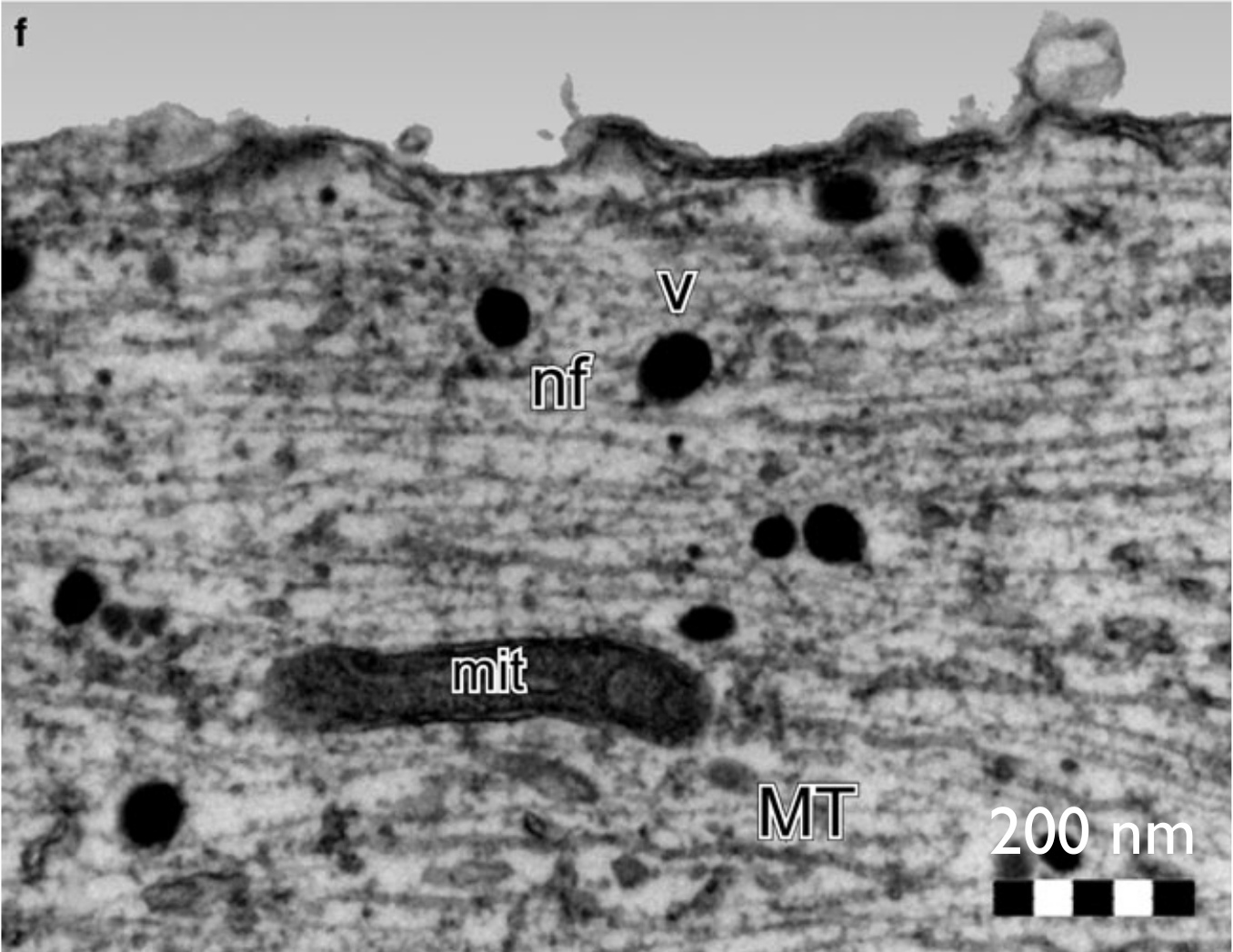}
\caption[]{Highly magnified view of a small part of the axon
obtained from electronic microscopy. Microtubules (MT),
neurofilaments (nf), vesicles (v) and mitochondria (mit)
are visible.
    Reprinted from~\cite{shemesh_s2010b}.
}
\protect\label{fig:axon_crowded}
\end{center}
\end{figure}

Many results obtained \emph{in vitro} cannot be directly
applied to \emph{in vivo} transport. One reason is that
the interior of a cell is very crowded, and diffusion
is strongly limited by steric hindrance~\cite{luby-phelps1999}.
An example is given in Fig.~\ref{fig:axon_crowded}
which shows the internal structure of an axon.
The size of common vesicles ($\approx 100~\text{nm}$~\cite{forman_l_s1987})
is not negligible in front of
MT interspacing. Neurofilaments also limit the
available space~\cite{brown_j2012}.

The assumption made in section \ref{sec:PFF}
of an infinite diffusion rate is not justified
in this context.
One may argue, however, that between two episodes of microtubule
based transport, a given cargo may attach to the actin
network:
though each actin filement is polarized,
there is no global alignment of the polarization as for MTs,
so the cargo could undergo bidirectional motion
by changing filament.
Diffusion should then account for some motor-driven diffusive motion
with a diffusion constant that depends on the speed
of the cargo-motor complex and detachment rate on
actin~\cite{smith_s2001}.
Though this mechanism may increase the longitudinal
diffusion constant, first it does not make it infinite,
and second it should be relevant only for cargos
that can reach the actin network, located mostly
along the membrane of the axon, and not for those
travelling in the bulk part of the axon.

For cargos of intermediate size, confinement will limit
strongly the
diffusion around the MT (in the radial direction),
and diffusion will occur mostly in the longitudinal
direction, along the MT.
This is the case that will be considered in the next
subsections.

It must still be kept in mind that much larger
cargos can be transported,
e.g. mitochondria, which are too large to travel
in the empty space between MTs and require a local
rearrangement of MTs during their processive phase.
We shall address this issue later in section
\ref{sect:large_cargos}.

Confinement may also have an effect on the dynamics
of the MTs and on the overall structure of the axonal
MT network. Indeed, a MT can grow only if there is empty
space available for that. An example of such a steric
effect on MT growth will be
given in section~\ref{sec:dyn_on_networks}.

\subsection{Finite diffusion: a two-species TASEP coupled to a diffusive lane}
\label{sec:model2}

In a living cell, the cytoplasm is
filled with structures of different sizes such that diffusion is
limited and actually dependent on the size of the diffusing
object~\cite{luby-phelps_t_l1986,popov1992}. Furthermore,
vesicles in axonal transport usually originate at the synapse or
in the cell body, so that conservation of transported particles
is provided within the axon.
Eventually, in real cells, more than one motor species is present.
Especially in the axon, different motor families ensure
anterograde and retrograde transport.
Thus at least two types of motors have to be considered,
moving in opposite directions on the MTs to capture
the bi-directional nature of axonal transport.

In the next sections, we shall address this question of
finding a minimal model that could lead to efficient
bi-directional transport while retaining some basic
constraints imposed by the axonal environment,
namely confinement and particle number conservation.

A first attempt to include all these ingredients in
a single model was performed in~\cite{ebbinghaus_s2009}
and we shall now summarize the results.

\subsubsection{Model definition}

Instead of describing the full diffusion reservoir, it can be
instructive to consider a simplified version of it, as proposed
in~\cite{ebbinghaus_s2009,grzeschik_h_s2010}.

\begin{figure}[tbp]
\begin{center}
\includegraphics[width=0.7\linewidth]{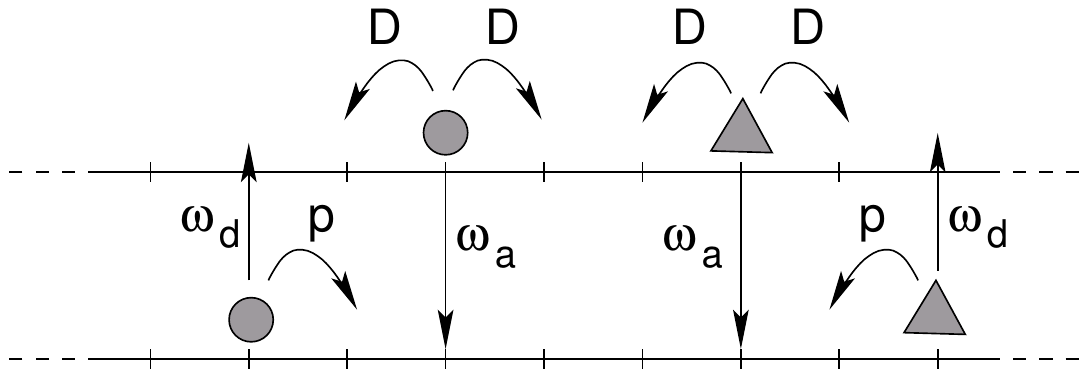}
\caption[Two-species TASEP coupled to a diffusive
lane.]{Two-species TASEP coupled to a diffusive lane. The model
consists of two periodically closed one-dimensional lattices. The
lower lattice serves as track for directed motion of the two
oppositely moving particle species (represented respectively
as disks and triangles) which interact via hard-core
exclusion on this lane. The upper lane is a diffusive environment
with no biased motion and without any interaction between
particles. As a result, one site of the diffusive lane
can be occupied by several particles at the same time.
From~\cite{ebbinghaus_a_s2010a}.}
\protect\label{fig:standard_model}
\end{center}
\end{figure}

The model consists of two parallel one-dimensional lattices with
periodic boundary conditions (figure~\ref{fig:standard_model}).
While one lattice represents the filament along which transport
is carried out, the other mimics a diffusive environment in which
no directed transport takes place. This corresponds to the
cytoplasm which surrounds the microtubules and within which
axonal cargos diffuse if not attached to one of the MTs.
Particles interact via hard-core
exclusion on the filament along which they hop in their preferential direction with rate $p$.
On the diffusive lattice, particles hop
in both directions with equal rate $D$\abk{$D$}{diffusion rate}
and do not interact with each other such that a single site on
the diffusive lattice can be occupied by multiple particles at
the same time. Particle exchanges between the two lattices occur
at rate $\oma$ and $\omd$, corresponding to attachment and
detachment moves to and from filament, respectively.

Whereas the motivation for the hopping rules on the filament has
been discussed in previous sections, the particular choice of
the diffusive lattice needs some clarification: As discussed in
section~\ref{sec:confin}, diffusion of axonal cargos in the
cytoplasm is limited due to the size of the cargos and to the
confinement imposed by the amount of structures present in the
axon. After detachment from the filament, an organelle might
therefore not be able to diffuse away from the filament in radial
direction.
The diffusive lattice can be seen as representing a narrow
cylindrical region of cytoplasm surrounding the MT and within
which unbound particles are confined.
The hopping rate on the diffusive lattice is related
to longitudinal diffusion, i.e. in the direction parallel
to the MT filament.
The limitation of diffusion in the transverse direction (i.e.
turning around the MT) is not
yet included into the model, so that the model allows multiple
particles to occupy the same site on the diffusive lattice.
Therefore, organelles do not feel the local density of axonal
cargo and can pass in front of each other if they are not
attached to the filament. The model consequently overestimates
this bypassing in the narrow environment and the results are
likely to underestimate the jamming in a real system with strong
spatial confinement.

The generic result of this model is that a huge cluster is formed
which absorbs almost all the particles of the system. This effect
is due to the highly inefficient bypassing of obstacles on the
filament. A particle which encounters an obstacle such as a
cluster of particles on the filament would have to detach to the
diffusive lane and then diffuse over the obstacle in order to
re-attach to the filament on the other side. Being unbiased, the
diffusive motion is quite inefficient, especially when the
obstacle becomes longer.
As a result, a cluster on the
filament is extremely stable for large system sizes.

This cluster formation does not depend on the density but on the
number of particles as only a limited number of particles is
necessary to form a nucleating cluster which will then start to
grow indefinitely.
It therefore has been reasoned in~\cite{ebbinghaus_s2009} that
the model exhibits macroscopic clustering at any finite density
$\rhototp=\rhototm>0$ in the limit of infinite system sizes
$L\to\infty$ (See~\cite{ebbinghaus2011} for details).

High values of the diffusion rate $D$ decrease the tendency to
form macroscopic clusters and actually totally suppress it in the
limit $D\to\infty$.
Indeed, this limit case correspond to the models of
section~\ref{sec:PFF}, for which diffusing particles loose memory
of their last attachment site.
This underlines that systems where particles have some sort of
memory of their last site of attachment to the filament
behave quite differently and require a specific study.

The effect of multiple transport filaments in parallel has also
been considered by enlarging the model to have several subsystems
as in figure~\ref{fig:standard_model} put next to each other.
Particles may change from one subsystem to the other either via
the diffusive lanes or by moves from filament to filament.
Counter-intuitively, this coupling of multiple filaments has
turned out to {\em promote} the formation of clusters although
additional ways for the bypassing of obstacles are available. The
moves from subsystem to subsystem transport locally high
densities from one filament to the other, leading to a system
which is blocked on all filaments at approximately the same
horizontal position.

Taken together, these results indicate that a simple bypassing
mechanism via a diffusive lane does not help in making
bidirectional transport efficient and overcoming the blocking of
particles of opposite charge on the filament. In real systems,
the consequences should be even worse, since these systems are
open at their ends so that the
number of particles in the system is not limited as for the
periodically closed system considered here.

\subsubsection{Varying confinement}

The jamming described in the previous section~\ref{sec:model2}
strongly relies on the fact that radial diffusion is inhibited.
Actually the degree of confinement in the axon is not precisely
known, and may depend on the type (and size) of the cargos under
consideration.

It is possible to increase effectively the radial extension of
the diffusive reservoir just by lowering the attachment rate. If
we assume that the diffusive environment has the shape of a
cylinder of radius $R$ centered around the MT, a simple scaling
argument indicates that the effective attachment rate should vary
as $\frac{\oma}{R^2}$.
When the diffusive reservoir is made larger, the total number
of particles must also be increased so as to keep a constant
density.

In~\cite{ebbinghaus_a_s2010b}, the effect of a larger cytoplasmic
environment was considered in this way.  We could show a
crossover from a jamming regime for strong confinement, to a free
flow regime in the case of a large diffusive reservoir.  However,
in the latter case, as particles spend most of their time
unbound, the overall net transport is not very efficient.
Besides, we expect confinement to be quite severe in the axon.
Thus it is still crucial to understand how bi-directional
transport in a confined environment can be regulated to avoid
desastrous jamming.

A naturally arising question therefore is: What are the minimal
prerequisites in order to maintain bidirectional stochastic
transport of interacting particles in small volumes if site
exchange on the track is not possible?

Several mechanisms can be considered to provide efficient
bi-directional transport. In the following sections, we shall
explore two possible families of scenarios.  One of them would be
that interparticle interactions lead to lane formation. We shall
consider this scenario in section~\ref{sec:interact}.
However, we shall see that it is not obvious that this mechanism
alone could achieve efficient transport in the context of axonal
transport.
Another scenario, perhaps a priori less intuitive, based on the
dynamics of the underlying lattice, will be presented now.
We will show that it could also be a key ingredient for efficient
transport.
The advantages and drawbacks of each scenario will be discussed.

\subsection{Lattice dynamics}
\label{sec:dyn}

As presented in the previous section, generic problems occur when
considering bidirectional transport on a static lattice surrounded
by a confining diffusive reservoir, if
particles interact via exclusion.
On the other hand, real axonal transport is
very efficient although it is bidirectional and molecular motors
exclude each other from MT binding sites.

In the present section, we shall present a mechanism that can
lead to efficient bidirectional transport on a single track
through consideration of the filament
dynamics~\cite{ebbinghaus_a_s2010a,ebbinghaus_a_s2010b,ebbinghaus_a_s2011a}.
This extension of
the model of section \ref{sec:model2} has been inspired by the
experimentally observed dynamics of the cytoskeleton (see
section~\ref{sec:axon_mt_dyn}). In fact, the MTs on
which molecular motors move are themselves highly dynamic, due to
nucleation, polymerization and depolymerization, which occur on
time scales similar to those involved in motor transport and are
thus likely to interfere with the motor dynamics.

Beyond the
interest for intracellular traffic, the present section gives an
example where a dynamically driven jammed phase is hindered by
the lattice dynamics.

\subsubsection{Idealized MT dynamics: model definition}
\label{sec:simplified_dyn}

The model is a modified version of the two-lane model presented in
section~\ref{sec:model2}
(figure~\ref{fig:standard_model}). As a new feature we add some
lattice dynamics for the lower lane, i.e., some sites are
eliminated and recreated. The diffusive upper lane remains
unchanged. On the lower lane, particles can only occupy a site if
this binding site exists. The attachment moves (rate $\oma$) are
consequently rejected if the binding site has been eliminated.
Additionally, a particle will automatically switch to the upper
lane if it makes a forward step (rate $p$) onto an eliminated
site or if the site which is currently occupied by the particle
is eliminated.
The whole system is closed with periodic boundary conditions.
In the following,
the state of the filament lattice is given by
the local lattice states $\tau_i$ which take the value zero if
site $i$ is eliminated and one if the site exists.

Different types of lattice dynamics can be considered.
We shall as a starting point take a very simple one,
and discuss the robustness of our results against
other choices for the lattice dynamics later in
sections~\ref{sec:diff_lattice_dyn}
and~\ref{sec:realistic_MT_dyn}.

\paragraph{Uncorrelated lattice dynamics}

\begin{figure}[tbp]
\begin{center}
\includegraphics[width=0.7\linewidth]{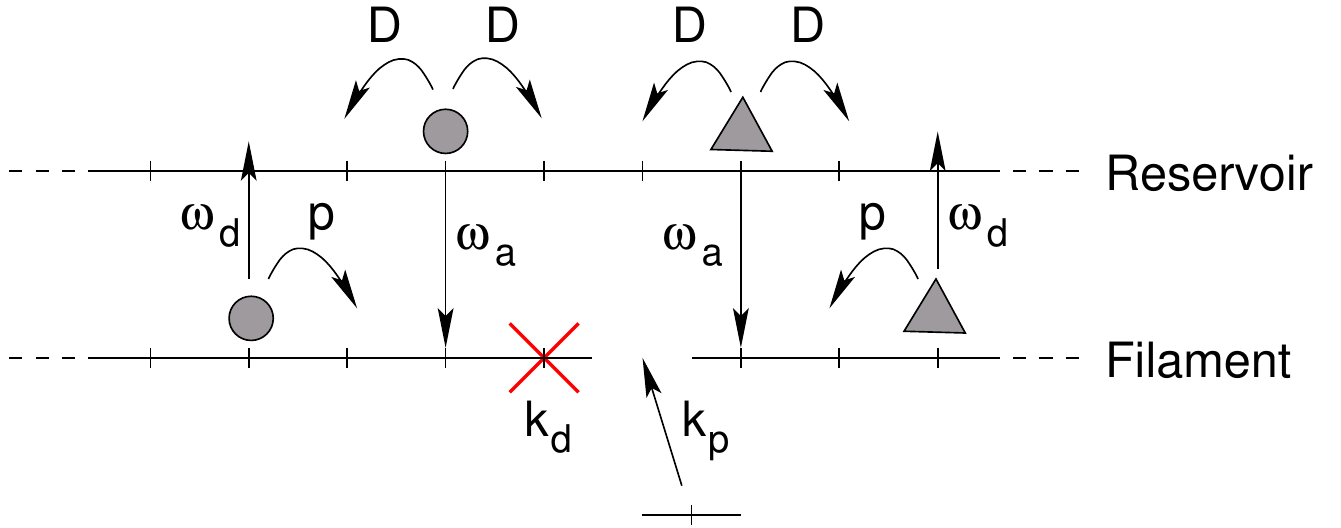}
\caption[Bidirectional model with simplified lattice dynamics.]{
Model for bidirectional transport with simplified uncorrelated
filament dynamics. Same legend as Fig. \ref{fig:standard_model},
with the dynamics of the lower filament added.
A filament site is eliminated at rate $\kd$ and recreated at rate
$\kp$.
From~\cite{ebbinghaus_a_s2011a}.}
\label{fig:basic_model_dynamic}
\end{center}
\end{figure}

This dynamics consists of eliminating randomly a site of the
filament with rate $\kd$
and recreating eliminated sites with rate
$\kp$
(figure~\ref{fig:basic_model_dynamic}) where the indices are
reminiscent of the processes of MT polymerization and
depolymerization.
Here, the evolution of the lattice state $\tau_i$
is not influenced by the particle dynamics and
can be easily expressed as:
\begin{align}
\frac{\partial\langle\tau_i\rangle}{\partial t}=\kp(1-\langle \tau_i\rangle)-\kd\langle\tau_i\rangle.
\end{align}
In the stationary state, one therefore has a homogenous density
of filament sites given by $\langle\tau_i\rangle=\kp/(\kp+\kd)$.

Though this dynamics does not pretend to be realistic,
the link with axonal transport can be thought by considering
that the lower filament does not represent only one MT,
but several MTs following each other end-to-tail.
The elimination of one lattice site would then be equivalent
to the depolymerization of a plus-end MT tip. 
This lattice dynamics could also model
localized obstacles on the microtubules that oblige the
motors to detach \cite{telley_b_s2009}.

It is obvious that a much more complex model should be
developped to model realistically axonal transport.
Here, we rather take this simplistic uncorrelated lattice
dynamics as a minimal tool to demonstrate some possible
generic effects of lattice dynamics.

\subsubsection{Phase transition to a homogenous state}
\label{sec:dyn_model_phase_transition}

We shall first explore numerically the properties of the
model with uncorrelated
lattice dynamics.

The results presented on this model therefore come from numerical
Monte Carlo simulations
obtained over at least $10^6$ Monte Carlo sweeps for
$p=1$, $D=0.33$, $\omd=0.02$ and $\oma=0.33$. Again, $p\gg\omd$
is chosen in order to capture the processivity of molecular
motors. The polymerization is fixed at $\kp=1$ if not stated
otherwise and $\kd$ is usually varied in order to observe the
effects of the lattice dynamics with $\kd=0$ corresponding to a
static lattice. The density of particles is fixed at
$\rhotot=2\rhototp=2\rhototm=1$ in order to have clustering on
static lattices in numerically accessible system lengths.

\begin{figure}[tbp]
 \begin{center}
 \includegraphics[scale=0.3, clip]{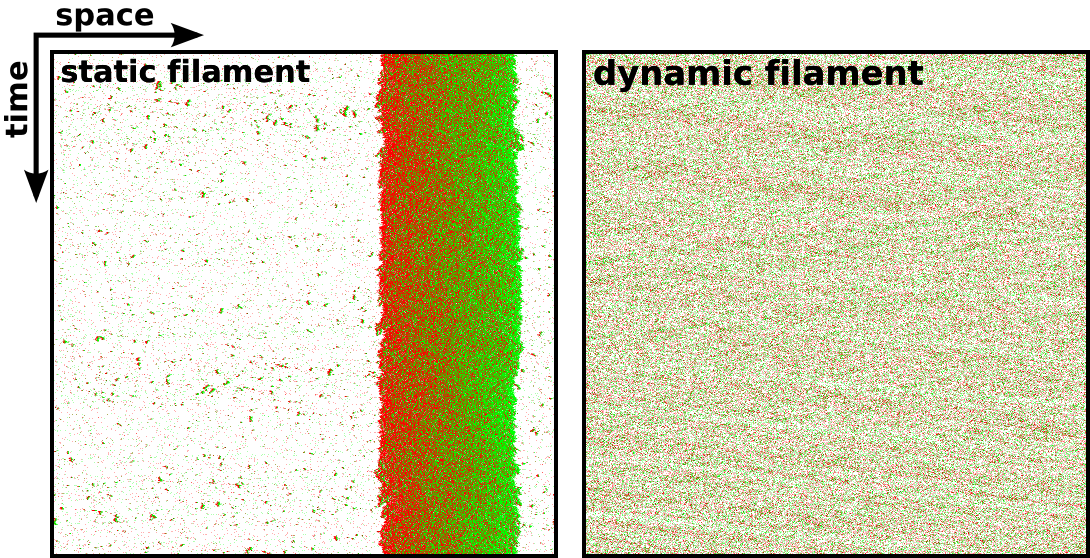}
\caption[Space-time plots on a static and a dynamic
filament.]{Space-time plots on a static and a dynamic filament.
Every line in the images corresponds to the configuration of the
filament lattice in a system of length $L=1000$ with positive
particles in red and negative particles in green. Vacant lattice
sites are white. The static lattice plot has been recorded at
$\kd=0$ and the dynamic lattice plot at $\kd=0.2$,
with uncorrelated lattice dynamics.
From~\cite{ebbinghaus_a_s2010a}.}
\label{fig:dyn_model_space_time}
\end{center}
\end{figure}

\begin{figure}[tbp]
 \begin{center}
 \includegraphics[scale=0.29, clip]{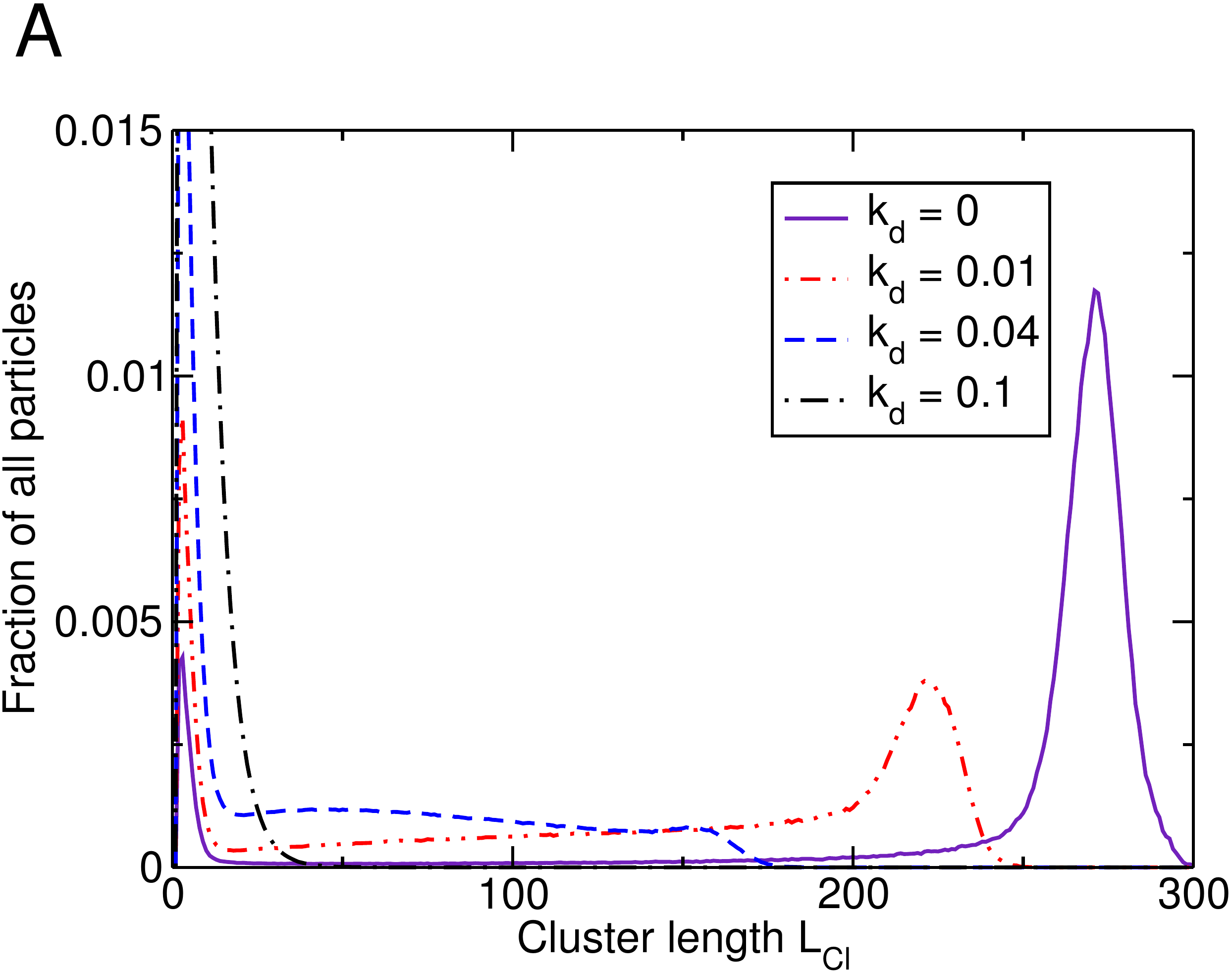}\hspace{0.3cm}
 \includegraphics[scale=0.29, clip]{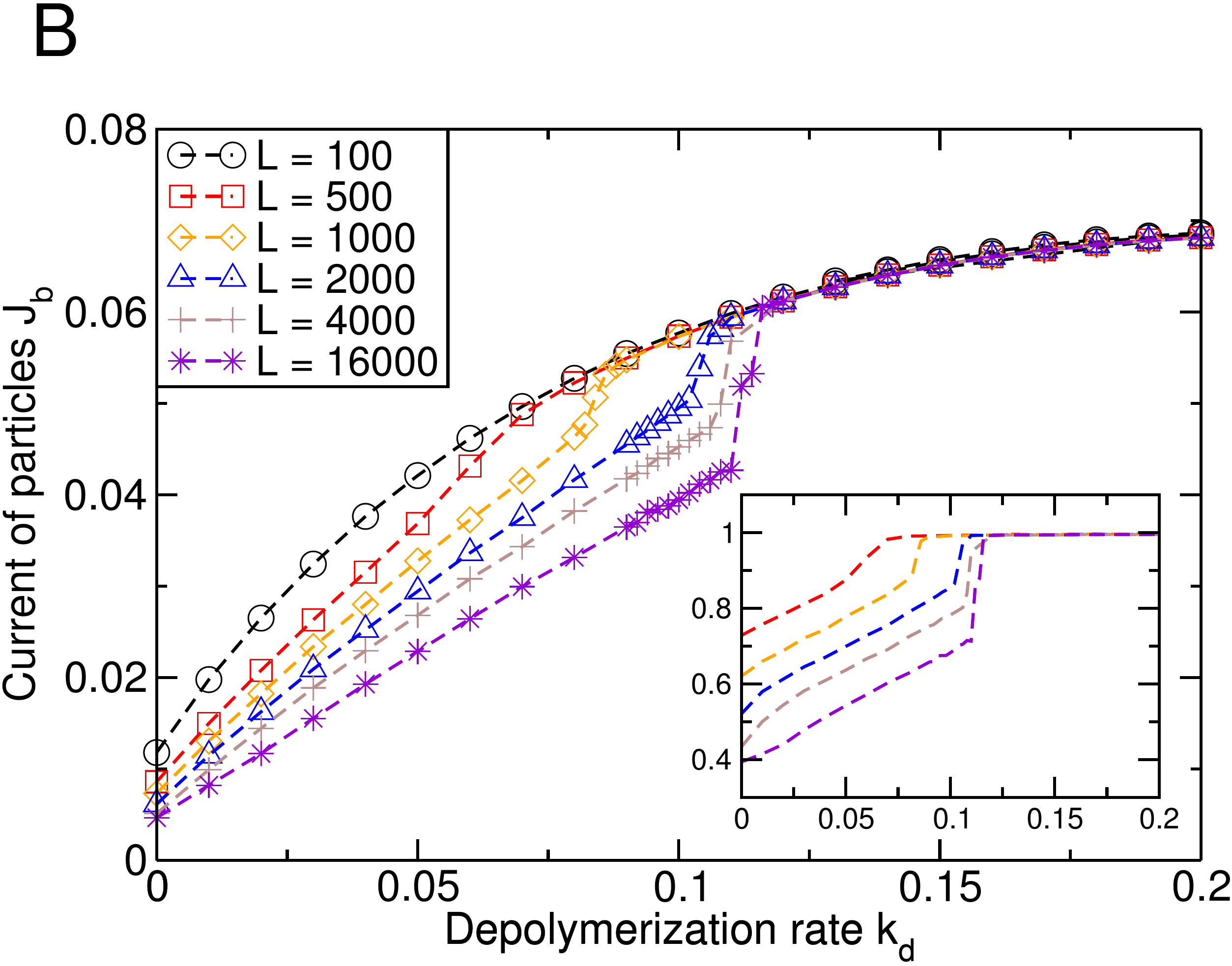}
\caption[Cluster distribution and current with lattice
dynamics.]{Cluster distribution and \textit{(B)} current in the
basic model with additional lattice dynamics. \textit{(A)} The
cluster distribution shows the fraction of all particles in the
system which are found in a cluster of a given size $\LCl$.
\textit{(B)} The current of positive particles bound to the
filament $\Jb$ is shown for different system sizes $L$ as a
function of the depolymerization rate $\kd$. Since both particle
species behave symmetrically, the current of negative particles
takes an identical form. The inset shows the same data divided by
the current in the smallest system $L=100$.
From~\cite{ebbinghaus_a_s2010a}.}
\label{fig:dyn_model_cluster_current}
  \end{center}
\end{figure}

The effect of lattice dynamics is very visible on
the spatio-temporal plots of 
Fig.~\ref{fig:dyn_model_space_time}.
For a static lattice ($\kd=0$),
a macroscopic cluster is formed which seems
quite stable in time.
No such cluster is observed in the simulation
with lattice dynamics (with $\kd=0.2$).
Instead, the spatio-temporal plot 
is rather uniform.
Indeed, the cluster distribution of Fig.~\ref{fig:dyn_model_cluster_current}~\textit{A}
confirms that for $\kd$ sufficiently large,
large clusters disappear.

The central observation is seen on
Fig.~\ref{fig:dyn_model_cluster_current}~\textit{B}, showing the
particle current as a function of $\kd$, which can be seen as the
dynamics strength.
For $\kd=0$, i.e. for a static lattice, the particle
current is quite low and decreases when the lattice size
increases. Actually, as explained in section~\ref{sec:model2},
the flow depends on the total number of particles
in the system. This damaging feature persists for low
$\kd$ values, though for a given system size the current
increases with $\kd$.
The interesting point is that beyond a certain $\kd$ threshhold,
the current becomes system size independent.
Actually it depends then rather on density.

The entire curve is not shown; of course, for very large $\kd$
values, the filament is so much destroyed that the current
must decrease again. So there is an optimal value of $\kd$
somewhere in the middle, which does not depend strongly
on the value for $\kp$.
The disappearance of the large cluster is accompanied by a transition from a size-dependent to a size-independent state (figure~\ref{fig:dyn_model_cluster_current}~\textit{B}).

\subsubsection{Mean-field approximation}

For the stationary state,
in the frame of a mean-field approximation assuming
factorization of correlation functions and
translational invariance,
one obtains the relation between bound and unbound densities
for each motor type
\begin{align}
\oma\frac{\kp}{\kp+\kd}\rhoupm(1-\rhobp-\rhobm)=\left(\omd+p\frac{\kd}{\kp+\kd}+\kd\right)\rhobpm.\label{eq:dyn_model_mf}
\end{align}
The l.h.s. (respectively r.h.s.) gives the flux of particles
attaching to the filament (resp. away from the filament).
One can thus define effective detachment and attachment rates
\begin{align}
\omega_\text{a,eff}:=\oma\frac{\kp}{\kp+\kd};\qquad\omega_\text{d,eff}:=\omd+p\frac{\kd}{\kp+\kd}+\kd,
\end{align}
where the case of the static lattice is recovered as a limit case.
The effective attachment rate is thus
lowered by the lattice dynamics while the effective detachment
rate is increased. The global effect is that the particles have a
lower effective affinity to the filament.

It should be noted however that for a static lattice,
considering only a
lower affinity to the filament has been shown to reduce but not
eliminate the tendency of cluster
formation~\cite{ebbinghaus_s2009}.
The mean-field approach cannot capture the main effect
of lattice dynamics, which prevents the growth of clusters.
Indeed, for the value of $\kd$ that optimizes the flow,
blocking situations with at least two particles of different
species still occur frequently. As a consequence, the maximum
currents that are obtained are about one third of the flux in a
comparable single-species system~\cite{klumpp_l2003}, which is
nevertheless a great improvement compared to the static filament
case.

Further understanding of this system could be obtained
from an analysis similar to the one in \cite{nagar_m_b2006} for
the symmetric motion of non-interacting particles in fluctuating
energy landscapes.
Here the effective `potential' landscape
emerges spontaneously from the particle jamming.

\subsubsection{Rubustness against different lattice dynamics}\label{sec:diff_lattice_dyn}

It is an open question whether the main result of
section~\ref{sec:dyn_model_phase_transition}, i.e., the phase
transition to a density-dependent state with efficient transport
in a finite system is independent of the exact choice of lattice
dynamics.
In this section, in order to test the robustness of
our results, we consider different simplistic lattice dynamics
which incorporate different generic ingredients.

\paragraph{Lattice dynamics correlated with particle occupation}

It was shown that some motors do have an influence
on the dynamics of MTs and contribute to the regulation
of the MT length~\cite{hunter_w2000}, though the
mechanisms of such a regulation are not fully understood yet.
Though it is not clear whether this effect is
relevant for axons,
and though the MT-depolymerizing
kinesins are usually not responsible for intracellular long-range
transport,
this observation inspired another simplified lattice
dynamics correlated now with particle occupation.

The lattice dynamics rules are the same as for
the uncorrelated lattice dynamics, except that now,
a site is eliminated with rate $\kd$
\emph{only} if that site is occupied by a particle.

The behavior of the system with this correlated lattice
dynamics is qualitatively the same as for uncorrelated
lattice dynamics.
The main difference is that the flux in the homogenous phase is
higher than in the first scenario. Indeed, when a motor is
moving freely, it encounters less holes than in the case of the
uncorrelated lattice dynamics, as empty lattice sites cannot be
eliminated.

A variant in which several successive lattice sites must
be occupied to allow depolymerization exhibits also
the transition to a density-dependent efficient state.

\paragraph{Treadmilling lattice dynamics}

We have also considered some idealization of treadmilling
(see section~\ref{sec:mt}), though this type of motion is
actually of minor importance in axons, but not necessarily in
other biological systems: 
In the initial configuration, 
some lattice sites are suppressed so as to create
equally spaced holes on the filament.
During the time evolution, these holes
propagate synchronously but stochastically through the system.
The aforementioned transition toward efficient transport is still
observed. It is to be noted that, since the holes move only in
one direction, the two species of particles are affected
differently.
For both species, there is a considerable increase of the
flux and a maximum current comparable as for previous lattice
dynamics is reached.

\subsubsection{Toward more realistic MT network and dynamics}
\label{sec:realistic_MT_dyn}

The models presented so far are very simplistic
in the description of the microtubule network.
Some more realistic description of the geometrical
organization of the microtubules would be useful
to check whether the results obtained on isolated
filaments still hold.

We have seen that in the axon, transport occurs on
many parallel microtubules.
In some preliminary work, we have implemented the model of
Fig.~\ref{fig:standard_model} for bidirectional transport
on a network of parallel MTs placed on a hexagonal lattice
similar to the axonal MT structure of
Fig.~\ref{fig:neuron_section}.
Besides, the simplified uncorrelated lattice
dynamics of Fig.~\ref{fig:basic_model_dynamic} is applied only
on a finite fraction of each MT length.
Still, the transition towards efficient transport described in
section~\ref{sec:dyn_model_phase_transition}
is still observed.
The next step, under development, is to consider
more realistic MT dynamics, with explicit description
of the dynamic instability and of MT transport.

\subsection{Motor-motor interactions}
\label{sec:interact}

Some other scenarios could be considered to account
for the efficiency of bidirectional intracellular trafic.
A quite natural one would be to include motor-motor interactions.
In particular, one could expect that if these interactions
can lead to lane formation, bidirectional transport could be
quite efficient.

We shall first review some experimental facts about motor-motor
interactions, and then some modeling efforts along these lines.

\subsubsection{Experimental observations}
\label{sec:exp_obs}

There are much more results on mutual interactions for kinesins
than for dyneins, as the latter are more difficult to handle
experimentally.
Still, it is established that 
dynein can quite easily side step to a neighbouring
proto-filament~\cite{wang_k_s1995,wang_s1999,ross2006}, while
there is no clear agreement on side-stepping for kinesin.

In some experiments, it has been shown that kinesins stay tightly
bound to the MT and
wait when being blocked by an obstacle~\cite{seitz_s2006}.
Although the general direction of kinesin motion is along
individual protofilaments~\cite{ray1993,block2007},
recent experiments seem to show that occasional
side steps
onto neighboring protofilaments of the same MT are
possible.
Telley et al~\cite{telley_b_s2009} have decorated some
MTs with obstacles that occupy exactly one binding site
of a kinesin. They observe that
the average run length of kinesins along the MT is much longer
than the average interdistance between obstacles.
Thus kinesins have a low but finite probability to
pass obstacles.
The authors suggest as the most probable interpretation of this
fact that kinesin can change protofilament, though
another explanation could be that kinesins detach and reattach
shortly after, while being kept weekly bound with the obstacle.
Some experiments reported by Dreblow et al~\cite{dreblow_k_b2010}
confirm the ability of kinesin to
bypass obstacles (fixed or slowly moving).
The width of side steps seems compatible with the single
kinesin heads steps~\cite{dreblow_k_b2010,yildiz2004}.
Side-stepping still remains a rare event for kinesin-1,
while some representatives of the kinesin-2 class were
found in vivo to perform left-handed spiraling around
microtubules~\cite{brunnbauer2012}.
Anyhow, the general conclusion is that kinesin side-stepping,
if it occurs at all, 
is far less frequent than for dynein.

Another type of interaction has been evidenced between kinesins.
Indeed, for conventional kinesin, there is evidence for a mutual
attractive interaction between motor proteins, so that a kinesin
preferentially binds to the MT in the vicinity of other
kinesins~\cite{vilfan2001,muto_s_k2005,roos2008}. The strength of
the interaction between two kinesin-1 molecules has been
estimated at $1.6\pm0.5~k_BT$~\cite{roos2008}.

\subsubsection{Modelling modified attachment rate}
\label{sec:modifrate}

\begin{figure}[tbp]
\begin{center}
\includegraphics[width=0.5\linewidth]{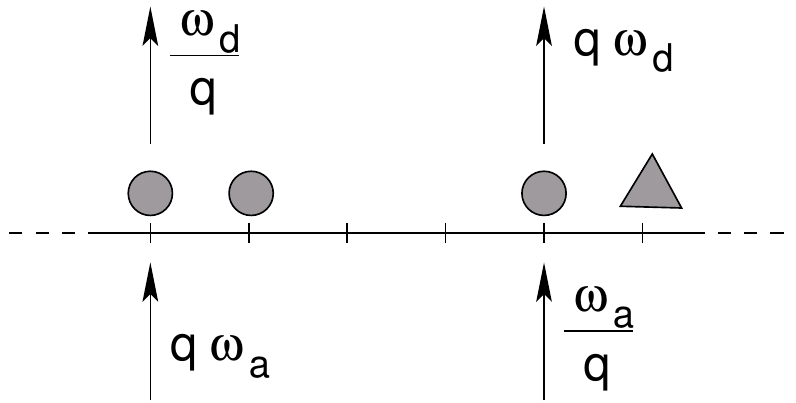}
\caption[Particle-particle interactions in a two-species
TASEP.]{Particle-particle interactions in a two-species TASEP,
as studied in~\cite{klumpp_l2004a}.
The attachment (detachment) rate increases (decreases) by a
factor $q$ if the next site is occupied by a particle of the same
species. In the case of a particle of the opposite species,
detachment is enhanced and attachment is lowered.}
\protect\label{fig:KL_interaction}
\end{center}
\end{figure}

Based on the aforementioned preferential binding of kinesin
near other kinesins, Klumpp and Lipowsky~\cite{klumpp_l2004a}
have proposed a model with modified attachment and
detachment rates along the filament
(figure~\ref{fig:KL_interaction}).
While the motivation for taking this modified rates comes from
the decoration experiments with kinesin mentioned in
section~\ref{sec:exp_obs},
this concept is generalized here to apply
for both particle species which are furthermore supposed to
repulse each other reciprocally.

Klumpp and Lipowsky take as a starting point a two--species
version of the model described in Fig.~\ref{fig:PFF_model} of
section~\ref{sec:PFF} (considered here with periodic boundary
conditions). Without specific attachment/detachment rates,
this two-species version would lead to jamming on the filament.

When the modified rates of Fig.~\ref{fig:KL_interaction} are
used, spontaneous symmetry breaking occurs for $q>\qc$,
for which one particle species is
completely excluded from the filament while the other
undergoes one-species directed motion.
If one considers the possibility of multiple filaments
in the system, some filaments are be populated with
one motor species, and the others by the other motor species.

As we said, this interesting lane formation effect was
found~\cite{klumpp_l2004a} for a reservoir with infinite
diffusion rates, and with unbound number of particles
(grand-canonical reservoir).
However, we found later that for this type of reservoir,
lane formation occurs at such high densities that
transport is not efficient, because of intra-species
hindrance on the filaments~\cite{ebbinghaus_a_s2011a}.

Actually, the behavior of such models depends strongly
on the type of diffusion reservoir that is considered,
as we already showed in previous sections.
Thus we explored the behavior of the model
also for a canonical reservoir (infinite diffusion
but fixed total number of particles) and for single-lane
diffusion reservoir as in Fig.~\ref{fig:standard_model}
(finite diffusion and particle conservation).

\begin{table}
\centering
\begin{tabular}{|c|c|c|}
\hline
 & Static lattice & Dynamic lattice \\
\hline
Infinite diffusion rate & Lane formation~\cite{klumpp_l2004a}. & No lane formation \\
& & for strong dynamics. \\
Grand-canonical reservoir & Not efficient & Current increase only \\
& & due to lattice dynamics.\\
\hline
Infinite diffusion rate & Lane formation. & No lane formation \\
& & for strong dynamics. \\
Canonical reservoir & Very efficient & Then current increase only \\
& at low densities & due to lattice dynamics.\\
\hline
Finite diffusion & No lane formation. & No lane formation.\\
& Very weak current increase & Strong current increase \\
& due to the interaction. & due to lattice dynamics.\\
\hline
\end{tabular}
\caption{Summary of the model behavior
 with the modified
attachment and detachment rates of Fig.~\protect\ref{fig:KL_interaction}.
The central column corresponds to static filaments,
while for the right column, both modified attachment/detachment
rates and uncorrelated lattice dynamics were implemented.
}
\label{tab:summary}
\end{table}

Table.~\ref{tab:summary} summarizes the results that we found~\cite{ebbinghaus_a_s2011a}.
We confirm that, for modified attachment/detachment rates,
lane formation is obtained both for grand canonical
and canonical reservoir.
Lane formation implies efficient transport only in the
case of a canonical reservoir, and if the number of motors is
low enough so as to avoid the filaments to become
crowded.

Lane formation can be obtained only if the interaction strength
$q$ is large enough.
There is no direct measurement of this interaction strength
yet.
Very crude estimates made from the experiments
in~\cite{muto_s_k2005} and \cite{roos2008}
would give too low values.
However, more experimental data would be necessary before
giving a conclusive statement.

The most important drawback of this scenario based on lane
formation is that it does not seem to survive under
confinement~\cite{ebbinghaus_a_s2011a}. However, infinite
diffusivity seems an unrealistic assumption for intracellular
traffic.
Thus there is a interest in finding alternative scenarios
that would work in confined environments.
We explore in the next section some other kind of motor-motor
interactions.

\subsubsection{Modelling side-stepping}
\label{sec:side}

As seen in section~\ref{sec:exp_obs},
dyneins may side-step to a neighboring proto-filament
and it is likely that kinesin can
also do so when meeting an obstacle, though it remains
a rare event.

\begin{figure}[tb]
\begin{center}
\includegraphics[scale=1.0, clip]{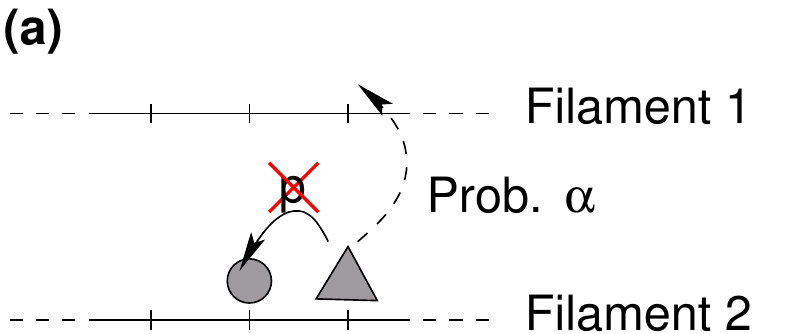}\hspace{3 mm}
\includegraphics[scale=1.0, clip]{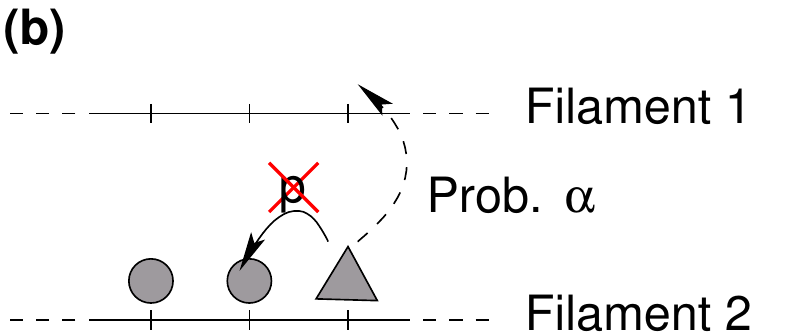}
\caption[Modelling side stepping in front of an obstacle.]{
Modelling side stepping in front of an obstacle.
A particle changes lane with rate $\alpha$ if it is blocked
by
(a) one or
(b) two particles of the other species. The reservoirs are not shown.
From~\cite{ebbinghaus_a_s2011a}.
}
\label{fig:side_steps}
\end{center}
\end{figure}

We consider a variant of the model of Fig.~\ref{fig:standard_model}
with finite diffusion, with several parallel
two-lane subsystems, and with the possibility for motors
to change filament when they meet motors of another species
(see Fig.~\ref{fig:side_steps} for
details of the rules)~\cite{ebbinghaus_a_s2011a}.

When the rule of Fig.~\ref{fig:side_steps}(a) is used,
lane exchanges are too numerous, and no great current improvement
is obtained.
By contrast, the more constrained lane changing rule of
Fig.~\ref{fig:side_steps}(b) does lead to lane formation
and efficient transport.

Interestingly, when uncorrelated lattice dynamics 
is added to these side-stepping rules, it promotes
lane formation such that lane formation can be
obtained even for the rule of Fig.~\ref{fig:side_steps}(a).
Besides, as we already saw in
section~\ref{sec:dyn_model_phase_transition},
lattice dynamics by itself can improve current,
with a higher efficiency than obtained by one-obstacle
side-stepping.

\subsection{Discussion}

The examination of the models in this chapter yields the general
result that it is not trivial to observe bidirectional transport
of particles which move along the same track, if motor exclusion
and finite diffusion are taken into account.

Though the nature of motor-motor interactions
still has to be explored experimentally, we have shown
that it is possible to find some interactions that
make bidirectional transport efficient even under confinement
- an appealing scenario as it proposes a very easy solution to
achieve long-range transport by local interactions.
It should be possible to design experiments with molecular
motors and microtubules to demonstrate lane formation - if the
interactions between molecular motors turn out to be strong
enough.

We also reported in section~\ref{sec:dyn} that a transition
towards efficient transport is obtained when the underlying
lattice itself is dynamical. This rather counter-intuitive
result - transport is more efficient when parts of the
roads are suppressed - still needs to be confirmed
under more realistic modeling efforts of axonal transport.

However, several experimental observations seem to confirm
the strong link between MT dynamics and axonal transport.
In medicine, a possible neuropathy following the treatment of
cancer by MT-stabilizing drugs such as paclitaxel or docetaxel is
a well-known phenomenon~\cite{keime-guibert_n_d1998} and
presumably linked to the breakdown of axonal
transport~\cite{fazio1999}.
Recently, \emph{in vivo} experimental imaging has shown that the
breakdown of MT dynamics induced by the injection of paclitaxel
indeed coincides with the breakdown of vesicle
transport~\cite{shemesh_s2010a}.
Though this does not prove that one is the cause of the other,
these experiments are strong indications of an interplay between
MT dynamics and transport.

An important remark is that scenarios for efficient transport
based on motor-motor interactions or on lattice dynamics should
not be thought as competing scenarios.
On the contrary, we found in many cases that both
contribute in a cooperative way to the global efficiency of
transport.

In this chapter, we have discussed that efficient transport is
based on the control of jammed configurations which naturally arise in
bidirectional transport under strong confinement. 
A full modeling of motor driven transport
in axons is not accessible yet since much information is still
lacking, concerning for example motor interactions or the
details of axonal MT dynamics. A few of the open questions will
be listed in section~\ref{sec:challenges}.
However, many informations should become available in
the coming years and allow for more accurate modeling.

\section{Challenges for the development of transport oriented models}
\label{sec:challenges}
In the present chapter we try to address some important open problems
that may have strong impact on the transport of cargos in cells.
Though these problems are general for all cells,
they are particularly relevant for axons, 
which are so long and so crowded that active transport is essential.
Therefore, our discussion will be oriented towards axonal transport
but the reader should keep in mind
that most problems that are considered here are also encountered
in other cell types.

\subsection{Cargo transport in axons and protein synthesis}
\label{sec:proteins}

The axon provides a crowded environment where a large variety of cargos has to be transported.
A cargo is often not transported by a
single but by teams of molecular motors. 
This is in particular true for large objects.

We have seen in section~\ref{sec:tug} that even the transport of a
{\em single} cargo by teams of motors is 
not fully understood yet. 
Once the characteristics of bidirectional cargo transport will be established 
the next task will be to consider the {\em collective effects} of such
cargo-motors complexes.

These have been partially considered in chapter~\ref{sec:vivo},
in which a particle can be seen as a coarse grained description
for a whole motor-cargo complex.
However, the bidirectional nature of the cargos' trajectories
and its role in collective motion has to be explored further.
In particular, the tendency to have jamming can depend on the
variability of trajectory bias between different cargo types.

The dynamics of cargos can also be considered from a functional
point of view. The cell needs a certain number of proteins
to be synthesized, transported at the right place where they
will be used, and then recycled or degradated.
Miller and Heidemann \cite{miller_h2008} list
a number of open questions related to transport and synthesis of proteins. In particular,
it is not clear yet how very long axons are
supplied with proteins. It is also not known
where protein degradation occurs along the axon, if it is 
relevant at all.

The original picture was that all proteins
were synthesized in the soma, and then transported
where needed. But several experiments have indicated
that mRNA replication
also occurs locally along the
axon~\cite{si2003,piper_h2004,sinnamon_c2011}.
There are still many open questions about the
spatial localization of axonal protein synthesis - e.g., does
it occur in specific axonal compartments? - and
about the triggering of mRNA translation~\cite{piper_h2004}.
The tip of the axon in particular should be a place
where the translation of mRNA occurs, to comply
with the needs of axonal growth.
Protein synthesis at the distal end
cannot be completely autonomous from the cell body,
and some retrograde signaling transport is needed,
for example to inform the cell body that the
axon has been injured~\cite{vuppalanchi_w_t2009}.
Thus protein synthesis and transport are necessarily
highly linked, and much more understanding of
these coupled phenomena is needed.

\subsection{The transport network: microtubule dynamics in axons}
The recent discovery that a large number of MT plus-ends are dynamic in axons,
even in  mature neurons, considerably changed 
the picture of the axonal MT network, as we already pointed out in the previous chapter. This finding 
immediately raises questions about the global dynamics of the axonal MT network.

In axons this question is not easy to answer, since the minus-ends are not bound to a MTOC as in other 
mammalian cells. Therefore, it would be very interesting to visualize the spatio-temporal evolution of 
the MT network in axons. However, in contrast with cells where the MT network is very sparse and
where a simple marking of tubulin allows to track the network structure and dynamics,
this is not possible in axons due to the high density of MTs.
A possible solution to this problem would be labeling of the plus- and minus-ends rather than 
the entire MTs.
Until now, most of the {\em in vivo} experiments 
concentrate on the growing ends of MTs, which can be traced by a number of 
standard labels.
By contrast, the information about the spatial and temporal distribution of 
microtubule minus-ends in axons is very sparse.
Indeed, it is still an experimental challenge to find how to track
the minus-ends or the shrinking plus-ends of MTs.
The recent finding of families of proteins that specifically
bind to the MT minus-ends opens new perspectives for the
marking of minus-ends~\cite{jiang2014}.
In the future, the simultaneous tracing of plus- and minus-ends would allow to get
much more insight to the actual distribution of static and dynamic microtubules in axons.

Another issue is related to the role of steric interactions
in MT network dynamics. It is known that obstacles
can bend MTs, or limit their growth by inducing catastrophes.
As the axon is very crowded a better knowledge of its inner structure
would be necessary to take these steric interactions into account
in a model of MT networks dynamics in axons.

The polymerization of MTs relies obviously on the supply of tubulin subunits. As growing 
MT plus-ends have been observed in every part of the  axon, there is a need to provide 
new material to sustain the MT-structure over the whole lifespan of the axon. 
There is, however, an ongoing debate about how and under which form tubulin is  transported.
In order to illustrate possible mechanisms of tubulin supply we discuss some of the 
mechanisms that are presently discussed.

As already mentioned in section~\ref{sec:proteins},
the historical view is that proteins can only be synthesized in the cell body,
and then transported to a location where they are needed.
The length of axons prevents transport to be performed only by diffusive means, and
active transport has to be organized~\cite{morfini2012}. It is possible that short assembled 
MTs are transported to the plus-end of growing MTs in the axon. Alternatively only tubulin 
subunits could be transported and then assembled at their arrival point.
It is unclear whether there is a nucleating structure allowing for free tubulin to assemble
into MTs directly in the axon. However, it is known from in vitro experiments that
tubulin can spontaneously assemble even in the absence of a nucleating site if the tubulin 
concentration is high enough~\cite{janke_b2011}.
More recently it was found that minus-ends can be stabilized
even if not connected to any MT nucleating site,
through decoration by a specific protein family
(CAMSAP)~\cite{jiang2014}.

Some experiments seem to indicate that MTs can be transported as a 
whole~\cite{baas2002,ahmad_b1995,ahmad1999}.
Indeed, one knowns from other biological systems that
molecular motors are able to displace MTs (e.g. in
the mitotic spindle). The mechanism is the following:
motors produce some relative motion between the cargo they carry and the MT
they are stepping on.
If the MT is strongly anchored, the cargo will move.
However, if the cargo is fixed, for example 
if the cargo domain of the motor is attached
to the actin cytomatrix of the axon,
then it is the MT that could move~\cite{baas2002}.
Under such a scenario, cytoplasmic dynein
could move the MTs either anterogradely or
retrogradely depending whether its cargo domain
is attached to the actin cytomatrix or to other
MTs, though anterograde motion should dominate.
Indeed, experiments have confirmed the crucial
role of cytoplasmic dynein in MT movement~\cite{ahmad2006}.

The so-called \emph{cut-and-run} model has been conceived by Baas
\emph{et al.}~\cite{baas_k_q2005} to explain the maintenance of
the stationary state of the axonal MT network: MTs
nucleate and polymerize at the centrosome in the soma and
are then cut by a MT-breaking protein such as katanin. 
Katanin has indeed
been shown to be located in higher concentrations near the
centrosome~\cite{ahmad1999}. Short MT pieces are then
transported through the axon with their plus-end pointing
in distal direction~\cite{ahmad_b1995,baas1997}. The motility is
achieved through molecular motors, which are anchored on
actin filaments or longer MTs and exert forces on the
transported MT by walking on it. It has in fact been shown
that dynein and the actin cortex are responsible for half
of the anterograde MT transport in axons while the rest
depends on kinesin~\cite{myers2006,ahmad2006}. One thus finds
two populations of MTs in the axon: short and mobile MTs
which are transported and perform bidirectional motion and
long and immobile MTs which serve as structures along
which the short MTs can be transported. The long MTs are
imagined to be immobile because of a large number of
crosslinks to other axonal structures such as the
neighboring MTs via tau~\cite{ahmad2006,baas_n_m2006}, though
actually they are still active at their plus-end and might
depolymerize at any moment. Experiments are contradictory
whether the short MTs undergo dynamic instability during
transport~\cite{baas1997,wang2002} or whether they are
stable~\cite{ma2004,ahmad2006}.

Furthermore, it is an open
question if and how tubulin is transported retrogradely.
Such motion has not been observed yet, but might be needed
to balance the constant influx of tubulin in form of
anterogradely transported MTs. Alternatively, tubulin
might degrade at the synapse or be released to the
extracellular space from there~\cite{paggi_l1987}.

The conclusion of this section is that 
we are still lacking some global image of the
living cycle of MTs in the axon, although it is confirmed 
that the plus-ends of axonal MTs are dynamic~\cite{shemesh_s2010a}.
A more conclusive picture of the axonal MT-network  
would greatly improve the understanding of motor-driven transport in axons 
and support the further development of transport models.

\subsection{Motors action on MTs}

We have seen in chapter~\ref{sec:vivo} how the MT dynamics could
play a role in motor transport. Reciprocally, motors can
influence MT dynamics.

Some other kinesins than kinesin-1 have been shown to be
less implicated in intracellular transport but to
have an influence on the MT dynamics. For example,
members of the kinesin-8 and kinesin-13 family are able to find
the plus-end of MTs (and are therefore considered as
+TIPs~\cite{akhmanova_s2008}) where they destabilize
MTs and promote depolymerization by using the energy
from ATP hydrolysis to deform the MT lattice until
it breaks~\cite{desai1999,moores2002,tan2006}.

Exclusion processes have been used to model
the collective behavior of such
depolymerases~\cite{reese_m_f2011}.
The model predicts some collective effects
through the jamming of polymerases, 
that would affect the depolymerization speed.

Other exclusion process based models
have demonstrated how the activity of plus-end tracking
molecular motors acting as depolymerases
can regulate the length of filaments,
which becomes then less sensitive
to variations of the surrounding medium.
These length regulation mechanisms were found both
for treadmilling filaments~\cite{johann_e_k2012},
or for filaments attached to 
an MTOC~\cite{melbinger_r_f2012}.

Molecular motors can also have a stabilizing effect.
For example, cortical dynein can bind to MT plus-ends
and stabilize them, with a length regulation effect,
as was confirmed by several \emph{in vitro} experiments~\cite{hendricks2012,laan_r_d2012,laan2012}.

If MT network dynamics plays a crucial role in transport
as proposed in chapter~\ref{sec:vivo}, the features
presented in this subsection provide a feedback loop
on MT dynamics. The transport of tubulin by motors
 - needed for MT polymerisation - is another one.

\subsection{Modification of MTs by MAPs : The example of tau proteins}
\label{sec:tau}

The MTs that we have considered until now were ideal ones, without
any defect or any variation in their surface structure. This is not the case {\em in vivo},
where enzymes can modify tubulins units~\cite{janke_k2010},
or where MAPs can associate to the MT.
The cell could possibly use inhomogeneities
or MT decoration factors
to control the motor transport properties.
All these aspects should be integrated for a full description
of the interplay between transport and tracks.
In this section we shall discuss more thoroughly a specific
type of protein that binds on the MTs, namely tau proteins.
A similar role is played in dendrites by MAP2
proteins~\cite{bernhardt_m1984,binder_f_r1985}.

The structure of both tau and MAP2 proteins has a MT binding
domain and a hydrophobic part which extends into the
surrounding cytosol~\cite{lewis1989}. The MT binding
domain is able to bind at the outside of the MT
cylinder as well as across
protofilaments~\cite{al-bassam2002,santarella2004}.
The interactions between the hydrophobic parts
of MAPs induce constraints
on the structure of MTs (see Fig.~\ref{fig:neuron_section}).
It is still debated whether these MAPs are
``crosslinkers or spacers of microtubules''
\cite{marx2000}.
On the one hand, the hydrophobic
parts of bound tau molecules are able to dimerize and
to induce bundling of
MTs~\cite{rosenberg2008}.
Cross-links have been observed both \emph{in vitro}
\cite{aamodt_c1986} and \emph{in vivo} \cite{hirokawa1982,lewis1989}.
On the other hand steric constraints between
MT-MAPs are responsible for repulsive forces
between MTs \cite{shahpasand_a_r2008}.
The hydrophobic part is very flexible and can be
described in terms of a polymer brush model
\cite{mukhopadhyay_k_h2004}. The
interdistance between MTs in the axon is directly
related to the effective length of tau proteins
\cite{chen1992,frappier1994}.

Phosphorylation plays a key role to regulate
the various functions of tau or MAP2 proteins.
For example, highly phosphorylated tau or MAP2 proteins are more rigid.
Then, as a consequence of higher repulsive forces,
the interdistance between MTs is increased both in
axons and dendrites
\cite{mukhopadhyay_k_h2004}, a feature which can even result
in dendritic branching \cite{friedrich_a1991}.
It is also known that 
phosphorylation of tau may reduce its affinity to MTs~\cite{drechsel1992,drewes_e_m1998}.

Since kinesins and tau proteins both bind on the
outside of the MT cylinder, it is not unlikely that
both proteins interfere with each other. In fact, it
has been shown \emph{in vitro} that kinesin's
attachment to the MT is hindered by tau, whereas
kinesin is not affected in its motion along the
filament once it is bound, i.e., processivity and
walking speeds are unaffected~\cite{seitz2002,trinczek1999,yuan2008a}. On the
other hand, another study found that kinesin had a
higher probability of detachment when encountering tau
patches on the MT~\cite{dixit2008}.

As a consequence of the fact that tau proteins may prevent locally kinesins
to bind, tau proteins have been also shown to favor switches of cargos from one
microtubule to another one \cite{ross_a_w2008}, a feature which may be helpful
in crowded environments.

The interaction of dyneins with tau proteins are similar to those reported for kinesin: Run lengths are decreased at high concentrations of tau on the MT~\cite{seitz2002}. In contrast to kinesin, dynein does not detach at encounters of tau but tends to reverse direction, maybe by switching to the diffusive state of motion described above~\cite{dixit2008}. Altogether, dynein is much less affected by the presence of tau than kinesin and is only seriously impeded in its motion at overexpression of tau~\cite{dixit2008,vershinin2008}.

Tau furthermore has a stabilizing effect on MTs by modifying parameters of the dynamic instability of MTs~\cite{drechsel1992}.  In the presence of tau, the speed of polymerization is increased while that of depolymerization is decreased. Also, the frequency of catastrophes is lowered and the frequency of rescues is increased. \emph{In vitro} grown MTs thus reach much higher maximum lengths as a consequence of the action of tau.
Still, in normal axons, the MT network is highly dynamic and it is only in
some pathologies that the tau concentration becomes high enough to
prevent network dynamics.

MTs in axons have been found to be particularly densely decorated by tau proteins at regions of high curvature, suggesting an influence of the mechanical properties~\cite{samsonov2004}.  It was also hypothesized that tau has a protective function for MTs against the action of the MT-severing protein katanin~\cite{qiang2006}.

\subsection{Structural defects of the MT network }

Until now, we have not considered that, actually,
structural defects of the MT network appear spontaneously
in normal axons. Though a large majority of
 MTs are oriented with their plus-end pointing anterogradely,
it seems that $15$ to $20$ \% of the MTs 
have their plus-end pointing retrogradely~\cite{shemesh_s2010a}.
Besides, the whole structure inside the axon includes
not only MTs but also neurofilaments and actin filaments.
It is not clear whether a minority of filaments
oriented differently
could not have a strong effect on the overall
transport properties, for example by allowing
blocked vesicles to go backwards to try another path
forward.

Structural defects and axonal transport failure
also appear in many neurodegenerative diseases (see~\cite{roy2005}
for a review).
The most prominent example in this
respect is Alzheimer's Disease~\cite{cash2003,goldstein2001,hirokawa_t2004,seog_l_l2004}.
Axonal swellings caused by
accumulations of MAPs, molecular motors, organelles and vesicles
have been found in early stages of Alzheimer's Disease, suggesting failure of
axonal transport to be linked to the progress of the
disease~\cite{stokin2005}.
In particular, tau accumulations in neurofibrillary tangles
is a hallmark pathology of Alzheimer's Disease so that
overexpression of tau is hypothesized to be implicated in the
failure of axonal transport.

As mentioned in the previous section, one of the effects of tau proteins is to lower the attachment rate of kinesins. 
We have discussed in section~\ref{sec:disorder} that binding defects can affect
collective motion of particles only if present at almost all sites.
This remark may brings a first clue to explain why moderate tau concentrations are not deleterious while tau overexpression is.
Indeed, since tau has important stabilizing functions for MTs, it has to be present
in normal axons and thus it is desirable that its action as binding defect for kinesins does not affect the overall transport properties if tau is not expressed at extreme concentrations.

A model of the cascade leading from tau overexpression to neuronal
degeneration has been proposed by Baas and Qiang~\cite{baas_q2005},
based on some experimental results by Mandelkow \emph{et al.}~\cite{mandelkow2004}.
As already mentioned,
MT-bound tau interferes with the attachment of both motor families,
kinesin and dynein, but the effect is more pronounced for
kinesin~\cite{trinczek1999,seitz2002}.
At normal tau concentrations, this may allow tau to regulate
the directionality of axonal transport~\cite{dixit2008}.
In the case of
abnormal tau overexpression, this proposed regulation fails and
anterograde transport is almost completely suppressed.
Baas and Qiang~\cite{baas_q2005},
following Mandelkow \emph{et al.}~\cite{mandelkow2004},
propose that the neuron reacts to this blocking of the anterograde
transport by hyperphosphorylation of tau which causes it to
dissociate from the MT filament and thus clears a path for
kinesin-driven transport. The hyperphosphorylated tau eventually
forms the abnormal filaments mentioned above as a hallmark of Alzheimer's Disease.
Additionally to restoring the axonal transport, the loss of tau on
the MT makes it more susceptible to severing by katanin 
causing a disruption of the MT network and
neuronal degeneration.

In line with this model, a massive reorientation of the MT network was identified under overexpression of tau in \emph{Aplysia} neurons~\cite{shemesh2008}. The reorientation might be caused by the elongation of shorter MTs which have been cut by katanin and are brought out of the parallel orientation by steric interaction or molecular motors which pull on the MT fragments. As a consequence of the MT network reorientation, axonal transport is impaired in both directions as swirls in the MT network act as attractive regions for molecular motors so that accumulations of axonal cargos occur as is observed in Alzheimer's Disease.

The fact that it is possible to reproduce in experiments
some swellings similar to those of taupathies, 
either by using drugs~\cite{shemesh_s2010a} or mutations~\cite{shemesh_s2010b} that impair MT dynamics
and vesicular transport, should allow
learn more not only for the understanding of neuronal
diseases, but more generally of the regulation
of intracellular transport.

The knowledge of how an axon regenerates itself when transected
brings also some information about how the MT network
organizes itself.
In~\cite{erez2007}, an experiment shows the transformation
of a transected axonal tip. The experiment allows to visualize
the supply in vesicles - a crucial ingredient for the regeneration
of the neuron - and the orientation of newly growing MTs.
The self-organization of a vesicle trap has been modeled
in~\cite{greulich_s2011}, where the 
active transport of particles on a filament networks under confinement has been considered. 
It has been shown that, by applying confining boundary
conditions, a self-organization of the network towards a
polarized structure is induced, similar to the experimental one,
even without explicit
regulation and interactions.

\subsection{Global view of model/experiments interplay}
\label{sec:scheme_conclusion}

\begin{figure}[htbp]
\begin{center}
\includegraphics[height=0.6\linewidth]{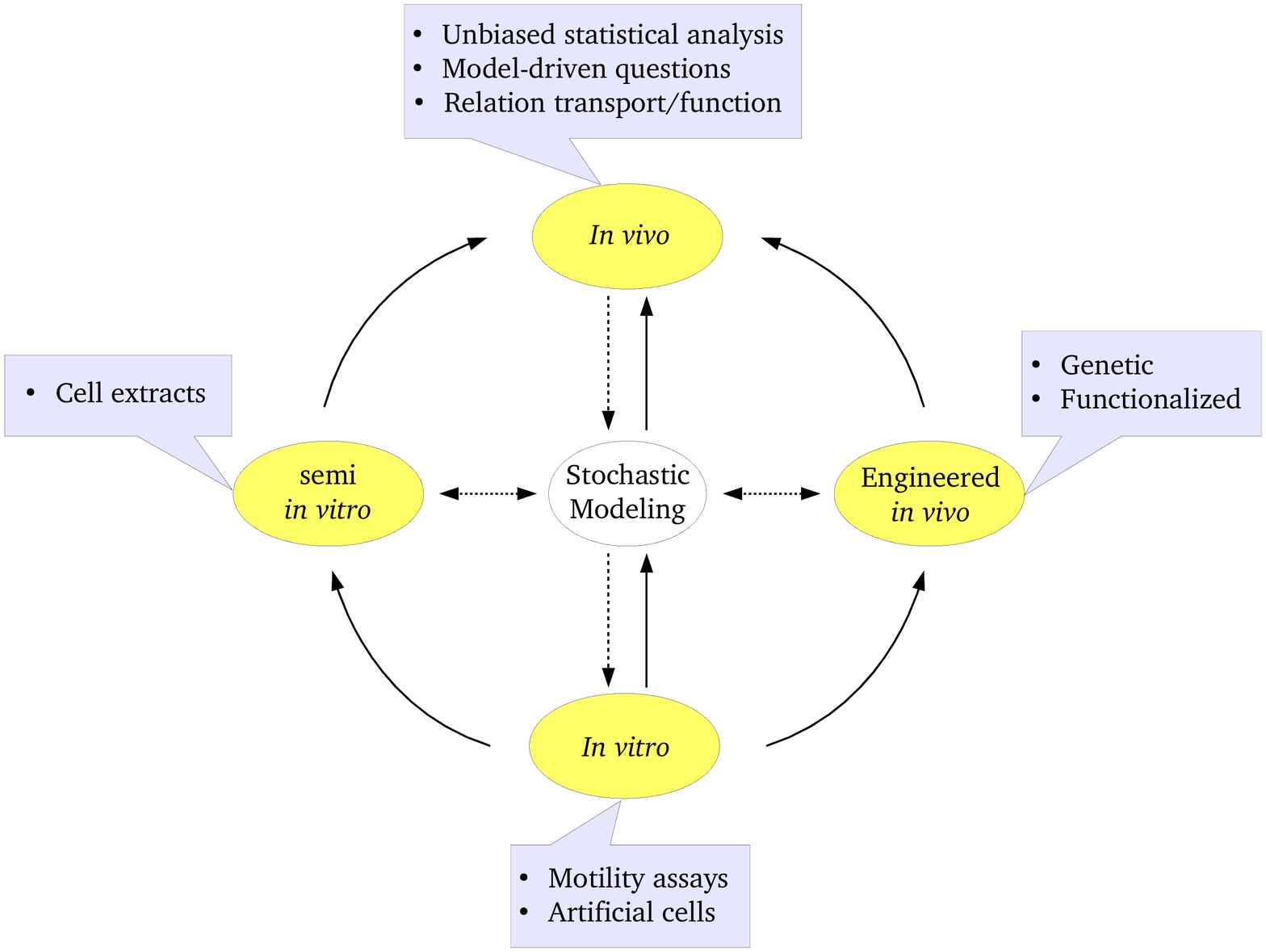}
\caption[sketch experiments and models]{
Schematic representation of the different types
of experiments, with their interplay with theoretical
modeling.
}
\protect\label{fig:scheme_conclusion}
\end{center}
\end{figure}

To conclude this review, we would like to give a picture of
the current interplay between modeling approaches
\footnote{
Note that throughout this review, {\em model} or {\em modeling}
refers to theoretical modeling and simulation of models,
in the sense of physicists.
In biology, a {\em model} can also refer to
animal models (for example the mouse is widely considered as
a good model to study human diseases), and more generally
to all kind of model systems. 
An example in the frame of the current review
is melanophores, which turn out to be a good model system
to study organelle transport~\cite{aspengren_h_w2007}.
}
and experiments.
The aim is to understand transport in real cells,
but intermediate steps are needed for that.
Fig.~\ref{fig:scheme_conclusion} summarizes various
types of approaches to tackle this problem.

{\em In vitro} assays are useful to fully characterize
the elementary units of transport, for example motors and MTs.
Compared to {\em in vivo} experiments, they allow for a better control. 
One may argue that {\em in vivo}, systems are much
more complex and cannot be reduced to a sum of elementary
units. Still, {\em in vitro} experiments are ideal to test
the predictive power of models: building blocks are controlled,
and the resulting behavior can be compared between experiments
and model simulations.

While motility assays have allowed to fully characterize
purified kinesin since a long time,
similar results should soon be available for dynein,
as it has been recently discovered how to activate
{\em in vitro} purified mammalian dynein~\cite{mckenney2014c}.
This also opens the way for building simple
cargo-motor complexes~\cite{derr2012,furuta2013,arpag2014}
on which tug-of-war models' predictions
could be tested.

One may argue that {\em in vivo}, motors are not isolated
but supplemented by many factors which may change their
behavior. In order to include these factors, one may not
start from purified motors, but directly work with
the whole complexes obtained from cell extracts,
as done for example in~\cite{barak2014,mallik2013} or~\cite{hendricks2010}
in {\em semi in vitro} experiments.

Of course the final aim is to understand {\em in vivo}
transport inside cells.
Though some measurements can be performed directly in
the cell, the interpretation of such measurements is
in general difficult as one cannot fully control
the setting.

Here we would like to suggest two directions where progress
could be made.
The first one concerns measurements of cargo trajectories.
These trajectories are often cut into runs and pauses.
A lot of information is then lost which is of particular 
relevance for the competition of motors in case of 
bidirectionally transported cargo. Besides, the definition
of runs and pauses requires some threshold-based sorting
that may introduce some sensitivity of the results to
unwanted features. Another frequent bias of data analysis
is to extract extreme events which in fact are rare and
not representative.
Systematic unbiased statistical study as the one performed
for melanosome velocity in~\cite{snezhko2010}
would be highly valuable.

Another promising direction of research is to link
transport and functions in a more thorough way.
Logistics impose some constraints on transport
and signaling. Some cell regions may be specialized,
e.g., through specific decoration of MTs
having an impact on transport properties.
More generally, even if models lack firm ground 
{\em in vivo}, they may still help to orient experimental
measurements by providing relevant questions.

Another interesting perspective for the understanding
of {\em in vivo} systems is to modify them.
A classical way to do it is {\em via} genetic manipulation,
that allows either to test a mechanism by modifying
one of the actors involved, or to produce cell lines
mimicking for example the phenomenology of a given
disease.

Genetics can also provide living systems with simpler
characteristics, as for example the mutant filamentous
fungus {\em Ustilago Maydis} studied by G.~Steinberg's group~\cite{schuster2011a,schuster2011b},
which has a much simpler and sparse cytoskeleton than
most cells, and besides is quasi one-dimensional
(a feature helpful for the modeling task).

New types of engineered cells can now be obtained as well.
L. Kapitein et al~\cite{kapitein2010,kapitein2013} managed to
control the attachment of motors to some cargo (peroxisome),
by inserting some molecular switch between the motor
and the cargo.
Then a simple light exposure can induce attachment
of given motors to the cargo.
Focusing light on a precise region allows to perform
this attachment only locally.
Cargos can thus be displaced (using kinesin or dynein)
or anchored (using myosin V) at will in specific regions.
This opens the possibility to test {\em in vivo}
various hypothesis, concerning transport itself or
its relation with various functions.

Coming back to {\em in vitro} systems, we finish
our "tour" of the sketch of Fig~\ref{fig:scheme_conclusion}
by noting that, aside the classical motility assays,
it is now possible to build artificial cells, i.e.
systems obtained from purified units that are combined
to build a specific machinery. 
In M. Dogterom's group for example, some {\em in vitro} experiments
have been realized, to study how dyneins attached
to an artificial cortex can capture MT plus-ends,
stabilize them, and mediate forces~\cite{laan_r_d2012,laan2012}.

Numerous experimental results of all the types described
above are expected to come out in the coming years, 
generating a need for further modeling,
and we hope this review will trigger more interest for these
systems.

\section{Conclusion}
\label{sec:conclusion}

The application of stochastic transport models to intracellular transport driven
by molecular motors stimulated a new and very active branch of research. 
Compared to typical applications 
of stochastic transport models, as for example vehicular traffic, several fundamental 
differences exist: The motion of motor proteins is stepwise such that lattice models are 
not an approximation but a natural description of the system. Moreover, the number 
of particles is not conserved - a particle property which is influencing the macroscopic 
behavior of the system.
While stochastic transport models are often defined
on a one-dimensional lattice, intracellular transport requires
a better understanding of transport on several parallel lanes
and on a whole variety of networks.
Finally, the networks which carry the active particles themselves
are dynamic, a feature that modifies the transport properties.
Conversely,
the network dynamics and structure can be influenced by the activity of molecular motors. 
There is growing evidence that the interplay between the cytoskeletal structure and transport is crucial.
Implementing into theoretical models 
the aforementioned characteristics of intracellular transport systems
has shown their large impact on the transport capacity and particle
distribution. However, there is still a need to understand the
mechanisms by which this occurs.

Indeed, despite the progress already made, the theoretical understanding of intracellular transport is only in
its infancy and still several open problems of fundamental interest exist.
Gaining insight in those would help to address questions of strong
social impact. 
The most prominent 
one is perhaps the relation between perturbation of axonal transport and
many neurodegenerative diseases (see~\cite{roy2005} for a review). Perturbation of axonal transport 
may be induced by altering the structure of the MT network or the cargo dynamics.
For example, mutations of the protein huntingtin, which influences the directionality of vesicles transport in 
neurons~\cite{colin2008}, may cause Huntington's disease~\cite{borrell-pages2006}. Similarly, transport 
defects have also been related to  Alzheimer's Disease~\cite{cash2003,goldstein2001,hirokawa_t2004,seog_l_l2004}.

In order to fully establish the mechanisms that are responsible for neuropathology observations it is necessary 
to gain a better understanding of unperturbed axonal transport or,
to phrase it in a more focused way,
of bidirectional 
motor-driven transport under strong confinement.

We expect that the understanding of intracellular transport processes will dramatically increase by recent progress on the experimental characterization of single motor properties. It is now possible to design rather complex transport systems in vitro in a very controlled way. The results of these experiment can be used to parametrize and validate theoretical models of intracellular transport. 

From our point of view the interplay between lattice dynamics and transport is a key issue for
biological applications of stochastic transport. Future contributions to this subject may have also 
strong impact beyond biological applications and involve systems as different
as internet or electric networks~\cite{holme_s2012}.

\section*{Acknowledgments}

We would like to thank our collegues that
have contributed to our better knowledge of
the field - we thank in particular
Philip Greulich, Sarah Klein
- and those who have accepted to spend their
time to read the manuscript and help us with
their valuable advises - in particular
Ksenia Astanina, Rosemary Harris and Gunther Sch\"utz.
We also thank Daniel Bahr for his help,
in particular for some of the figures.

This work was supported by the Deutsche
Forschungsgemeinschaft (DFG) within the collaborative
research center SFB 1027 and the research training group GRK 1276.

\section*{References}

\bibliographystyle{elsarticle-harv}

\end{document}